\documentclass[journal]{IEEEtran}
\usepackage{cite}
\usepackage{verbatim}

  \usepackage[pdftex]{graphicx}
  \usepackage{caption}
  \usepackage{subcaption}
  \usepackage[font=scriptsize]{subcaption}
  \usepackage[font=scriptsize]{caption}
  \usepackage{epstopdf}

\usepackage[cmex10]{amsmath}
\usepackage{mathrsfs}
\usepackage{amssymb}
\usepackage{amsfonts}
\usepackage{relsize}
\usepackage[capitalise]{cleveref}
\usepackage{cleveref}

\usepackage{amsthm}

\newtheorem{lemma}{Lemma}
\newtheorem{remark}{Remark}
\Crefname{figure}{Fig.}{Figs.}

\usepackage{array}

\usepackage{mdwmath}
\usepackage{mdwtab}

\usepackage{stfloats}

\usepackage{amsfonts}
\usepackage[fleqn,tbtags]{mathtools}
\usepackage{psfrag}

\usepackage{multirow}

\def\dashfill{\cleaders\hbox{-~-}\hfill}

\usepackage[]{algorithm2e}

\newcommand{\blue}{\color{blue}}
\usepackage{color,setspace}
\definecolor{blue}{rgb}{0.2,0.3,0.8}
\newcommand{\red}{\color{red}}

\def\tr{\mbox{${\mbox{tr}}$}}
\def\dashfill{\cleaders\hbox{-~-}\hfill}
\makeatletter
\newcommand*{\rom}[1]{\expandafter\@slowromancap\romannumeral #1@}
\makeatother

\begin{document}
\title{{\blue On Bayesian} Fisher Information Maximization for Distributed Vector Estimation}
\author{Mojtaba~Shirazi, 
        Azadeh~Vosoughi,~\IEEEmembership{Senior~Member,~IEEE}
        \thanks{Parts of this research were presented at 
        the IEEE 25th Annual International Symposium on Personal, Indoor, and Mobile Radio Communication, 2014, and the 48th Asilomar Conference on Signals, Systems and Computers, 2014 \cite{Shirazi_PIMRC2014}, \cite{Shirazi_asilomar2014}. 
        This research is supported by the NSF under grants CCF-1341966 and CCF-1319770.}}

\maketitle

\begin{abstract}
In this paper we consider the problem of {\blue bandwidth-constrained} distributed estimation of a Gaussian vector with linear observation model. Each sensor {\blue makes a scalar noisy observation of the unknown vector,} employs a multi-bit {\blue scalar} quantizer to quantize its observation, maps it to a digitally modulated symbol. {\blue Sensors transmit their symbols} over {\blue orthogonal} power-constrained {\blue fading} channels to a fusion center (FC). The FC is tasked with fusing the received signals {\blue from sensors} and estimating the unknown vector. We derive the Bayesian Fisher Information Matrix (FIM) for three types of receivers: (i) coherent receiver (ii) noncoherent receiver with known channel envelopes (iii) noncoherent receiver with known channel statistics only. {\blue We also derive the Weiss-Weinstein bound (WWB).} We formulate two constrained optimization problems, namely maximizing trace and log-determinant of {\blue Bayesian} FIM under network transmit power constraint, with sensors' transmit powers being the optimization variables (we refer to as FIM-max schemes). We show that for coherent receiver, these problems are concave. However, for noncoherent receivers, they are not necessarily concave. The solution to the trace of {\blue Bayesian} FIM maximization problem can be implemented in a distributed fashion, in the sense that each sensor calculates its own transmit power using its local parameters. On the other hand, the solution to the log-determinant of {\blue Bayesian} FIM maximization problem cannot be implemented in a distributed fashion and the FC needs to find the powers (using parameters of all sensors) and inform the active sensors of their transmit powers. 
We numerically investigate how the FIM-max power allocation across sensors depends on the sensors’ observation qualities and physical layer parameters as well as the network transmit power constraint. Moreover, we evaluate the system performance in terms of MSE using the solutions of FIM-max schemes, and compare it with the solution obtained from minimizing the MSE of the LMMSE estimator (MSE-min scheme), and that of uniform power allocation. 
{\blue These comparisons illustrate that, although the WWB is tighter than the inverse of Bayesian FIM, it is still suitable to use FIM-max schemes, since the performance loss in terms of the MSE of the LMMSE estimator is not significant. Furthermore, comparing} the performance of different receivers, our numerical results reveal that coherent receiver and noncoherent receiver with known channel statistics have the best and the worst performance, respectively.
\end{abstract}
%
\IEEEpeerreviewmaketitle
\vspace{-0.2cm}
\section{Introduction} \label{Introduction}
%
%
%
%
The plethora of wireless sensor network (WSN) applications, with practical constraints on network power and bandwidth raises a series of challenging technical problems for system-level engineers \cite{Matin_1,Matin_2}. One of these problems is {\blue bandwidth-constrained} distributed parameter estimation {\blue problem}, where geographically distributed battery-powered sensors are deployed over a sensing field to monitor physical or environmental conditions \cite{Alireza_azadeh_2014_asilomar}. 
{\blue Each sensor makes a noisy observation of the unobservable parameter to be estimated, and transmits its locally processed observation to a fusion center (FC). The FC is tasked with estimating the unknown parameter, via fusing the received data from the sensors with the WSN. 

In this work, we consider bandwidth-constrained distributed estimation of a Gaussian vector $\boldsymbol{\theta}$, where each sensor makes a {\it scalar} observation $x_k\!=\!\mathbf{a}_k^T \boldsymbol{\theta}\!+\!n_k$, with $\mathbf{a}_k^T$ and $n_k$ being respectively, the observation vector and the {\it scalar} observation noise. We model the bandwidth constraint as limiting the number of quantization bits per observation period that a sensor can send to the FC. Each sensor applies a {\it multi-bit scalar quantizer} to quantize its observation, and maps it to a digitally modulated symbol. Sensors transmit their symbols to the FC over orthogonal power-constrained fading channels. 

Bandwidth-constrained distributed estimation problem has a long and rich history in both signal processing and information theory literature. Depending on how the bandwidth constraint is modeled, these works can be classified into two classes: the works in the first class model the bandwidth constraint as limiting the number of quantization bits per observation period that a sensor can send to the FC. On the other hand, the works in the second class model the bandwidth constraint as limiting the number of real-valued messages per observation period that a sensor can send to the FC}\footnote{{\blue In these works, each sensor makes a noisy observation vector of (the entire or part of) vector $\boldsymbol{\theta}$ and locally compresses its observation vector. The focus in these works is finding the optimal compression matrices such that the mean square error (MSE) of reconstruction of $\boldsymbol{\theta}$ at FC is minimized.}}{\blue \cite{Leshem_TSP_2010,Leshem_Elsevier_2017,Giannakis_Luo_TSP_2007,Vetterli_TIT_2008,Li_TSP_2010}. While quantization is important in the works of the first class, compression is the critical component in the works of the second class. With respect to this classification, our work belongs to the first class. 

The works in the first class mentioned above can be further categorized into several subclasses. The two most related subclasses to our work are the works that consider optimal quantization design strategies (dubbed subclass I) and the works that, given quantizers, optimize a network performance metric with respect to energy or power consumption during transmission (dubbed subclass II). Most of the works in subclass I assume} 
that sensors' {\blue quantized} observations are sent over bandwidth constrained  {\it error-free} communication channels. For example, \cite{Varshney_2010,Varshney_2012,Giannakis_2006_part1,Giannakis_2006} studied this problem for estimating a {\it deterministic scalar} {\blue unknown parameter}. 
The authors in \cite{Goldsmith_2006,Vandendorpe_2012,Valenti_2014,Collings_2013} studied this problem for {\it erroneous} bandwidth constrained channels. In particular, this problem was investigated for estimating a {\it deterministic scalar} in \cite{Goldsmith_2006,Valenti_2014,Collings_2013} and for estimating a zero-mean {\it Gaussian scalar} in \cite{Vandendorpe_2012}. 
When addressing the problem, these works {\blue have focused on the} linear estimator at the fusion center (FC) and studied the MSE distortion pertaining to this linear estimator. 
\begin{figure*}[t]
	\centering
	\includegraphics[width=5.6in]{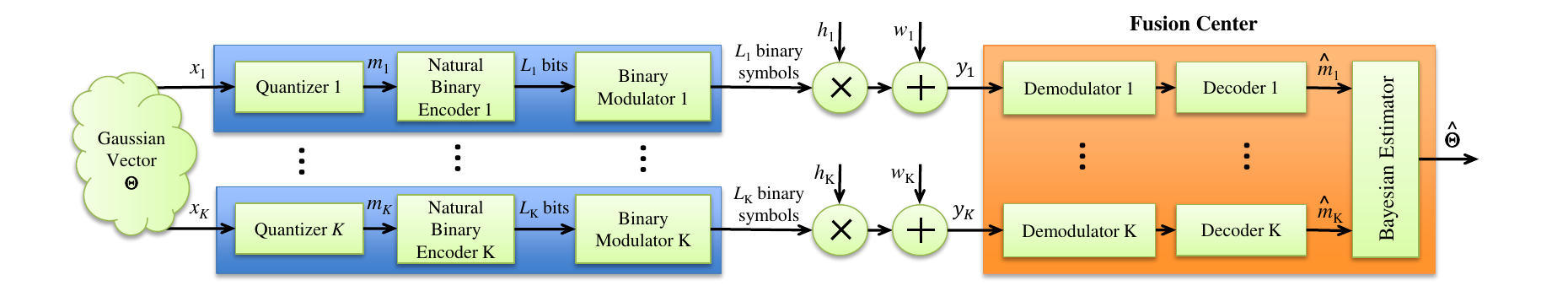}
	\vspace{-0.1cm}
	\caption{Our system model consists of $K$ sensors and a FC, that is tasked with estimating a Gaussian vector {\boldmath$\theta$}, via fusing collective received signals.}
	\label{system-model}
	\vspace{-.5cm}
\end{figure*}

{\blue Among the works in subclass II,} \cite{Goldsmith_2006,khandani_TC_2008} explored the optimal power allocation scheme that minimizes network transmission power subject to a target MSE constraint. On the contrary, for estimating a {\it deterministic scalar} \cite{AlRegib_2009, Giannakis_2008} minimized the MSE of the best linear unbiased estimator (BLUE) subject to a network transmit power constraint. 
The authors in \cite{Vandendorpe_2012, Vosoughi_Sani_2016}, proposed joint transmit power and rate allocation schemes for estimating a {\it random scalar} \cite{Vandendorpe_2012} and a {\it random  vector} \cite{Vosoughi_Sani_2016}, where they minimized an upper bound on the MSE of the LMMSE estimator. 

As an alternative to the MSE of the best linear estimator (BLUE and LMMSE for estimating deterministic and random unknowns, respectively), one can consider the Cram\'{e}r-Rao bound (CRB) and its inverse Fisher information, which are widely employed to explore the fundamental limits of a parameter estimation problem, to optimize the power consumption of a resource constrained WSN tasked with distributed estimation. According to the Cram\'{e}r-Rao inequality \cite{Van_Trees_estimation_book}, maximizing Fisher information minimizes the CRB and Bayesian (classic) CRB sets a lower bound on the MSE of any Bayesian (unbiased) estimator \cite{Vosoughi2006sp2}. Within the context of distributed estimation, maximizing {\blue Bayesian} Fisher information has been adopted before to address sensor selection \cite{Varshney_sensor_selection_tsp2016} and optimal quantization design \cite{Varshney_quant_design_tsp2014,Reibman}. In particular, \cite{Varshney_sensor_selection_tsp2016} investigated the optimal sensor activation strategy with linear observation model, via maximizing trace of {\blue Bayesian} Fisher information matrix (FIM) subject to energy constraints. \cite{Varshney_quant_design_tsp2014} derived the optimality conditions of quantizers that maximize the Bayesian Fisher information for conditionally independent and dependent observations. \cite{Reibman} studied the quantizer designs that minimize the MSE of minimum mean square error (MMSE) and maximum a posteriori (MAP) estimators, and compared their performances with the quantizer design that maximizes Fisher information. 
In \cite{Shirazi_PIMRC2014}\cite{Shirazi_asilomar2014}, we presented our preliminary results on deriving Bayesian CRB and studied its behavior with respect to the system parameters for distributed estimation of a Gaussian vector with linear and nonlinear observation models.

{\bf Our Contributions}: Considering the distributed estimation of a Gaussian vector with linear observation model \cite{Vosoughi_Sani_2016,Chang_TSP_2011}, we formulate two constrained optimization problems, namely, maximization of trace and log-determinant of Bayesian FIM, subject to network transmit power constraint, where sensors’ transmit powers are the optimization variables. We link {\blue log-determinant of} Bayesian FIM to the mutual information between the unknown vector and its Bayesian estimator. We derive Bayesian FIM 
{\blue and the Weiss-Weinstein bound (WWB), which is known to be one of the tightest Bayesian bounds \cite{WWB}.} 
We develope two transmit power allocation schemes from solving the two formulated problems (which we refer to as FIM-max schemes). 
We derive the MSE corresponding to the LMMSE estimator at the FC for coherent and noncoherent receivers. 
Our {\blue numerical} results demonstrate the {\blue effectiveness} of FIM-max schemes, as
these power allocations perform close to the power allocation obtained from minimizing the MSE of LMMSE estimator, and 
outperform uniform power allocation. 
{\blue Based on these results, we draw the conclusion that although the WWB is tighter than the Bayesian CRB in our problem (and Bayesian CRB is not attainable), it is still appropriate to use FIM-max schemes, since the performance loss in terms of the MSE of the LMMSE estimator is not significant.}

{\bf Notations}: Matrices are denoted by bold uppercase letters, vectors by bold lowercase letters, and scalars by normal letters. $\mathbb{E}$ denotes the mathematical expectation operator, $||.||$ and $[.]^T$ represent the $L^2$ norm of a vector and the matrix-vector transpose operation, respectively. tr(.) and $|.|$ indicate trace and determinant of a matrix, respectively, and $|\cal A|$ is the cardinality of set $\cal A$. {\blue $\boldsymbol{A}\!\succ\!\boldsymbol{0}$ ($\boldsymbol{A}\!\succeq\!\boldsymbol{0}$) means that $\boldsymbol{A}$ is a (semi-)positive definite matrix} The definition of Q-function is $Q(x)\!\!=\!\!\frac{1}{\sqrt{2\pi}}\!\int_{x}^{\infty}\!e^{-\frac{u^2}{2}}du$, the Marcum-Q function of nonnegative real numbers $a$ and $b$ , denoted as ${\cal Q}(a,b)$, is defined as \cite{marcum_Q_function_book} ${\cal Q}(a,b)\!=\!\int_{b}^{\infty}\!xe^{-\frac{x^2+a^2}{2}}I_0(ax)dx$,
and the two dimensional Gaussian Q-function, denoted as ${\mathfrak Q}\left(x,y;\rho\right)$, is defined as \cite{two_dim_Q_function} ${\mathfrak Q}\left(x,y;\rho\right)\!=\!\frac{1}{2\pi\sqrt{1\!-\!\rho^2}}\!\int_{x}^{\infty}\!\!\int_{y}^{\infty}\!e^{-\frac{u^2+v^2-2\rho uv}{2\left(1-\rho^2\right)}}\!dudv$.
The notations $\mathcal{N}$ and $\mathcal{CN}$ represent Gaussian distribution and complex Gaussian distribution, respectively. 
\vspace{-0.2cm}
\section{System Model and Problem Formulation} \label{System Model}
\vspace{-0.1cm}
Suppose there are $K$ spatially-distributed and inhomogeneous sensors, each making a noisy observation of a common unobservable zero-mean Gaussian vector {\boldmath$\theta$}= $[\theta_1, \theta_2,..., \theta_q]^T\!\in\!\mathbb{R}^q$ with covariance matrix $\boldsymbol{\mathcal C}_{\boldsymbol{\theta}}=\mathbb{E}\{\boldsymbol{\theta}\boldsymbol{\theta}^T\}$. 
Let $x_k$ denote the scalar noisy observation of sensor $k$ (see Fig.~\ref{system-model}). Our linear observation model is:
\vspace{-0.25cm}
\begin{equation} \label{obs_model}
x_k=\mathbf{a}_k^T \boldsymbol{\theta}+n_k, \ \ \ \ \ \ k=1,..., K
\end{equation}
where $\mathbf{a}_k\!=\![a_{k_1}, a_{k_2},..., a_{k_q}]^T\!\in\!\mathbb{R}^q$ is {\blue the} known observation vector and $n_k$ denotes zero-mean Gaussian observation noise with variance $\sigma_{n_k}^2$. We assume that $n_k$'s are uncorrelated across the sensors and also are uncorrelated with {\boldmath$\theta$}. Sensor $k$ employs a 
scalar quantizer
with $M_k\!=\!2^{L_k}$ quantization levels $m_{k,l},\ l\!=\!1,...,M_k$ where $l$ is the index of the quantization level.
In particular, the quantizer maps $x_k$ to one of the quantization levels $m_k \in \{m_{k,1},...,m_{k,M_k}\}$ as the following:
%
\vspace{-.2cm}
\begin{equation*} 
m_k=m_{k,l},\ \ \ \text{for}\ \ x_k\in [u_{k,l},u_{k,l+1}],\ \ l=1,...,M_k\\
\end{equation*}
where 
$u_{k,l},\ l\!=\!1,...,M_k\!+\!1$, are the quantization boundaries. 
Following quantization, sensor $k$ employs a fixed length encoder, which encodes the index $l$ corresponding to the quantization level $m_{k,l}$ to a binary sequence of length $L_k=\log_2 M_k$ according to natural binary encoding\footnote{Natural binary encoding is needed {\blue for} the derivations of {\blue Bayesian} FIM.} 
\cite{Vandendorpe_2012,Vosoughi_Sani_2016}, and finally modulates these $L_k$ bits into $L_k$ binary symbols. Let $P_k$ denote the average transmit power corresponding to $L_k$ symbols from sensor $k$, which is equally distributed among $L_k$ symbols. We consider two types of modulators, Binary Phase Shift Keying (BPSK) modulator, which maps each bit of $L_k$-bit sequence into one symbol with transmit power $P_k/L_k$, and On-Off Keying (OOK) modulator, which maps each ``1'' bit of $L_k$-bit sequence into one symbol with transmit power $2P_k/L_k$ and sends no carrier for ``0'' bit.

Sensors send their modulated symbols to the FC over orthogonal flat fading channels, with fading coefficient $h_k=|h_k|e^{j\phi_k}$. 
We assume that channel $h_k$ remains constant during the transmission of $L_k$ symbols. Denote $w_{k,i}$ as communication channel noise during the transmission of $i$-th symbol of $L_k$ symbols corresponding to sensor $k$. We assume $w_{k,i}$'s are independent across $k$ channels and independent and identically distributed (i.i.d.) across $L_k$ transmitted symbols, $w_{k,i}\sim \mathcal{CN}\left(0,2\sigma_{w_k}^2\right)$. We further assume that there is a constraint on the network average transmit power, i.e., $\sum_{k=1}^{K}P_k\leq P_{tot}$.

To describe the estimation operation at the FC, let $\hat{m}_{k}$ denote the recovered quantization level corresponding to sensor $k$, where in general, $\hat{m}_{k}\neq m_{k}$ due to communication channel errors. The FC processes the channel output corresponding to sensor $k$ to recover the transmitted quantization levels $\hat{m}_k \in \{\hat{m}_{k,1},...,\hat{m}_{k,M_k}\}$. We consider coherent and noncoherent receivers, corresponding to BPSK and OOK modulation schemes, respectively. For noncoherent receiver, we consider two scenarios: a) channel envelopes $|h_k|$'s are available at the FC \cite{Biao_Chen_2005}, b) only statistics of complex Gaussian channel $h_k$'s are available at the FC \cite{Varshney_2009}. Having $\{\hat{m}_1,...,\hat{m}_K\}$, the FC applies a Bayesian estimator to form the estimate $\hat{\boldsymbol{\theta}}$. We define vector $\boldsymbol{m}=[m_1,...,m_K]^T$ which consists of transmitted quantization levels, and vector $\boldsymbol{\hat{m}}=[\hat{m}_1,...,\hat{m}_K]^T$ that includes recovered quantization levels at the FC. Let $p(\boldsymbol{\hat{m}},\boldsymbol{\theta})$ denote the joint probability distribution function (pdf) of the {\blue recovered} quantization levels and the unknown vector $\boldsymbol{\theta}$. Under certain regularity conditions that are {\blue satisfied} by Gaussian vectors, the $q\times q$ {\blue Bayesian} FIM, denoted as $\boldsymbol{J}$, is defined based on the joint pdf $p(\boldsymbol{\hat{m}},\boldsymbol{\theta})$ as \cite{Van_Trees_estimation_book,Vosoughi2006sp2,Vosoughi2006sp1}:
\vspace{-0.15cm}
\begin{equation} \label{FIM_Van_Trees}
\boldsymbol{J}=\mathbb{E}\{(\frac{\partial \ln p(\boldsymbol{\hat{m}},\boldsymbol{\theta})}{\partial \boldsymbol{\theta}}){(\frac{\partial \ln p(\boldsymbol{\hat{m}},\boldsymbol{\theta})}{\partial \boldsymbol{\theta}})}^T\},
\end{equation}
where the expectation is taken over $p(\boldsymbol{\hat{m}},\boldsymbol{\theta})$.

Our goals are to characterize $\boldsymbol{J}$ and study the transmit power allocation schemes that maximize either tr($\boldsymbol{J}$) \cite{Varshney_sensor_selection_tsp2016} or $\textnormal{log}_{\textnormal{2}}(|\boldsymbol{J}|)$ \cite{Li_IET_2014}, subject to the network average transmit power constraint (which we refer to as FIM-max schemes). In other words, we are interested in solving the following constrained optimization problems\footnote{\blue Let CRB denote the Bayesian CRB matrix. We have tr$(\mbox{CRB})\!= \!\text{tr}(\mathbf{J}^{-1}) \!\geq\! \frac{q^2}{ \text{tr}(\mathbf{J})}$ \cite{Vosoughi_Sani_2016} and $\text{log}_2(|\mbox{CRB}|)\!=\!\text{log}_2(|\mathbf{J}^{-1}|)\!=\! -\text{log}_2(|\mathbf{J}|)$. Therefore, maximizing tr(J) is equivalent to minimizing the lower bound on tr$(\mbox{CRB})$ and maximizing $\text{log}_2(|\mathbf{J}|)$ is equivalent to minimizing $\text{log}_2(|\mbox{CRB}|)$.}:
\vspace{-0.3cm}
\begin{align} \label{maximization problem of tr(J)}
\mathop{\text{maximize}}_{P_k,\forall k}\ \ \ \ &\text{tr}\left(\boldsymbol{J}\left(\{P_k\}_{k=1}^K\right)\right)\nonumber\\
\vspace{-0.1cm}
\text{s.t.}\ \ \ \ &\sum_{k=1}^{K}P_k\leq P_{tot},\ P_k\in \mathbb{R}^{+},\ \forall k
\end{align}
\vspace{-0.2cm}
and
\vspace{-0.25cm}
\begin{align} \label{maximization problem of logdet(J)}
\mathop{\text{maximize}}_{P_k,\forall k}\ \ \ \ &\text{log}_2\left(\left|\boldsymbol{J}\left(\{P_k\}_{k=1}^K\right)\right|\right)\nonumber\\
\vspace{-0.1cm}
\text{s.t.}\ \ \ \ &\sum_{k=1}^{K}P_k\leq P_{tot},\ P_k\in \mathbb{R}^{+},\ \forall k
\end{align}
%
Interestingly, the constrained maximization problem in \eqref{maximization problem of logdet(J)} can be linked to the constrained maximization of mutual information between the unknown $\boldsymbol{\theta}$ and its Bayesian {\blue estimator} $\hat{\boldsymbol{\theta}}$. Let $\tilde{\boldsymbol{\theta}}=\boldsymbol{\theta}-\hat{\boldsymbol{\theta}}$, where $\tilde{\boldsymbol{\theta}}$ is the corresponding estimation error vector. Suppose $\boldsymbol{\mu}=\mathbb{E}\{\tilde{\boldsymbol{\theta}}\}$ and $\boldsymbol{\mathcal D}\!=\!\mathbb{E}\{\tilde{\boldsymbol{\theta}}\tilde{\boldsymbol{\theta}}^T\}$ are the error mean vector and the MSE matrix, respectively. According to inequality (6) in \cite{Vosoughi2006sp2} and using the fact that $\boldsymbol{\theta}$ is Gaussian, we can write:
\vspace{-0.25cm}
\begin{align} \label{MI lower bound}
I(\boldsymbol{\theta};\hat{\boldsymbol{\theta}})\geq\frac{1}{2}(\text{log}_{\text{2}}(|\boldsymbol{\mathcal C}_{\boldsymbol{\theta}}|)-\textnormal{log}_{\textnormal{2}}(|\boldsymbol{\mathcal D}|)).
\end{align}
On the other hand, under the regularity conditions \cite{Van_Trees_estimation_book}, the inverse of {\blue Bayesian} FIM establishes a lower bound on the MSE matrix $\boldsymbol{\mathcal D}$. The Bayesian Cram\'{e}r-Rao inequality states {\blue that} $\boldsymbol{\mathcal D}\succeq \boldsymbol{J}^{-1}$ \cite{Van_Trees_estimation_book}.
Using the concavity of the function log$(|.|)$ on the cone of positive definite Hermitian matrices \cite{Cover_Book}, we conclude that {\blue $\textnormal{log}_{\textnormal{2}}(|\boldsymbol{\mathcal D}|)\!\geq\!\textnormal{log}_{\textnormal{2}}(|\boldsymbol{J}^{-1}|)\!=\!-\textnormal{log}_{\textnormal{2}}(|\boldsymbol{J}|)$. Therefore,} the lower bound on $I(\boldsymbol{\theta};\hat{\boldsymbol{\theta}})$ is {\it maximized}
if we substitute {\blue $\textnormal{log}_{\textnormal{2}}(|\boldsymbol{\mathcal D}|)$} in (\ref{MI lower bound}) with {\blue $-\textnormal{log}_{\textnormal{2}}(|\boldsymbol{J}|)$}. In other words:
\vspace{-.15cm}
\begin{align} \label{lower bound on the achievable rates}
I(\boldsymbol{\theta};\hat{\boldsymbol{\theta}})\geq\frac{1}{2}(\text{log}_{\text{2}}(|\boldsymbol{\mathcal C}_{\boldsymbol{\theta}}|)+\textnormal{log}_{\textnormal{2}}(|\boldsymbol{J}|)).
\end{align}
%
Based on \eqref{lower bound on the achievable rates}, we observe that the problem in \eqref{maximization problem of logdet(J)} is equivalent to {\blue constrained} maximization of the mutual information lower bound.
\vspace{-.2cm}
\section{Characterization of {\blue Bayesian} FIM} \label{FIM}
\vspace{-0.1cm}
In this section, we characterize $\boldsymbol{J}$ in terms of the optimization parameters $P_k, \forall k$. The matrix $\boldsymbol{J}$ in \eqref{FIM_Van_Trees} can be expressed as \cite{Vosoughi2006sp2,Vosoughi2006sp1}:
%
\vspace{-0.25cm}
\begin{equation*} 
\boldsymbol{J}=\mathbb{E}\{\mathbb{E}\{(\frac{\partial \ln p(\boldsymbol{\hat{m}},\boldsymbol{\theta})}{\partial \boldsymbol{\theta}})(\frac{\partial \ln p(\boldsymbol{\hat{m}},\boldsymbol{\theta})}{\partial \boldsymbol{\theta}})^T|\boldsymbol{\theta}\}\},
\end{equation*}
%
where the first and second expectations are taken over {\blue the pdf of $\boldsymbol{\theta}$, denoted as $f(\boldsymbol{\theta})\!=\!\frac{1}{\sqrt{{(2\pi)}^q|\boldsymbol{\cal C}_{\boldsymbol{\theta}}|}}\exp{(-\frac{1}{2}{\boldsymbol{\theta}}^T{\boldsymbol{\cal C}_{\boldsymbol{\theta}}}^{-1}\boldsymbol{\theta})}$} and {\blue the conditional distribution} $p(\boldsymbol{\hat{m}}\arrowvert\boldsymbol{\theta})$, respectively. Using the Bayes' rule $p(\boldsymbol{\hat{m}},\boldsymbol{\theta})\!=\!p(\boldsymbol{\hat{m}}\arrowvert\boldsymbol{\theta})f(\boldsymbol{\theta})$, we {\blue can decompose $\boldsymbol{J}$ into two terms}:
\vspace{-0.35cm}
\begin{align} \label{FIM in decomposed form}
\boldsymbol{J}&=\mathbb{E}\{\overbrace{(\frac{\partial \ln f(\boldsymbol{\theta})}{\partial \boldsymbol{\theta}})(\frac{\partial \ln f(\boldsymbol{\theta})}{\partial \boldsymbol{\theta}})^T}^{=\boldsymbol{\Omega}(\boldsymbol{\theta})}\}\nonumber\\
&+\mathbb{E}\{\underbrace{\mathbb{E}\{(\frac{\partial \ln p(\boldsymbol{\hat{m}}\arrowvert\boldsymbol{\theta})}{\partial \boldsymbol{\theta}})(\frac{\partial \ln p(\boldsymbol{\hat{m}}\arrowvert\boldsymbol{\theta})}{\partial \boldsymbol{\theta}})^T\}}_{=\boldsymbol{\Lambda}(\boldsymbol{\theta})}\},
\end{align}
%
%
%
in which the {\blue outer} expectations are taken over $\boldsymbol{\theta}$.
The $q\!\times \!q$ matrix $\boldsymbol{\Omega}(\boldsymbol{\theta})$ only depends on $f$({\boldmath$\theta$}) \cite{Vosoughi2006sp2}. In particular, let $[\boldsymbol{\Omega}(\boldsymbol{\theta})]_{ij}$ denote the $(i,j)$-th entry of matrix $\boldsymbol{\Omega}(\boldsymbol{\theta})$. We have \cite{Van_Trees_estimation_book}:
\vspace{-.2cm}
\begin{equation*} 
[\boldsymbol{\Omega}(\boldsymbol{\theta})]_{ij}=-\frac{\partial^2 \ln f(\boldsymbol{\theta})}{\partial \theta_i\partial \theta_j},\ \ i,j=1,...,q
\end{equation*}
Since $\boldsymbol{\theta}$ is Gaussian with covariance matrix $\boldsymbol{\cal C}_{\boldsymbol{\theta}}$, we obtain $\mathbb{E}\{\boldsymbol{\Omega}(\boldsymbol{\theta})\}={\boldsymbol{\cal C}_{\boldsymbol{\theta}}^{-1}}$.
Let $[\boldsymbol{\Lambda}(\boldsymbol{\theta})]_{ij}$ represent the $(i,j)$-th entry of matrix $\boldsymbol{\Lambda}(\boldsymbol{\theta})$. We can write \cite{Van_Trees_estimation_book}:
\vspace{-0.2cm}
\begin{equation} \label{ij-th element of second term of FIM}
[\boldsymbol{\Lambda}(\boldsymbol{\theta})]_{ij}\!=\!-\mathbb{E}\{\frac{\partial^2 \ln p(\boldsymbol{\hat{m}}\arrowvert\boldsymbol{\theta})}{\partial \theta_i\partial \theta_j}\},\ i,j\!=\!1,...,q
\end{equation}
We note that the entries $[\boldsymbol{\Lambda}(\boldsymbol{\theta})]_{ij}$ depend on the parameters of the observation model as well as the physical layer parameters (e.g., modulation scheme, receiver type, channel gain, channel noise, transmit power, and quantization bits). To find $[\boldsymbol{\Lambda}(\boldsymbol{\theta})]_{ij}$ in \eqref{ij-th element of second term of FIM}, {\blue we need Lemma \ref{p(m_hat|theta)} below, which shows} that, given $\boldsymbol{\theta}$, the entries of vector $\boldsymbol{\hat{m}}$ are conditionally independent.
{\blue 
	\vspace{-.2cm}
\begin{lemma} \label{p(m_hat|theta)}
\textnormal{
Given our system model we have $p(\boldsymbol{\hat{m}}\arrowvert\boldsymbol{\theta})=\prod_{k=1}^{K}p(\hat{m}_k\arrowvert\boldsymbol{\theta})$.
}
\end{lemma}
\vspace{-.3cm}
\begin{proof}
	See Appendix \ref{Proof of Lemma p(m_hat|theta)}.
\end{proof}
}
\vspace{-0.2cm}
{\red Combining the result of Lemma \ref{p(m_hat|theta)} and \eqref{ij-th element of second term of FIM}}
{\blue and recalling that the expectation in \eqref{ij-th element of second term of FIM} is taken with respect to $p(\boldsymbol{\hat{m}}\arrowvert\boldsymbol{\theta})$, we reach}:
\vspace{-0.35cm}
\begin{align*} 
[\boldsymbol{\Lambda}(\boldsymbol{\theta})]_{ij}&=-\sum\limits_{\hat{m}_1}\dots\sum\limits_{\hat{m}_{K}}\{\sum\limits_{k=1}^{K}[\frac{\partial^2 p(\hat{m}_k\arrowvert\boldsymbol{\theta})}{\partial \theta_i\partial \theta_j}\notag\\
&-\frac{1}{p(\hat{m}_k\arrowvert\boldsymbol{\theta})}\frac{\partial p(\hat{m}_k\arrowvert\boldsymbol{\theta})}{\partial \theta_i}\frac{\partial p(\hat{m}_k\arrowvert\boldsymbol{\theta})}{\partial \theta_j}]\}\prod_{\underset{n\neq k}{n=1}}^{K}p(\hat{m}_n\arrowvert\boldsymbol{\theta}).
\end{align*}
Using the following two facts:
\vspace{-0.35cm}
\begin{align} \label{two fact for deriving FIM}
\sum\limits_{\hat{m}_1}\dots\sum\limits_{\hat{m}_{k-1}}\sum\limits_{\hat{m}_{k+1}}\dots\sum\limits_{\hat{m}_K}\prod_{\underset{n\neq k}{n=1}}^{K}p(\hat{m}_n\arrowvert\boldsymbol{\theta})=1,
\end{align}
\vspace{-.4cm}
\begin{align*}
\sum_{k=1}^{K}\sum_{t=1}^{M_k}\frac{\partial^2 p(\hat{m}_{k,t}\arrowvert\boldsymbol{\theta})}{\partial \theta_i\partial \theta_j}=\sum_{k=1}^{K}\frac{\partial^2}{\partial \theta_i\partial \theta_j}(\underbrace{\sum_{t=1}^{M_k}p(\hat{m}_{k,t}\arrowvert\boldsymbol{\theta})}_{=1})=0,\quad\quad
\end{align*}
%
where index $t$ indicates the quantization level corresponding to $\hat{m}_k$, we find that $[\boldsymbol{\Lambda}(\boldsymbol{\theta})]_{ij}$ reduces to:
\vspace{-0.25cm}
\begin{equation} \label{ij-th element of lambda_theta 1}
\![\boldsymbol{\Lambda}(\boldsymbol{\theta})]_{ij}\!=\!\!\sum_{k=1}^{K}\!\sum_{t=1}^{M_k}\!(\frac{1}{p(\hat{m}_{k,t}\arrowvert\boldsymbol{\theta})}\frac{\partial p(\hat{m}_{k,t}\arrowvert\boldsymbol{\theta})}{\partial \theta_i}\frac{\partial p(\hat{m}_{k,t}\arrowvert\boldsymbol{\theta})}{\partial \theta_j}\!)\!\!.
\end{equation}
Examining \eqref{ij-th element of lambda_theta 1} we realize that we need to find two terms in order to fully characterize $[\boldsymbol{\Lambda}(\boldsymbol{\theta})]_{ij}$: the probability term $p(\hat{m}_{k,t}\arrowvert\boldsymbol{\theta})$, and its first derivative with respect to $\theta_i$, i.e., $\partial p(\hat{m}_{k,t}|\boldsymbol{\theta}) / \partial \theta_i$. In the following, we derive these two terms.
According to the Bayes' rule and the fact that $\boldsymbol{\theta}, m_k, \hat{m}_k$ form a Markov chain, we have:
\vspace{-0.2cm}
\begin{align} \label{P(m_hat given theta) based on alpha and beta} 
p(\hat{m}_{k,t}\arrowvert\boldsymbol{\theta})\!=\!\sum_{l=1}^{M_k}\underbrace{p(\hat{m}_{k,t}\arrowvert m_{k,l})}_{=\alpha_{k,t,l}}
\underbrace{p(m_{k,l}\arrowvert\boldsymbol{\theta})}_{=\beta_{k,l}(\boldsymbol{\theta})}\ \ t\!=\!1,...,M_k.
\end{align}
\vspace{-0.1cm}
Considering $p(\hat{m}_{k,t}\arrowvert\boldsymbol{\theta})$ in (\ref{P(m_hat given theta) based on alpha and beta}) we realize that each term inside the sum is the product of two probabilities: the first probabilty $\alpha_{k,t,l}$ does not depend on $\boldsymbol{\theta}$; it depends on the modulation scheme (BPSK or OOK) and the receiver type at the FC (coherent or noncoherent) as well as the physical layer parameters, i.e., channel errors due to fading and noise, transmit power $P_k$, and number of transmitted bits $L_k$. On the other hand, the second probability $\beta_{k,l}(\boldsymbol{\theta})$ depends on $\boldsymbol{\theta}$, the observation model and its parameters as well as quantizer. In other words, the contributions of the observation model and quantization in each term inside the sum in \eqref{P(m_hat given theta) based on alpha and beta} are {\it decoupled} from those of communication system.  

{\blue The probability} $\beta_{k,l}(\boldsymbol{\theta})$ in 
{\blue \eqref{P(m_hat given theta) based on alpha and beta} becomes:
	\vspace{-0.15cm}
	\begin{align} \label{beta_theta}
	\beta_{k,l}(\boldsymbol{\theta})&=\int_{u_{k,l}}^{u_{k,l+1}}f\left(x_k|\boldsymbol{\theta}\right)dx_k\nonumber\\
	&\overset{(a)}{=}Q(\frac{u_{k,l}-\mathbf{a}_k^T\boldsymbol{\theta}}{\sigma_{n_k}})-
	Q(\frac{u_{k,l+1}-\mathbf{a}_k^T\boldsymbol{\theta}}{\sigma_{n_k}}),
	\end{align}
	in which ($a$) follows from the fact that the conditional pdf of $x_k$ given $\boldsymbol{\theta}$ is ${\cal N}(\mathbf{a}_k^T\boldsymbol{\theta},\sigma_{n_k}^2)$.
}
Next, we find ${\partial p(\hat{m}_{k,t}\arrowvert\boldsymbol{\theta})}/{\partial\theta_i}$ in \eqref{ij-th element of lambda_theta 1}. Since $\alpha_{k,t,l}$ does not depend on $\boldsymbol{\theta}$, from \eqref{P(m_hat given theta) based on alpha and beta} we have:
\vspace{-0.5cm}
\begin{align} \label{der of P(m_hat given theta) wrt theta_i}
&\!\!\frac{\partial p(\hat{m}_{k,t}\arrowvert\boldsymbol{\theta})}{\partial\theta_i}\!=\!\sum_{l=1}^{M_k}\!{\blue \frac{a_{k_i}}{\sqrt{2\pi}\sigma_{n_k}}\alpha_{k,t,l}\dot{\beta}_{k,l}(\boldsymbol{\theta})},~~i=1,...,q,\\
&\!\!\dot{\beta}_{k,l}(\boldsymbol{\theta})\!=\!\exp{\!(\!-\frac{\left(u_{k,l}\!-\!\mathbf{a}_k^T\boldsymbol{\theta}\right)^2}{2\sigma_{n_k}^2})}\!\!-\!
\exp{\!(\!-\frac{(u_{k,l+1}\!-\!\mathbf{a}_k^T\boldsymbol{\theta})^2}{2\sigma_{n_k}^2})}.\notag
\end{align}
%
%
%

{\blue Now we characterize} $\alpha_{k,t,l}$ in $p(\hat{m}_{k,t}\arrowvert\boldsymbol{\theta})$.
As we mentioned before, $\alpha_{k,t,l}$ depends on the modulation scheme and the receiver type at the FC. In this section we derive $\alpha_{k,t,l}$ for BPSK modulation with coherent receiver and OOK modulation with noncoherent receiver. For OOK modulation with noncoherent receiver, we consider two scenarios: {\it a)} channel envelopes are available at the FC, {\it b)} only channel statistics are available {\blue at the FC}. We assume that the FC performs a symbol-by-symbol demodulation. To enable derivations of $\alpha_{k,t,l}$, we let indices $l$ and $t$, respectively, indicate the quantization levels corresponding to $m_k$ and $\hat{m}_k$, and $[b_{k,l,1},\dots b_{k,l,L_k}]$ and $[\hat{b}_{k,t,1},\dots \hat{b}_{k,t,L_k}]$, respectively, be the transmitted bit sequence and recovered (received) bit sequence of sensor $k$.
\vspace{-.15cm}
\subsection{Coherent Receiver} \label{derivations of Coherent Receiver}
\vspace{-.1cm}
Suppose the Hamming distance between two bit sequences $[b_{k,l,1},\dots b_{k,l,L_k}]$ and $[\hat{b}_{k,t,1},\dots \hat{b}_{k,t,L_k}]$ is $N_{e_{k,t,l}}=\sum_{i=1}^{L_k}\hat{b}_{k,t,i}\oplus b_{k,l,i}$, in which $\oplus$ is the Boolean sum operator.
We define $\gamma_k$ as the channel signal to noise ratio (SNR) of sensor $k$, where:
\vspace{-0.2cm}
\begin{equation} \label{SNR for k-th channel}
\gamma_k=\frac{P_k|h_k|^2}{2L_k\sigma_{w_k}^2}.
\end{equation}
We can model the channel between sensor $k$ and the FC as a binary symmetric channel (BSC) with {\blue the} probability of flipping a bit ${\cal E}_k=Q\left(\sqrt{2\gamma_k}\right)$, where ${\cal E}_k$ does not depend on the bit index. Hence, the probability $\alpha_{k,t,l}$ in (\ref{P(m_hat given theta) based on alpha and beta}) becomes:
\vspace{-0.15cm}
\begin{equation} \label{alpha_{k,t,l} for coherent}
\alpha_{k,t,l}={\cal E}_k^{N_{e_{k,t,l}}}(1-{\cal E}_k)^{L_k-N_{e_{k,t,l}}}.
\end{equation}
%
\vspace{-0.6cm}
\subsection{Noncoherent Receiver} \label{derivations of noncoherent Receiver}
The channel between sensor $k$ and the FC can no longer be modeled as a BSC. 
Instead, we can model it as a binary asymmetric channel, where ${\cal E}_{1_k}$ is the probability that ``0'' bit is flipped into ``1'' bit, and ${\cal E}_{2_k}$ is the probability that ``1'' bit is flipped into ``0'' bit. 
Therefore, the probability $\alpha_{k,t,l}$ {\blue in \eqref{P(m_hat given theta) based on alpha and beta} becomes}:
\vspace{-0.15cm}
\begin{align} \label{alpha_{k,t,l} for noncoherent}
&\alpha_{k,t,l}\!=\!\!\prod_{i=1}^{L_k}\!\left[\mathbf{1}_{\{b_{k,l,i}=\hat{b}_{k,t,i}=0\}}(1\!-\!{\cal E}_{1_k})\!+\!\mathbf{1}_{\{b_{k,l,i}=0,\hat{b}_{k,t,i}=1\}}({\cal E}_{1_k})\right.\nonumber\\
&\!\!\left.+\mathbf{1}_{\{b_{k,l,i}=1,\hat{b}_{k,t,i}=0\}}({\cal E}_{2_k})\!+\!\mathbf{1}_{\{b_{k,l,i}=\hat{b}_{k,t,i}=1\}}(1\!-\!{\cal E}_{2_k})\right]\!\!,
\end{align}
where $\mathbf{1}_{\{X\}}$ is indicator function with subscript $X$ describing the event of inclusion. Next, we compute probabilities ${\cal E}_{1_k}$ and ${\cal E}_{2_k}$ in \eqref{alpha_{k,t,l} for noncoherent}. Note that ${\cal E}_{1_k}$ and ${\cal E}_{2_k}$ do not depend on the bit index. 
The problem of demodulating $L_k$ symbols (bits) sent by sensor $k$, based on $L_k$ received signals, $y_{k,1}, \dots, y_{k,L_k}$ can be cast into $L_k$ binary hypothesis testing problems, in which the channel output corresponding to each problem is:
\vspace{-0.2cm}
\begin{equation*} 
y_{k,i}=\left\{
\begin{array}{lr}
B_kh_k+w_{k,i},&{\cal H}_{1,i}:b_{k,l,i}\!=\!1\\
w_{k,i},&{\cal H}_{0,i}:b_{k,l,i}\!=\!0
\end{array}
\right.
\end{equation*}
for $i\!=\!1,\dots L_k$, where $B_k$ is transmitted signal amplitude for sensor $k$. Denoting $r_{k,i}$ as the test statistics, the optimal likelihood ratio test (LRT) at the FC can be expressed as:
\vspace{-0.25cm}
\begin{equation} \label{binary hyp test problem}
\frac{f\left(r_{k,i}|{\cal H}_{1,i}\right)}{f\left(r_{k,i}|{\cal H}_{0,i}\right)}\ \underset{{\cal H}_{0,i}}{\overset{{\cal H}_{1,i}}{\mathlarger{\mathlarger{\gtrless}}}}\ \frac{p\left({\cal H}_{0,i}\right)}{p\left({\cal H}_{1,i}\right)},~~~i=1,\dots L_k,
\end{equation}
%
where the probabilities $p\left({\cal H}_{1,i}\right)\!=\!p(b_{k,l,i}=1)$ and $p\left({\cal H}_{0,i}\right)\!=\!p(b_{k,l,i}=0)$. Lemma \ref{p(H_0)=p(H_1)} shows that for our system model, $p({\cal H}_{0,i})=p({\cal H}_{1,i})=1/2$.
\vspace{-0.1cm}
\begin{lemma} \label{p(H_0)=p(H_1)}
\textnormal{We have $p({\cal H}_{0,i})=p({\cal H}_{1,i})=1/2$ under the following two assumptions:
1) the pdf of noisy observation $x_k$ is smooth and symmetric,
2) sensor $k$ uses a symmetric mid-rise quantizer and encodes the quantization level $m_k$ according to natural binary encoding rule. Both assumptions hold true for our system model.}
\end{lemma}
\vspace{-0.15cm}
%
$\!\!\!\!\!\!${\it Proof}. See Appendix \ref{Proof of Lemma p(H_0)=p(H_1)}.\\
%
According to Lemma \ref{p(H_0)=p(H_1)}, we can state that $\mathbb{E}\{B_k^2\}=2P_k/L_k$, where $P_k$ is the average transmit power of sensor $k$. In the following, we find probabilities ${\cal E}_{1_k}$ and ${\cal E}_{2_k}$ for our two types of noncoherent receivers.

$\bullet$ \textbf{Noncoherent Receiver with Known Channel Envelopes:} 
For this receiver, the test statistics of LRT at the FC is the envelope of channel output, i.e., $r_{k,i}=|y_{k,i}|$ and $|h_k|$ is known to the FC. Hence, given $|h_k|$, the two conditional pdfs of the test statistics under hypotheses ${\cal H}_{0,i}$ and ${\cal H}_{1,i}$ are \cite{communication_schwartz}:
\vspace{-0.3cm}
%
\begin{equation*} 
\!\!f\left(r_{k,i}|{\cal H}_{0,i},|h_k|\right)\!=\!\frac{r_{k,i}}{\sigma_{w_k}^2}e^{-\frac{r_{k,i}^2}{2\sigma_{w_k}^2}},\qquad\qquad\qquad\qquad\qquad\quad
\end{equation*}
\vspace{-0.5cm}
\begin{equation*} 
f\left(r_{k,i}|{\cal H}_{1,i},|h_k|\right)\!=\!\frac{r_{k,i}}{\sigma_{w_k}^2}e^{-(\!\frac{r_{k,i}^2}{2\sigma_{w_k}^2}+2\gamma_k\!)}\!I_0(\!\sqrt{\frac{2P_k}{L_k}}\frac{|h_k|r_{k,i}}{\sigma_{w_k}^2}\!),
\end{equation*}
where $\gamma_k$ is defined in (\ref{SNR for k-th channel}) and $I_0(.)$ is the zeroth-order modified Bessel function of the first kind. Since $w_{k,i}$'s are independent across $L_k$ transmitted symbols, the random variables $r_{k,i}$ conditioned on each hypothesis and $|h_k|$ are i.i.d. for $i\!=\!1, \dots, L_k$. Therefore, the probabilities ${\cal E}_{1_k}$ and ${\cal E}_{2_k}$ do not depend on bit index $i$. Based on equations (7-4-7) and (7-4-11) in \cite{communication_schwartz}, probabilities ${\cal E}_{1_k}$ and ${\cal E}_{2_k}$ are:
\vspace{-.1cm}
\vspace{-0.15cm}
\begin{subequations} \label{epsilon1 and epsilon2 for known channel envelope receiver}
\begin{flalign} 
{\cal E}_{1_k}\!\!&=\!p\left(r_{k,i}>\zeta_k|{\cal H}_{0,i},|h_k|\right)\!=\!e^{-\frac{\zeta_k^2}{2}}\label{eq:subeq_eps1_2_known_env_1},&\\
{\cal E}_{2_k}\!\!&=\!p\left(r_{k,i}<\zeta_k|{\cal H}_{1,i},|h_k|\right)\!=\!1\!-\!{\cal Q}\left(2\sqrt{\gamma_k},\zeta_k\right)\label{eq:subeq_eps1_2_known_env_2},&
\end{flalign}
\end{subequations}
where the decision threshold $\zeta_k$ depends on $p({\cal H}_{0,i})$ and $p({\cal H}_{1,i})$. For $p({\cal H}_{0,i})=p({\cal H}_{1,i})=1/2$, \cite{communication_schwartz} provides {\blue an} accurate approximation of $\zeta_k$ {\blue as} $\zeta_k=\sqrt{2+\gamma_k}$.\\
%
%
Finally, by substituting \eqref{epsilon1 and epsilon2 for known channel envelope receiver} in \eqref{alpha_{k,t,l} for noncoherent}, we compute $\alpha_{k,t,l}$ for noncoherent receiver with known channel envelopes.

$\bullet$ \textbf{Noncoherent Receiver with Known Channel Statistics:} For this receiver, the test statistics of LRT at the FC is the power of channel output, i.e., $r_{k,i}=|y_{k,i}|^2$. The FC only knows the {\blue channel} statistics $h_k\sim \mathcal{CN}\left(0,2\sigma_{h_k}^2\right)$. Let $\overline{\gamma}_k$ denote the average channel SNR of sensor $k$, where:
\vspace{-0.25cm}
\begin{equation} \label{average SNR for k-th channel}
\overline{\gamma}_k=\mathbb{E}\{\gamma_k\}=\frac{P_k\mathbb{E}\{|h_k|^2\}}{2L_k\sigma_{w_k}^2}=\frac{P_k\sigma_{h_k}^2}{L_k\sigma_{w_k}^2},
\end{equation}
in which we have used the knowledge of channel statistics to obtain $\mathbb{E}\{|h_k|^2\}=2\sigma_{h_k}^2$. Since $y_{k,i}$ is complex Gaussian, we have \cite{Varshney_2009}:
\vspace{-0.4cm}
\begin{equation*} 
f\left(r_{k,i}|{\cal H}_{0,i}\right)=\frac{1}{2\sigma_{w_k}^2}e^{-\frac{r_{k,i}}{2\sigma_{w_k}^2}},\qquad\qquad\qquad\quad
\end{equation*}
\vspace{-0.5cm}
\begin{equation*} 
f\left(r_{k,i}|{\cal H}_{1,i}\right)=\frac{1}{2\sigma_{w_k}^2\left(1+2\overline{\gamma}_k\right)}e^{-\frac{r_{k,i}}{2\sigma_{w_k}^2\left(1+2\overline{\gamma}_k\right)}}.
\end{equation*}
Note that $r_{k,i}$'s conditioned on each hypothesis are i.i.d. for $i\!=\!1, \dots, L_k$ and therefore the probabilities ${\cal E}_{1_k}$ and ${\cal E}_{2_k}$ do not depend on bit index $i$. Hence:
\vspace{-0.2cm}
\begin{subequations} \label{epsilon1 and epsilon2 for known channel statistics receiver}
\begin{flalign} 
{\cal E}_{1_k}&=p\left(r_{k,i}>\zeta_k|{\cal H}_{0,i}\right)=(\frac{1}{2\overline{\gamma}_k+1})^{\frac{2\overline{\gamma}_k+1}{2\overline{\gamma}_k}}\label{eq:subeq_eps1_2_known_stat_1},&\\
{\cal E}_{2_k}&=p\left(r_{k,i}<\zeta_k|{\cal H}_{1,i}\right)=1-(\frac{1}{2\overline{\gamma}_k+1})^{\frac{1}{2\overline{\gamma}_k}}\label{eq:subeq_eps1_2_known_stat_2},&
\end{flalign}
\end{subequations}
in which the decision threshold $\zeta_k$ for $p({\cal H}_{0,i})\!=\!p({\cal H}_{1,i})\!=\!1/2$ is $\zeta_k=2\sigma_{w_k}^2(1+\frac{1}{2\overline{\gamma}_k})\ln\left(1+2\overline{\gamma}_k\right)$.\\
Finally, by substituting \eqref{epsilon1 and epsilon2 for known channel statistics receiver} in \eqref{alpha_{k,t,l} for noncoherent}, we compute $\alpha_{k,t,l}$ for noncoherent receiver with known channel statistics\footnote{ 
When the ratio $\frac{p\left({\cal H}_{0,i}\right)}{p\left({\cal H}_{1,i}\right)}\!=\!\tau_{FC}\!\ne\!1$ in \eqref{binary hyp test problem}, the expressions for the decision threshold $\zeta_k$ change. For noncoherent receiver with known channel envelopes, one can analytically find for each $\gamma_k$ the value of $\zeta_k$ which minimizes the average error probability corresponding to demodulating the symbols of sensor $k$ given as $p_{e_k}=p\left({\cal H}_{0,i}\right){\cal E}_{1_k}+p\left({\cal H}_{1,i}\right){\cal E}_{2_k}$. Equivalently, $\zeta_k$ satisfies $e^{-2\gamma_k}I_0\left(2\zeta_k\sqrt{\gamma_k}\right)=\tau_{FC}$. For noncoherent receiver with known channel statistics, we obtain $\zeta_k=2\sigma_{w_k}^2(1+\frac{1}{2\overline{\gamma}_k})\ln\left(\tau_{FC}(1+2\overline{\gamma}_k\right))$.
}. 
\vspace{-.45cm}
\subsection{Finding {\blue Bayesian} \textnormal{FIM} $\boldsymbol{J}$ in \eqref{FIM in decomposed form} } \label{finding FIM final}
\vspace{-.1cm}
At this point, we have all the components to write the entries $[\boldsymbol{\Lambda}(\boldsymbol{\theta})]_{ij}$ in \eqref{ij-th element of second term of FIM}. Combining {\blue \eqref{ij-th element of lambda_theta 1}-\eqref{der of P(m_hat given theta) wrt theta_i}}, we find the following compact form representation of $[\boldsymbol{\Lambda}(\boldsymbol{\theta})]_{ij}$:
\vspace{-.2cm}
\begin{equation} \label{ij-th element of lambda_theta 2}
[\boldsymbol{\Lambda}(\boldsymbol{\theta})]_{ij}=\frac{1}{2\pi}\sum_{k=1}^{K} \frac{a_{k_i}a_ {k_j}}{\sigma_{n_k}^2}G_k(\boldsymbol{\theta}),
\end{equation}
where the scalar $G_k(\boldsymbol{\theta})$ is:
\vspace{-.2cm}
\begin{equation} \label{G_k(theta)}
G_k(\boldsymbol{\theta})=\sum_{t=1}^{M_k}\frac{{(\sum_{l=1}^{M_k}\alpha_{k,t,l}\dot{\beta}_{k,l}
(\boldsymbol{\theta}))}^2}{\sum_{l=1}^{M_k}\alpha_{k,t,l}\beta_{k,l}(\boldsymbol{\theta})}.
\end{equation}
%
Finally, we compute $\mathbb{E}\{\boldsymbol{\Lambda}(\boldsymbol{\theta})\}$ and substitute it in \eqref{FIM in decomposed form} to obtain matrix $\boldsymbol{J}$ as:
\vspace{-0.3cm}
\begin{align} \label{final formula for FIM}
\boldsymbol{J}&={\boldsymbol{\cal C}_{\boldsymbol{\theta}}^{-1}}+\frac{1}{2\pi}\sum_{k=1}^{K}\frac{\mathbf{a}_k\mathbf{a}_k^T}{\sigma_{n_k}^2}\mathbb{E}\{G_k(\boldsymbol{\theta})\}\\
&={\boldsymbol{\cal C}_{\boldsymbol{\theta}}^{-1}}+\frac{1}{2\pi}{\boldsymbol{\mathcal A}}\,\text{diag}(\frac{\mathbb{E}\{G_1(\boldsymbol{\theta})\}}{\sigma_{n_1}^2}, ..., \frac{\mathbb{E}\{G_K(\boldsymbol{\theta})\}}{\sigma_{n_K}^2}){\boldsymbol{\mathcal A}}^T,\notag
\end{align}
%
where the columns of $\boldsymbol{\mathcal A}\!=\![\mathbf{a}_1, ..., \mathbf{a}_K]$ are observation vectors in \eqref{obs_model} and the expectations over $\boldsymbol{\theta}$ in \eqref{final formula for FIM} are computed using numerical integration.

{\blue For $\boldsymbol{J}$ in \eqref{final formula for FIM} there exists two baselines. For the first baseline}, suppose all sensors' observations $x_k$'s are available at the FC with full precision (centralized estimation) and let {\blue $\boldsymbol{J}_0\!=\!{\boldsymbol{\cal C}_{\boldsymbol{\theta}}^{-1}}\!+\!\mathbb{E}\{{\boldsymbol{\Lambda}}_0(\boldsymbol{\theta})\}$} be the corresponding {\blue Bayesian} FIM. To find $[{\boldsymbol{\Lambda}}_0(\boldsymbol{\theta})]_{ij}$, we start from \eqref{ij-th element of second term of FIM} and replace $p(\hat{m}_{k,t}\arrowvert\boldsymbol{\theta})$ with $f\!\left(x_k|\boldsymbol{\theta}\right)$. Following the same procedure as we described to obtain \eqref{ij-th element of lambda_theta 1} from \eqref{ij-th element of second term of FIM}, we reach: 
\vspace{-0.15cm}
\begin{equation*}
[{\boldsymbol{\Lambda}}_0(\boldsymbol{\theta})]_{ij}=\sum_{k=1}^{K}\int_{x_k}\!\!(\frac{1}{f\left(x_k|\boldsymbol{\theta}\right)}\frac{\partial f\left(x_k|\boldsymbol{\theta}\right)}{\partial \theta_i}\frac{\partial f\left(x_k|\boldsymbol{\theta}\right)}{\partial \theta_j})dx_k.
\end{equation*}
Since $\frac{\partial f\left(x_k|\boldsymbol{\theta}\right)}{\partial\theta_i}=\frac{a_{k_i}(x_k-\mathbf{a}_k^T\boldsymbol{\theta})}{\sigma_{n_k}^2}f\left(x_k|\boldsymbol{\theta}\right)$,
%
%
it is straightforward to show $[{\boldsymbol{\Lambda}}_0(\boldsymbol{\theta})]_{ij}=\sum_{k=1}^{K}\!\frac{a_{k_i}a_ {k_j}}{\sigma_{n_k}^2}$. Therefore:
%
\vspace{-0.15cm}
\begin{equation} \label{clairvoyant FIM}
\boldsymbol{J}_0={\boldsymbol{\cal C}_{\boldsymbol{\theta}}^{-1}}+{\boldsymbol{\mathcal A}}\,\text{diag}(\frac{1}{\sigma_{n_1}^2}, ..., \frac{1}{\sigma_{n_K}^2}){\boldsymbol{\mathcal A}}^T.
\end{equation}

{\blue 
For the second baseline, suppose communication channels between sensors and the FC are error-free and hence vector $\boldsymbol{m}$ is available at the FC. Let $\boldsymbol{J}^{ideal}\!=\!{\boldsymbol{\cal C}_{\boldsymbol{\theta}}^{-1}}\!+\!\mathbb{E}\{{\boldsymbol{\Lambda}}^{ideal}(\boldsymbol{\theta})\}$ be the corresponding Bayesian FIM. To find $G_k^{ideal}(\boldsymbol{\theta})$ for entries $[{\boldsymbol{\Lambda}}^{ideal}(\boldsymbol{\theta})]_{ij}$ using \eqref{ij-th element of lambda_theta 2} we note that $\alpha_{k,t,l}\!=\!1$ for $t\!=\!l$ and $\alpha_{k,t,l}\!=\!0$ otherwise, since the channel error probabilities (${\cal E}_k$ for coherent receiver, ${\cal E}_{1_k}, {\cal E}_{2_k}$ for noncoherent receivers) are zero. Therefore, from \eqref{ij-th element of lambda_theta 2} we find $G_k^{ideal}(\boldsymbol{\theta})\!=\!\sum_{t=1}^{M_k}\frac{{(\dot{\beta}_{k,t}(\boldsymbol{\theta}))}^2}{\beta_{k,t}(\boldsymbol{\theta})}$. Clearly, $\boldsymbol{J} \preceq \boldsymbol{J}^{ideal} \preceq \boldsymbol{J}_0$.}
%
\begin{remark}
\textup{ If $\boldsymbol{\theta}$ has a known nonzero-mean $\boldsymbol{\mu}_{\theta}$, sensor $k$ subtracts $\mathbf{a}_k^T\boldsymbol{\mu}_{\theta}$ from its observation $x_k$, before quantization. At the FC, $\mathbf{a}_k^T\boldsymbol{\mu}_{\theta}$ is first added to $\hat{m}_k$ to generate $\tilde{m}_k\!=\!\hat{m}_k+\mathbf{a}_k^T\boldsymbol{\mu}_{\theta}$ and then the Bayesian estimator $\hat{\boldsymbol{\theta}}$ is formed using $\boldsymbol{\tilde{m}}=[\tilde{m}_1,...,\tilde{m}_K]^T$. Thus, the corresponding {\blue Bayesian} FIM matrix $\boldsymbol{\tilde{J}}$ becomes:
\vspace{-0.2cm}
\begin{equation*} 
\boldsymbol{\tilde{J}}=\mathbb{E}\{(\frac{\partial \ln p_{\boldsymbol{\tilde{m}}\boldsymbol{\theta}}(\boldsymbol{\tilde{m}},\boldsymbol{\theta})}{\partial \boldsymbol{\theta}}){(\frac{\partial \ln p_{\boldsymbol{\tilde{m}}\boldsymbol{\theta}}(\boldsymbol{\tilde{m}},\boldsymbol{\theta})}{\partial \boldsymbol{\theta}})}^T\},
\end{equation*}
where the joint pdf $p_{\boldsymbol{\tilde{m}}\boldsymbol{\theta}}(\boldsymbol{\tilde{m}},\boldsymbol{\theta})=p_{\boldsymbol{\hat{m}}\boldsymbol{\theta}}(\boldsymbol{\tilde{m}}-\boldsymbol{\mathcal A}^T\boldsymbol{\mu}_{\theta},\boldsymbol{\theta})$. Noting that $\boldsymbol{\tilde{m}}-\boldsymbol{\mathcal A}^T\boldsymbol{\mu}_{\theta}=\boldsymbol{\hat{m}}$, we follow the same procedure as we conducted before to obtain $\boldsymbol{J}$ in \eqref{final formula for FIM} and we find that $\boldsymbol{\tilde{J}}$ has the same expression as $\boldsymbol{J}$ with the only difference that $\boldsymbol{\mathcal C}_{\boldsymbol{\theta}}=\mathbb{E}\{\boldsymbol{\theta}\boldsymbol{\theta}^T\}-\boldsymbol{\mu}_{\theta}{\boldsymbol{\mu}_{\theta}}^T$ for nonzero-mean $\boldsymbol{\theta}$.}
\end{remark}
\vspace{-0.25cm}
{\blue 
\section{WWB Bound: Derivation and Computation} \label{WWB bound}
\vspace{-0.05cm}
The MSE matrix of any Bayesian estimator $\hat{\boldsymbol{\theta}}$ of random vector  $\boldsymbol{\theta} \in \mathbb{R}^q$ satisfies the following inequality \cite{WWB,WWB_chng_pnt_2017}:
\vspace{-0.1cm}
\begin{equation}\label{WWB-inequality}
\text{MSE}_{\hat{\boldsymbol{\theta}}} \succeq \boldsymbol{R}{\boldsymbol{G}}^{-1}{\boldsymbol{R}}^T,
\end{equation}
where the columns of $q\times q$ matrix $\boldsymbol{R}={[\boldsymbol{r}_1, \boldsymbol{r}_2, ..., \boldsymbol{r}_q]}^T$, so-called test points, lie in the parameter space and their choices are left to the user \cite{WWB,WWB_chng_pnt_2017}. The $q\times q$  matrix ${\boldsymbol{G}}$ is defined by its entries $[{\boldsymbol{G}}]_{ij}$, which are computed as follows \cite{WWB}:
\vspace{-.1cm}
\begin{equation}\label{entries-of-G}
[{\boldsymbol{G}}]_{ij}\!=\!2\frac{\exp{\left(\mu(\boldsymbol{r}_i\!-\!\boldsymbol{r}_j)\right)}\!-\!\exp{\left(\mu(\boldsymbol{r}_i\!+\!\boldsymbol{r}_j)\right)}}{\exp{\left(\mu(\boldsymbol{r}_i)\right)}+\exp{\left(\mu(\boldsymbol{r}_j)\right)}},\ \ \ i,j=1, ..., q
\end{equation}
The inequality in (\ref{WWB-inequality}) holds for any $\boldsymbol{R}$ such that $\boldsymbol{G}$ in invertible \cite{WWB,WWB_chng_pnt_2017}. 
%
%
Maximizing the right side of (\ref{WWB-inequality}) with respect to $\boldsymbol{R}$  leads to the tightest WWB, denoted as $\boldsymbol{W}\boldsymbol{W}\boldsymbol{B}$. In other words:
%
\begin{equation}\label{supremum}
\boldsymbol{W}\boldsymbol{W}\boldsymbol{B}=\mathop{\text{supremum}}_{\boldsymbol{R} }\ \ \boldsymbol{R}{\boldsymbol{G}}^{-1}{\boldsymbol{R}}^T,
\end{equation}
where the supremum operation is taken with respect to Loewner partial ordering \cite{WWB_chng_pnt_2017}.
To find $\boldsymbol{W}\boldsymbol{W}\boldsymbol{B}$ in our problem, first we need to derive the entries $[{\boldsymbol{G}}]_{ij}$, or equivalently scalar $\mu(\boldsymbol{r})$ in (\ref{entries-of-G}). After deriving $\mu(\boldsymbol{r})$, we discuss how to compute the supremum in (\ref{supremum}).
\vspace{-0.2cm}
\subsection{Deriving $\mu(\boldsymbol{r})$ in (\ref{entries-of-G}) Based on Our System Model} \label{Deriving mu(r)}
\vspace{-0.1cm}
Using equation (43) in \cite{WWB} and the Bayes' rule to write $p(\hat{\boldsymbol{m}},\boldsymbol{\theta}\!+\!\boldsymbol{r})\!=\!p(\hat{\boldsymbol{m}}|\boldsymbol{\theta}\!+\!\boldsymbol{r})f(\boldsymbol{\theta}\!+\!\boldsymbol{r})$ and $p(\hat{\boldsymbol{m}},\boldsymbol{\theta})\!=\!p(\hat{\boldsymbol{m}}|\boldsymbol{\theta})f(\boldsymbol{\theta})$
we find:
\vspace{-.25cm}
\begin{align} \label{mu-r-WWB-2}
\mu(\boldsymbol{r})&=\ln[\int_{V_{\theta}}f^{\frac{1}{2}}(\boldsymbol{\theta}+\boldsymbol{r})f^{\frac{1}{2}}(\boldsymbol{\theta})\sum_{\hat{m}_1}\dots\sum_{\hat{m}_K}p^{\frac{1}{2}}(\hat{\boldsymbol{m}}|\boldsymbol{\theta}+\boldsymbol{r})\nonumber\\
&\times p^{\frac{1}{2}}(\hat{\boldsymbol{m}}|\boldsymbol{\theta})d\boldsymbol{\theta}],
\end{align}
%
%
%
where $V_{\theta}$ denotes the $q$-dimensional volume over which we take integral and $p^{\frac{1}{2}}(.,.)$ is the square root of the joint pdf. 
%
%
%
%
To characterize $\mu(\boldsymbol{r})$ in (\ref{mu-r-WWB-2}) we need to find $p(\hat{\boldsymbol{m}}|\boldsymbol{\theta})$, $p(\hat{\boldsymbol{m}}|\boldsymbol{\theta}+\boldsymbol{r})$, and $f^{\frac{1}{2}}(\boldsymbol{\theta}+\boldsymbol{r})f^{\frac{1}{2}}(\boldsymbol{\theta})$.
Let index $t$ indicate the quantization level corresponding to $\hat{m}_k$.
According to Lemma \ref{p(m_hat|theta)}, the followings are evident:
\vspace{-.25cm}
\begin{equation} \label{p(mhat|theta) and p(mhat|theta+r)}
p(\hat{\boldsymbol{m}}|\boldsymbol{\theta})\!\!=\!\!\prod_{k=1}^{K}\!p(\hat{m}_{k,t}|\boldsymbol{\theta}), \ p(\hat{\boldsymbol{m}}|\boldsymbol{\theta}+\boldsymbol{r})\!\!=\!\!\prod_{k=1}^{K}\!p(\hat{m}_{k,t}|\boldsymbol{\theta}+\boldsymbol{r}),
\end{equation}
where $p(\hat{m}_{k,t}|\boldsymbol{\theta})$ is given in \eqref{P(m_hat given theta) based on alpha and beta}, and $p(\hat{m}_{k,t}|\boldsymbol{\theta}+\boldsymbol{r})$ can be computed with a simple substitution of $\boldsymbol{\theta}$ by $\boldsymbol{\theta}+\boldsymbol{r}$ in \eqref{P(m_hat given theta) based on alpha and beta}.
%
%
Moreover, some easy manipulations yield:
\vspace{-.2cm}
\begin{align} \label{f(theta)*f(theta+r)}
f^{\frac{1}{2}}(\boldsymbol{\theta}+\boldsymbol{r})f^{\frac{1}{2}}(\boldsymbol{\theta})&=\frac{\exp{(-\frac{{\boldsymbol{r}}^T\boldsymbol{\mathcal C}_{\boldsymbol{\theta}}^{-1}\boldsymbol{r}}{8})}}{\sqrt{{(2\pi)}^q|\boldsymbol{\mathcal C}_{\boldsymbol{\theta}}|}}\\
&\times\exp{(-\frac{1}{2}{(\boldsymbol{\theta}+\frac{1}{2}\boldsymbol{r})}^T\boldsymbol{\mathcal C}_{\boldsymbol{\theta}}^{-1}(\boldsymbol{\theta}+\frac{1}{2}\boldsymbol{r}))}.\nonumber
\end{align}
Substituting \eqref{p(mhat|theta) and p(mhat|theta+r)} and \eqref{f(theta)*f(theta+r)} in \eqref{mu-r-WWB-2} and some straightforward manipulations produce:
\vspace{-.4cm}
\begin{align*}
\ \ \ \mu(\boldsymbol{r})&=c_q(\boldsymbol{r})+\ln[\int_{V_{\theta}}\exp(-\frac{1}{2}{(\boldsymbol{\theta}+\frac{1}{2}\boldsymbol{r})}^T\boldsymbol{\mathcal C}_{\boldsymbol{\theta}}^{-1}(\boldsymbol{\theta}+\frac{1}{2}\boldsymbol{r}))\nonumber\\
&\times\prod_{k=1}^{K}\sum_{t_k=1}^{M_k}{p^{\frac{1}{2}}(\hat{m}_{k,t}|\boldsymbol{\theta}) p^{\frac{1}{2}}(\hat{m}_{k,t}|\boldsymbol{\theta}+\boldsymbol{r})}d\boldsymbol{\theta}],
\end{align*}
where $c_q(\boldsymbol{r})=-\frac{q}{2}\ln(2\pi)-\frac{1}{2}\ln|\boldsymbol{\mathcal C}_{\boldsymbol{\theta}}|-\frac{{\boldsymbol{r}}^T\boldsymbol{\mathcal C}_{\boldsymbol{\theta}}^{-1}\boldsymbol{r}}{8}$.

%

\vspace{-0.35cm}
\subsection{Computation of the Tightest WWB} \label{tightest WWB}
\vspace{-0.1cm}
%
%
%
In the following, we explain how we compute the supremum in (\ref{supremum}). 
We note that the method to compute the supremum in (\ref{supremum}) does not depend on the system model (it only depends on the parameter space). Therefore, we adopt the same method as in \cite{WWB_chng_pnt_2017}.
Let $\boldsymbol{W}(\boldsymbol{R})=\boldsymbol{R}{\boldsymbol{G}}^{-1}{\boldsymbol{R}}^T$ and define set: $${\cal W}=\{\boldsymbol{W}(\boldsymbol{R})|\boldsymbol{R}~\text{is~chosen~such~that}~ \boldsymbol{G}\succ\boldsymbol{0}\}.$$ Then $\boldsymbol{W}\boldsymbol{W}\boldsymbol{B}$ is the supremum of set ${\cal W}$, where the supremum operation is taken with respect to Loewner partial ordering \cite{WWB_chng_pnt_2017}.
%
%
%
%
%
%
It is worth mentioning the difference between the maximum and the supremum of the set $\cal W$. The largest element of $\cal W$, if it exists, is defined as $\boldsymbol{W}\preceq{\boldsymbol{W}}^{*},\forall \boldsymbol{W}\in{\cal W}$. On the other hand, the supremum of $\cal W$ is a minimal-upper bound on $\cal W$ that is not necessarily contained in $\cal W$. This implies that the largest element of $\cal W$ may not exist, but if it exists, it is also the supremum. 

According to Lemma 3 of \cite{Nehorai_TSP_2010} for any two positive definite matrices $\boldsymbol{A}$ and $\boldsymbol{B}$ we have $\boldsymbol{A}\!\succeq\!\boldsymbol{B}$ if and only if $\varepsilon(\boldsymbol{A})\!\supseteq\!\varepsilon(\boldsymbol{B})$,
in which the hyper-ellipsoid $\varepsilon(\boldsymbol{A})$ centered at the origin can be represented by the set $\varepsilon(\boldsymbol{A})\!=\!\{\boldsymbol{z}|{\boldsymbol{z}}^T\!{\boldsymbol{A}}^{-1}\boldsymbol{z}\!\leq \!1\}$. Consequently, the supremum in (\ref{supremum}) can be computed by finding the minimum volume hyper-ellipsoid $\varepsilon({\boldsymbol{W}}^{*})$ containing the set $\varepsilon_{\cal W}\!=\!\{\varepsilon(\boldsymbol{W})|\boldsymbol{W}\!\in\!{\cal W}\}$, where the set $\varepsilon_{\cal W}$ itself consists of the hyper-ellipsoids generated by all matrices in ${\cal W}$.
%
%
%
The problem of finding the minimum volume ellipsoid $\varepsilon$ that contains the ellipsoids $\varepsilon_1, ..., \varepsilon_m$ (and therefore the convex hull of their union) has been formulated as a convex problem in \cite{Boyd_Book}:
%
%
%
\vspace{-.2cm}
\begin{align*}
\mathop{\text{minimize}}_{\boldsymbol{W}, b_i,\forall i}\ \ \ \ &\text{log}(\text{det}(\boldsymbol{W}^{\frac{1}{2}}))\\
\vspace{-0.1cm}
\text{s.t.}\ \ \ \ &b_i\geq 0,\\ 
&\begin{bmatrix}
\boldsymbol{W}^{-1}-b_i{\boldsymbol{W}}_i^{-1} & \boldsymbol{0}\\
\boldsymbol{0} & b_i-1
\end{bmatrix}\preceq\boldsymbol{0}, \ \ i=1,..., |\cal W|,
\end{align*}
where ${\boldsymbol{W}}_i\in{\cal W}$ and $|\cal W|$ is the cardinality of the set ${\cal W}$.
This problem can be solved efficiently using semidefinite programming. In particular, we solve this problem using CVX.
}
\vspace{-.5cm}
\section{Power Constrained {\blue Bayesian} Fisher Information Maximization} \label{Power Allocation}
\vspace{-.1cm}
In this section, we address the constrained optimization problems formulated in \eqref{maximization problem of tr(J)} and \eqref{maximization problem of logdet(J)}. We denote the solutions obtained from solving these two power constrained Fisher information maximization problems as {\it FIM-max schemes}.
Note that due to the cap on the network average transmit power, only a subset of the sensors might be active during each task period, which we refer to as the set of active sensors $S_{\cal A}=\{k: P_k>0,\ k=1, \dots, K\}$.
\vspace{-.25cm}
\subsection{Solving Optimization Problem in \eqref{maximization problem of tr(J)}} \label{solving max of tr(J)} 
\begin{figure}[!t]
	
	\centering
	\hspace{-.5cm}
	\begin{subfigure}[b]{0.25\textwidth}
		
		\centering
		\subcaptionbox{ different $\delta_k$, $\sigma_{n_k}\!=\!1$\label{der-tr-J-vs-P-k-Chaud-1}}{\vspace{-.2 cm}\includegraphics[width=1.8in,height=1.1in]{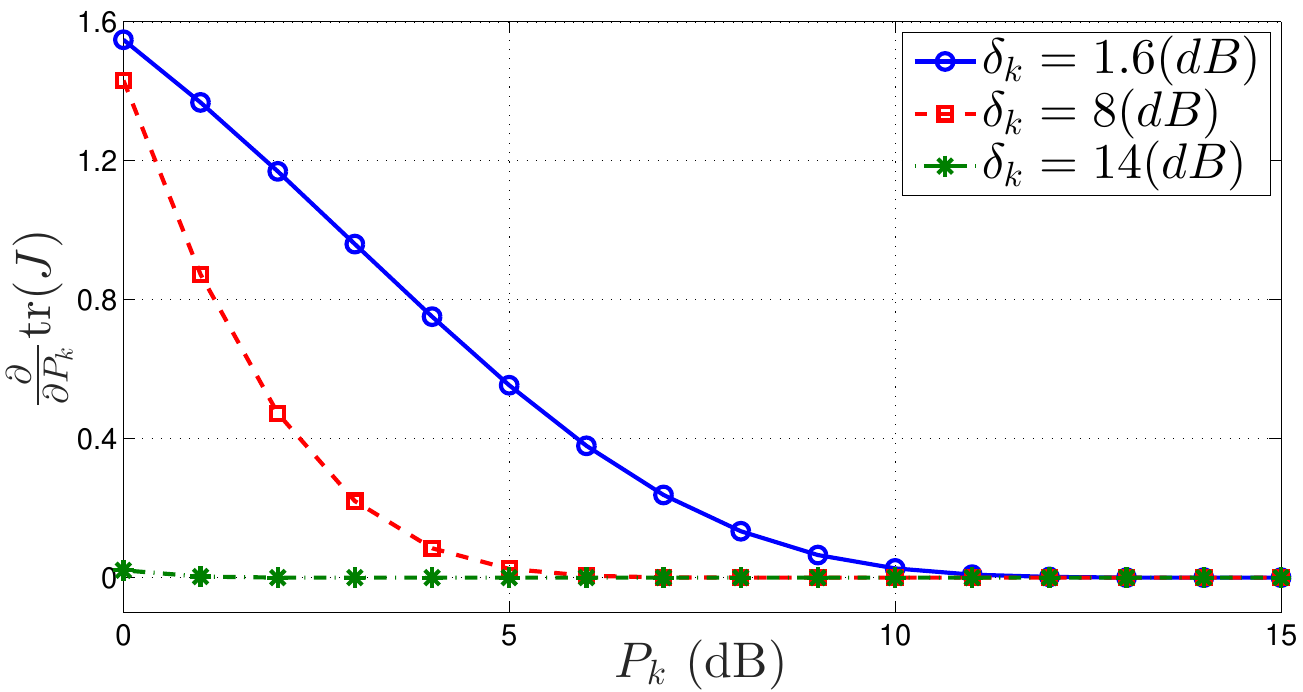}}
		
	\end{subfigure}%
	\begin{subfigure}[b]{0.25\textwidth}
		
		\centering
		\subcaptionbox{ different $\sigma_{n_k}$, $\delta_k\!=\!4$ dB\label{der-tr-J-vs-P-k-Chaud-2}}{\vspace{-.2 cm}\includegraphics[width=1.8in,height=1.1in]{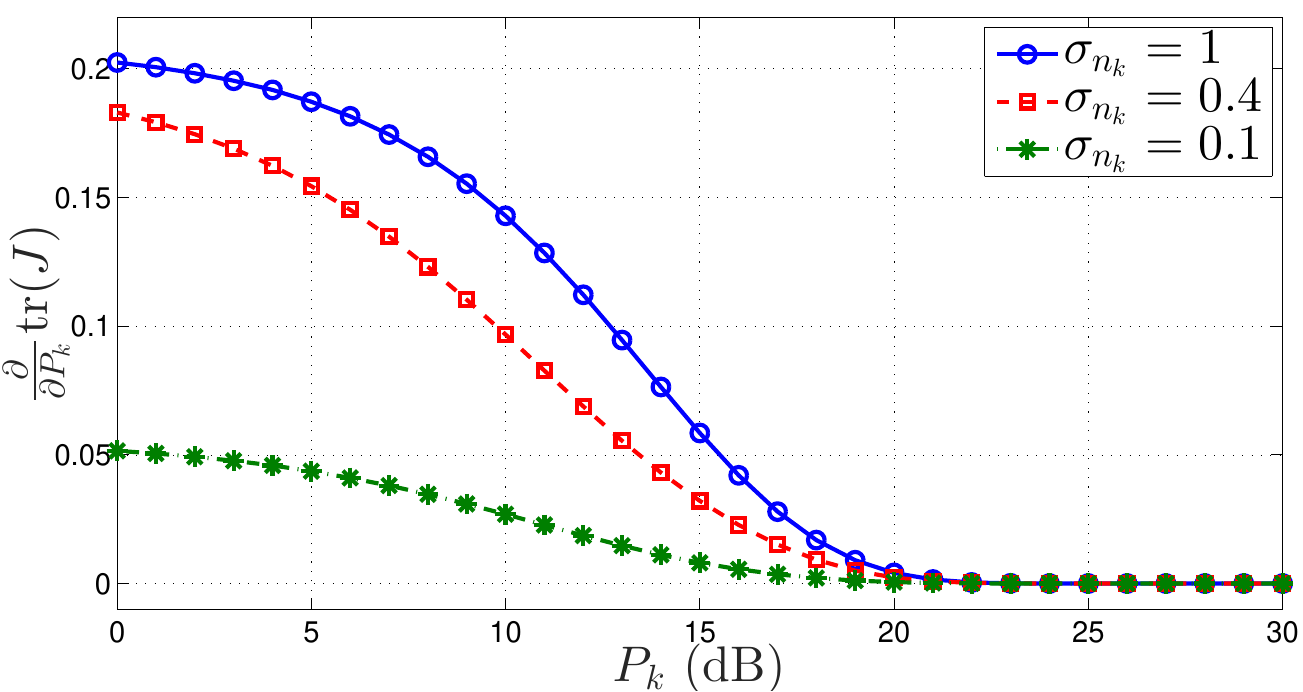}}
		
	\end{subfigure} \\
	
	\caption{coherent receiver: behavior of  $\frac{\partial\,\text{tr}(\boldsymbol{J})}{\partial P_k}$ versus $P_k$ (dB) for different values of (a) $\delta_k$ and (b) $\sigma_{n_k}$.}   
	\label{der-tr-J-vs-P-k-Chaud}
	
\end{figure}

\vspace{-.1cm}
We adopt the Lagrange multipliers method to solve the problem {\blue . The} Lagrangian $\mathcal L$ {\blue of this problem is}:
\vspace{-0.25cm}
\begin{equation} \label{Lagrangian}
\mathcal L(\lambda,\{\eta_k,P_k\}_{k=1}^K)\!=\!\text{tr}(\boldsymbol{J})\!-\!\sum_{k=1}^{K}P_k\left(\lambda-\eta_k\right)\!+\!\lambda P_{tot}.
\end{equation}
The Karush-Kuhn-Tucker (KKT) optimality conditions are:
\vspace{-.15cm}
\begin{align} \label{KKT cond. for problem in (3)}
&\frac{\partial \mathcal L}{\partial P_k}=\frac{\partial\,\text{tr}(\boldsymbol{J})}{\partial P_k}-\lambda+\eta_k=0,\ \forall k,\\
&\lambda (\sum_{k=1}^{K}P_k-P_{tot})=0,\ \lambda\geq 0,\ \sum_{k=1}^{K}P_k\leq P_{tot},\notag\\
&\eta_kP_k=0,\ \eta_k\geq 0,\ P_k\geq 0,\ \forall k,\notag
\end{align}
where $\lambda, \eta_k$'s are the Lagrange multipliers.
According to \eqref{final formula for FIM} we find:
\vspace{-.25cm}
\begin{align} \label{derivative of tr(J) wrt P_k}
\frac{\partial\,\text{tr}(\boldsymbol{J})}{\partial P_k}=\frac{\mathbf{a}_k^T\mathbf{a}_k}{2\pi\sigma_{n_k}^2}\mathbb{E}\{\frac{\partial\,G_k(\boldsymbol{\theta})}{\partial P_k}\}.
\end{align}
Thus, to show $\frac{\partial\,\text{tr}(\boldsymbol{J})}{\partial P_k}\!>\!0$, we need to show $\mathbb{E}\{\!\frac{\partial\,G_k(\boldsymbol{\theta})}{\partial P_k}\!\}\!>\!0$. {\blue Although we were not able to prove analytically, our extensive simulations for various system parameters} indicate that $\mathbb{E}\{\!\frac{\partial\,G_k(\boldsymbol{\theta})}{\partial P_k}\!\}\!>\!0$ and thus $\frac{\partial\,\text{tr}(\boldsymbol{J})}{\partial P_k}\!>\!0$. 
Fig.~\ref{der-tr-J-vs-P-k-Chaud} summarizes our extensive simulations to demonstrate $\frac{\partial\,\text{tr}(\boldsymbol{J})}{\partial P_k}\!>\!0$, for coherent receiver. 
To obtain this figure, we let $K\!=\!2$ and consider a zero-mean Gaussian vector $\boldsymbol{\theta}\!=\!\left[\theta_1,\theta_2\right]^T$ with $\boldsymbol{\cal C}_{\boldsymbol{\theta}}\!=\![4,0.5;0.5,0.25]$. We assume $L_k\!=\!3$, $\mathbf{a}_k\!=\![0.6,0.8]^T, \forall k$, and vary $|h_k|$, $\sigma_{w_k}$, $\sigma_{n_k}$ and use the uniform quantizer described in Section \ref{simulation}. 
Let $\delta_k\!=\!\frac{|h_k|^2}{2\sigma_{w_k}^2}$. For coherent receiver, Fig.~\ref{der-tr-J-vs-P-k-Chaud-1} and Fig.~\ref{der-tr-J-vs-P-k-Chaud-2} depict $\frac{\partial\,\text{tr}(\boldsymbol{J})}{\partial P_k}$ versus $P_k$ for different values of $\delta_k$  and $\sigma_{n_k}$, respectively. We observe that, for all different values of $\delta_k$ and $\sigma_{n_k}$, we have $\frac{\partial\,\text{tr}(\boldsymbol{J})}{\partial P_k}\!>\!0$, $\forall P_k$. 
{\blue Similar observations were made for} both types of noncoherent receivers. However, due to lack of space we have omitted those plots. 

\begin{figure}[!t]
	
	\centering
	\hspace{-.5cm}
	\begin{subfigure}[b]{0.25\textwidth}
		
		\centering
		\subcaptionbox{different $\delta_k$, $\sigma_{n_k}\!=\!1$\label{sec-der-tr-J-vs-P-k-Chaud-1}}{\vspace{-.2 cm}\includegraphics[width=1.8in,height=1.1in]{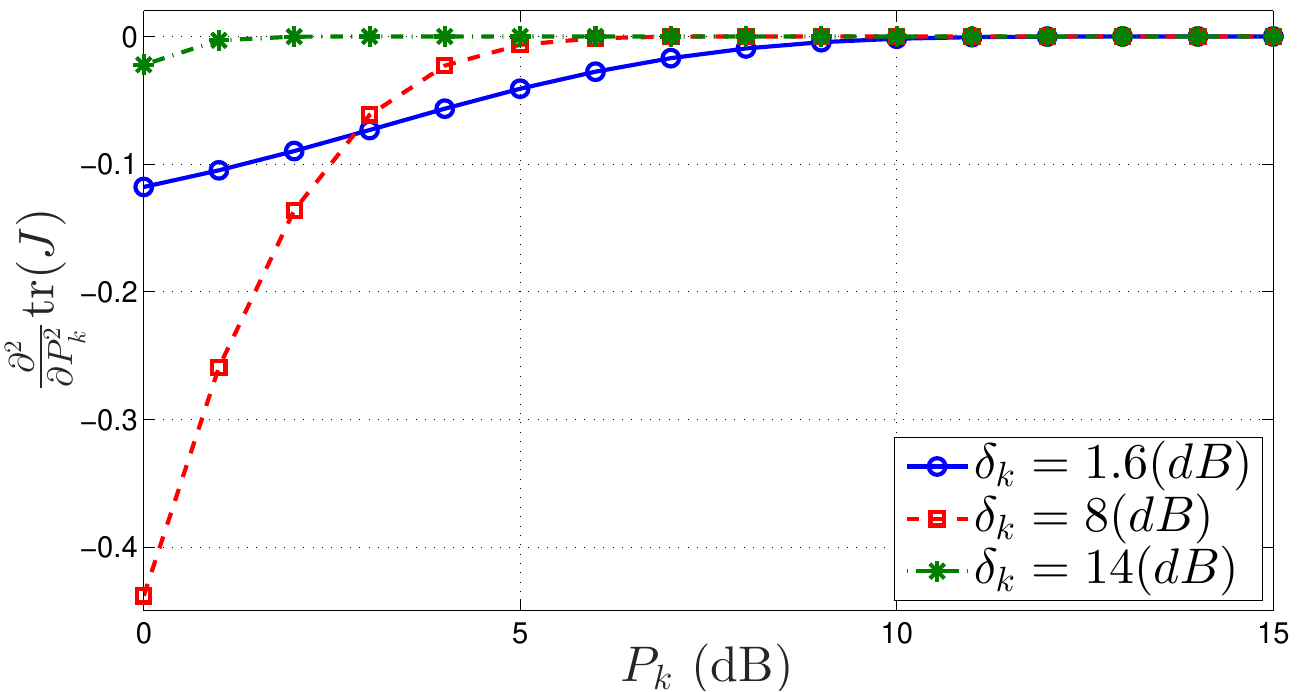}}
		
	\end{subfigure}%
	\begin{subfigure}[b]{0.25\textwidth}
		
		\centering
		\subcaptionbox{different $\sigma_{n_k}$, $\delta_k\!=\!4$ dB\label{sec-der-tr-J-vs-P-k-Chaud-2}}{\vspace{-.2 cm}\includegraphics[width=1.8in,height=1.1in]{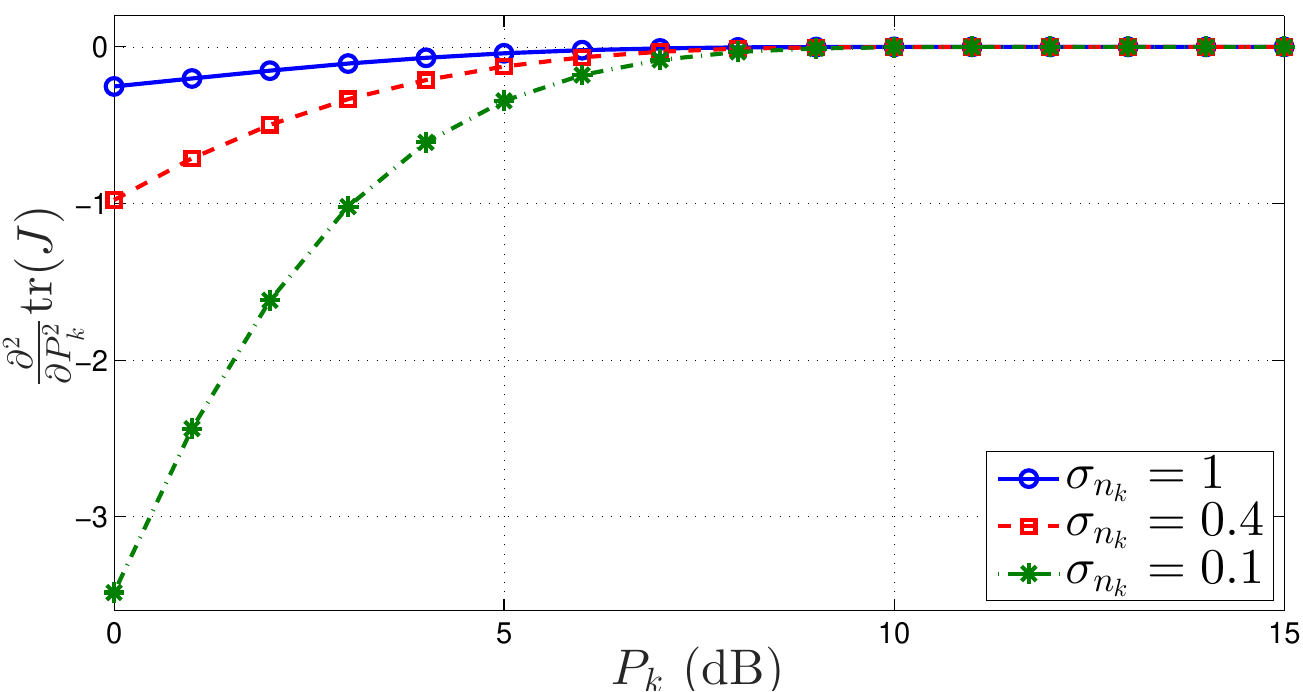}}
		
	\end{subfigure} \\
	
	\caption{coherent receiver: behavior of $\frac{\partial^2\,\text{tr}(\boldsymbol{J})}{\partial P_k^2}$ versus $P_k$ (dB) for different values of (a) $\delta_k$ and (b) $\sigma_{n_k}$.}  
	\label{sec-der-tr-J-vs-P-k-Chaud}
	
\end{figure}

Since tr$(\boldsymbol{J})$ is an increasing function of $P_k$'s, the Lagrange multiplier $\lambda$ in \eqref{KKT cond. for problem in (3)} should be determined such that it satisfies the network average transmit power constraint with equality, that is, $\sum_{k\in S_{\cal A}}P_k=P_{tot}$. Furthermore, for the set of active sensors $S_{\cal A}$ the Lagrange multiplier $\eta_k=0$. Hence, we can reformulate the KKT optimality conditions in \eqref{KKT cond. for problem in (3)} as:
\vspace{-0.2cm}
\begin{align} \label{KKT conditions for max. tr(J)}
\frac{\mathbf{a}_k^T\mathbf{a}_k}{2\pi\sigma_{n_k}^2}\mathbb{E}\{\frac{\partial\,G_k(\boldsymbol{\theta})}{\partial P_k}\}-\lambda&=0,\ \forall k\in S_{\cal A},\ \lambda>0,\notag \\
\sum_{k\in S_{\cal A}}P_k&=P_{tot}.
\end{align}
Let $\boldsymbol{P}\!=\![P_1, \dots, P_K]$ be the vector of sensors' transmit powers. The Hessian of $\text{tr}(\boldsymbol{J})$ with respect to $\boldsymbol{P}$ is a diagonal matrix, since using \eqref{derivative of tr(J) wrt P_k} we find $\frac{\partial^2 \text{tr}(\boldsymbol{J})}{\partial P_i\partial P_j}\!=\!0,\ i,j\!=\!1,\dots,K,\ i\!\ne\! j$. 
Fig.~\ref{sec-der-tr-J-vs-P-k-Chaud-1} and Fig.~\ref{sec-der-tr-J-vs-P-k-Chaud-2} depict $\frac{\partial^2 \text{tr}(\boldsymbol{J})}{\partial P_k^2}$ versus $P_k$ for different values of $\delta_k$ and $\sigma_{n_k}$, respectively, {\blue for coherent receiver, showing} that $\frac{\partial^2 \text{tr}(\boldsymbol{J})}{\partial P_k^2}<0$ {\blue , which implies the Hessian matrix is negative definite}. The negative definiteness of the Hessian matrix means that $\text{tr}(\boldsymbol{J})$ is jointly concave over $P_k$'s. Moreover, the constraints are linear, and thus, the problem in \eqref{maximization problem of tr(J)} is concave. For noncoherent receivers, unlike coherent receiver, our simulations show that the sign of $\frac{\partial^2 \text{tr}(\boldsymbol{J})}{\partial P_k^2}$ for various system parameters changes, and thus, $\text{tr}(\boldsymbol{J})$ is not necessarily a concave function over $P_k$'s.
{\blue The} optimal solutions for $\lambda$ and $P_k$ for $k\in S_{\cal A}$ cannot be obtained in closed-form expressions. Therefore, we resort to Newton-Raphson algorithm to solve the set of nonlinear equations in \eqref{KKT conditions for max. tr(J)}. For coherent receiver, since the problem is concave, it is guaranteed that the numerical solution obtained via the algorithm is globally optimal. Therefore, only one (carefully chosen) initial point suffices to run the algorithm. However, for noncoherent receivers, since the problem is not concave, we consider multiple initial points to run the algorithm. The description of this algorithm for noncoherent receivers follows.

Let $\boldsymbol{z}\coloneqq\left[\boldsymbol{P},\lambda\right]^T$ be the vector that contains the vector of sensors' transmit powers as well as the Lagrange multiplier $\lambda$. We let $\boldsymbol{f}$ and $\boldsymbol{\mathcal G}$, respectively be the gradient vector and the Jacobian matrix of the right side of the equality in \eqref{Lagrangian} with respect to $\boldsymbol{z}$. We have:
\vspace{-0.35cm}
\begin{align} \label{F and J for maximizing tr(J)}
\boldsymbol{f}=&\ [\frac{\partial\,\text{tr}(\boldsymbol{J})}{\partial P_1}-\lambda+\eta_1, ..., \frac{\partial\,\text{tr}(\boldsymbol{J})}{\partial P_K}-\lambda+\eta_K, P_{tot}\!-\!\sum_{k=1}^{K}P_k],\nonumber\\
\boldsymbol{\mathcal G}=&\begin{bmatrix}
\frac{\partial^2 \text{tr}(\boldsymbol{J})}{\partial P_1^2} & \cdots & 0 & -1 \\
\vdots & \ddots & \vdots & \vdots\\
0&\cdots &\frac{\partial^2 \text{tr}(\boldsymbol{J})}{\partial P_K^2} & -1\\
-1&\cdots &-1 & 0
\end{bmatrix}.
\end{align}  
Let $N_{i}$ be the total number of initial points. We choose $\boldsymbol{z}_i^{(j)},\ j\!=\!1,...,N_{i}$ initial points (solutions), where $j$ is the index of the initial points. The Newton-Raphson algorithm is carried out to obtain $\boldsymbol{z}_f^{(j)}$ and $T^{(j)}=\text{tr}(\boldsymbol{J}(\boldsymbol{z}_f^{(j)})),\ j\!=\!1,...,N_{i}$, which respectively are the final solution and the final value of the objective function obtained when the algorithm terminates, corresponding to the initial point $\boldsymbol{z}_i^{(j)}$. Suppose the algorithm runs for the initial point $\boldsymbol{z}_i^{(j)}$. {\blue We initialize the} iteration index $n\!=\!0$ and {\blue the initial point} $\boldsymbol{z}_{0}\!=\!\boldsymbol{z}_i^{(j)}${\blue . We} denote $\boldsymbol{z}_{n}$ as the solution at $n$-th iteration, and $\boldsymbol{f}\left(\boldsymbol{z}_{n}\right)$, $\boldsymbol{\mathcal G}\left(\boldsymbol{z}_{n}\right)$, respectively, as the gradient vector and the Jacobian matrix evaluated at $\boldsymbol{z}_{n}$. At iteration $n$, if the Jacobian matrix $\boldsymbol{\mathcal G}\left(\boldsymbol{z}_{n}\right)$ becomes singular, or $\sum_{k\in S_{\cal A}}P_k>P_{tot}$, the algorithm terminates. Otherwise, we let $\boldsymbol{z}_{n+1}=\boldsymbol{z}_{n}-\boldsymbol{\mathcal G}^{-1}\left(\boldsymbol{z}_{n}\right)\boldsymbol{f}\left(\boldsymbol{z}_{n}\right)$. As the stopping criterion, we check whether $\frac{\left|\left|\boldsymbol{z}_{n}-\boldsymbol{z}_{n-1}\right|\right|}{\left|\left|\boldsymbol{z}_{n}\right|\right|}\leq\epsilon_0$, where $\epsilon_0$ is a predetermined error tolerance, or whether the number of iterations exceeds a predetermined maximum $I_{max}$. {\blue Let} $\boldsymbol{z}^{*}\!=\!\left[\boldsymbol{P}^{*},\lambda^{*}\right]^T$ {\blue be} the optimal solution to this constrained optimization problem. After finding all $\{T^{(j)}\}_{j=1}^{N_i}$, $\boldsymbol{z}^{*}$ is $\boldsymbol{z}_f^{(j)}$ associated with the largest value among $T^{(j)},\ j\!=\!1,...,N_{i}$. 
\vspace{-0.35cm}
\subsection{Solving Optimization Problem in \eqref{maximization problem of logdet(J)}} \label{solving max of logdet(J)}
\vspace{-0.1cm}
We follow the same procedure as we described in Section \ref{solving max of tr(J)} to solve \eqref{maximization problem of tr(J)}. Specifically, we have:
\vspace{-0.2cm}
\begin{equation} \label{tr(abc)=tr(cab)}
\frac{\partial\,\text{log}_2(|\boldsymbol{J}|)}{\partial P_k}\!=\!\frac{1}{\ln2}\text{tr}(\!{\boldsymbol{J}}^{-1}\!\frac{\partial\boldsymbol{J}}{\partial P_k}\!)\!=\!\frac{\mathbb{E}\{\!\frac{\partial\,G_k(\boldsymbol{\theta})}{\partial P_k}\!\}}{2\pi\ln2\sigma_{n_k}^2}\mathbf{a}_k^T\!{\boldsymbol{J}}^{-1}\mathbf{a}_k,
\end{equation}
where we have used \eqref{final formula for FIM} and the fact $\text{tr}(\boldmath{A}\boldmath{B}\boldmath{C})\!=\!\text{tr}(\boldmath{C}\boldmath{A}\boldmath{B})$ to reach \eqref{tr(abc)=tr(cab)}. Since $\mathbb{E}\{\frac{\partial\,G_k(\boldsymbol{\theta})}{\partial P_k}\}\!>\!0$ and ${\boldsymbol{J}}^{-1}\!\succeq\!0$ we conclude $\frac{\partial\,\text{log}_2(|\boldsymbol{J}|)}{\partial P_k}>0$ and thus $\text{log}_2(|\boldsymbol{J}|)$ is an increasing function of $P_k$'s. The Lagrangian $\mathcal L$ of this problem is $\mathcal L(\lambda,\{\eta_k,P_k\}_{k=1}^K)\!=\!\text{log}_2(|\boldsymbol{J}|)\!-\!\sum_{k=1}^{K}P_k\left(\lambda-\eta_k\right)\!+\!\lambda P_{tot}$. 
The corresponding KKT optimality conditions are:
\vspace{-.15cm}
\begin{align} \label{KKT conditions for max. logdet(J)}
\frac{\mathbb{E}\{\frac{\partial\,G_k(\boldsymbol{\theta})}{\partial P_k}\}}{2\pi\ln2\sigma_{n_k}^2}\mathbf{a}_k^T{\boldsymbol{J}}^{-1}\mathbf{a}_k-\lambda&=0,\ \forall k\in S_{\cal A},\ \lambda>0,\notag \\
\sum_{k\in S_{\cal A}}P_k&=P_{tot}.
\end{align}
%
For coherent receiver our simulations show that the Hessian of $\text{log}_2(|\boldsymbol{J}|)$ with respect to $\boldsymbol{P}$ is diagonal and negative definite matrix, and thus, $\text{log}_2(|\boldsymbol{J}|)$ is jointly concave function over $P_k$'s. However, for noncoherent receivers the sign of $\frac{\partial^2 \text{tr}(\boldsymbol{J})}{\partial P_k^2}$ varies for different system parameters and hence $\text{log}_2(|\boldsymbol{J}|)$ is not necessarily concave function of $P_k$'s. We employ Newton-Raphson algorithm with multiple initial points as we described in {\blue Section} \ref{solving max of tr(J)} to solve the set of equations in \eqref{KKT conditions for max. logdet(J)}. 
A remark on the difference between power allocation schemes based on maximization of tr$(\boldsymbol{J})$ and $\text{log}_2(|\boldsymbol{J}|)$ follows.\\
\vspace{-0.55cm}
\begin{remark} \label{dist implementation for sol of tr and det}
\textup{Regarding the solution of \eqref{KKT conditions for max. tr(J)} on constrained maximization of tr$(\boldsymbol{J})$, we note that $\lambda^{*}$ is common and fixed for all active sensors and thus this power allocation scheme can be implemented in a distributed fashion, i.e., the FC sends $\lambda^{*}$ to the set of active sensors and each sensor calculates its own power $P_k^{*}$ using its local parameters. Unlike the solution of \eqref{KKT conditions for max. tr(J)}, the solution of \eqref{KKT conditions for max. logdet(J)} on constrained maximization of log$_2(|\boldsymbol{J}|)$ cannot be implemented in a distributed fashion. In other words, the FC needs to find $\{P_k^{*}\}_{k\in S_{\cal A}}$ and informs the active sensors of their transmit powers.}
\end{remark}
\vspace{-0.3cm}
\section{LMMSE Estimator and its MSE} \label{MSE}
\vspace{-.1cm}
Given $\boldsymbol{\hat{m}}$, finding the optimal MMSE estimate of $\boldsymbol{\theta}$ in a closed form is mathematically intractable, since it requires $q$ dimensional integrals that cannot be simplified. To curb computational complexity, we assume that the FC employs the LMMSE estimator to process $\boldsymbol{\hat{m}}$ and forms the estimate $\hat{\boldsymbol{\theta}}$. We derive the LMMSE estimator $\hat{\boldsymbol{\theta}}$ and its corresponding MSE matrix $\boldsymbol{\mathcal D}$. Let vector $\boldsymbol{\breve{m}}=\boldsymbol{\hat{m}}-\mathbb{E}\{\boldsymbol{\hat{m}}\}$. We have:
\vspace{-0.2cm}
\begin{align} \label{LMMSE and its MSE}
\hat{\boldsymbol{\theta}}&=\mathbb{E}\{\boldsymbol{\theta}\boldsymbol{\breve{m}}^T\}(\mathbb{E}\{\boldsymbol{\breve{m}}\boldsymbol{\breve{m}}^T\})^{-1}\boldsymbol{\breve{m}},\notag\\
\boldsymbol{\mathcal D}&=\boldsymbol{\mathcal C}_{\boldsymbol{\theta}}-\mathbb{E}\{\boldsymbol{\theta}\boldsymbol{\breve{m}}^T\}(\mathbb{E}\{\boldsymbol{\breve{m}}\boldsymbol{\breve{m}}^T\})^{-1}\mathbb{E}\{\boldsymbol{\theta}\boldsymbol{\breve{m}}^T\}^T.
\end{align}
Since $\boldsymbol{\theta}$ is zero-mean, we obtain $\mathbb{E}\{\boldsymbol{\theta}\boldsymbol{\breve{m}}^T\}=\mathbb{E}\{\boldsymbol{\theta}(\boldsymbol{\hat{m}}-\mathbb{E}\{\boldsymbol{\hat{m}}\})^T\}=\mathbb{E}\{\boldsymbol{\theta}\boldsymbol{\hat{m}}^T\}$.
The $k$-th column of the cross-covariance matrix $\mathbb{E}\{\boldsymbol{\theta}\boldsymbol{\hat{m}}^T\}$ describes the correlation between $\hat{m}_k$ and $\boldsymbol{\theta}$. Using the Bayes' rule we obtain:
\vspace{-.2cm}
%
%
\vspace{-0.05cm}
\begin{equation*} 
\mathbb{E}\{\boldsymbol{\theta}\hat{m}_k\}=\sum_{l=1}^{M_k}\mathbb{E}\{\boldsymbol{\theta}|m_{k,l}\}\mathbb{E}\{\hat{m}_k|m_{k,l}\}p(m_{k,l}),\\
\end{equation*}
\vspace{-.4cm}
\begin{equation*} 
\!\!\!\!\!\mathbb{E}\{\boldsymbol{\theta}|m_{k,l}\}=\frac{1}{p(m_{k,l})}\int_{V_{\theta}} \boldsymbol{\theta}p(m_{k,l}|\boldsymbol{\theta})f(\boldsymbol{\theta})d\boldsymbol{\theta},
\end{equation*}
where $V_{\theta}$ denotes the $q$-dimensional volume over which we take integral, and in the first equality we have used the fact that $\boldsymbol{\theta}$, $m_{k}$, $\hat{m}_{k}$ form a Markov chain and thus, given $m_{k}$, $\boldsymbol{\theta}$ and $\hat{m}_{k}$ are conditionally independent. Since $p(\hat{m}_{k,t}|m_{k,l})=\alpha_{k,t,l}$ and $p(m_{k,l}|\boldsymbol{\theta})=\beta_{k,l}(\boldsymbol{\theta})$, we reach:
\vspace{-.45cm}
\begin{equation} \label{E_theta_mhat_k_final}
\mathbb{E}\{\boldsymbol{\theta}\hat{m}_k\}=\sum_{t=1}^{M_k}\sum_{l=1}^{M_k}\hat{m}_{k,t}\alpha_{k,t,l}\overbrace{\int_{V_{\theta}} \boldsymbol{\theta}\beta_{k,l}(\boldsymbol{\theta})f(\boldsymbol{\theta})d\boldsymbol{\theta}}^{=\boldsymbol{\mathcal I}_{k,l}^{1}},
\end{equation}
%
and the expression for vector $\boldsymbol{\mathcal I}_{k,l}^{1}$ is given in \eqref{integrals I_1,I_2,I_3}. By definition, the $(i,j)$-th entry of matrix $\mathbb{E}\{\boldsymbol{\breve{m}}\boldsymbol{\breve{m}}^T\}$ is:
\vspace{-.2cm}
\begin{equation} \label{ij-th_entry_of_[E_m_breve][E_m_breve]^T}
\!\!\!\![\mathbb{E}\{\boldsymbol{\breve{m}}\boldsymbol{\breve{m}}^T\}]_{ij}\!=\!\mathbb{E}\{\!\hat{m}_{i}\hat{m}_{j}\!\}-\mathbb{E}\{\hat{m}_{i}\}\mathbb{E}\{\hat{m}_{j}\},i,j\!=\!1, ..., K.
\end{equation}
Similar to what we did in \eqref{E_theta_mhat_k_final}, to obtain $\mathbb{E}\{\hat{m}_k\}$ and the diagonal entries of $\mathbb{E}\{\boldsymbol{\hat{m}}\boldsymbol{\hat{m}}^T\}$ (i.e., $\mathbb{E}\{\hat{m}_k^2\}$), we condition on $m_k$; however, for the non-diagonal entries of $\mathbb{E}\{\boldsymbol{\hat{m}}\boldsymbol{\hat{m}}^T\}$ (i.e., $\mathbb{E}\{\hat{m}_{i}\hat{m}_{j}\}$), we condition on $\boldsymbol{\theta}$. Then using \eqref{P(m_hat given theta) based on alpha and beta}, we obtain:
\vspace{-.2cm}
\begin{subequations} \label{E_mhat_i_mhat_j}
\begin{align} 
&\mathbb{E}\{\hat{m}_{i}\hat{m}_{j}\}=\label{eq:sub_E_mhat_i_mhat_j_1}\\
&\left\{\begin{aligned}
&\sum_{t=1}^{M_k}\sum_{l=1}^{M_k}\hat{m}_{k,t}^2\alpha_{k,t,l}\overbrace{\int_{V_{\theta}} \beta_{k,l}(\boldsymbol{\theta})f(\boldsymbol{\theta})d\boldsymbol{\theta}}^{={\mathcal I}_{k,l}^{2}},\ \ \ \ \ \ \ \ \ \ \ \ \ i=j=k\\
&\sum_{t_1=1}^{M_{i}}\sum_{t_2=1}^{M_{j}}\sum_{l_1=1}^{M_{i}}\sum_{l_2=1}^{M_{j}}\hat{m}_{i,t_1}\hat{m}_{j,t_2}\alpha_{i,t_1,l_1}\alpha_{j,t_2,l_2}\times\\
&\underbrace{\int_{V_{\theta}} \beta_{i,l_1}(\boldsymbol{\theta})\beta_{j,l_2}(\boldsymbol{\theta})f(\boldsymbol{\theta})d\boldsymbol{\theta}}_{={\mathcal I}_{i,j,l_1,l_2}^{3}},\ \ \ \ \ \ \ \ \ \ \ \ \ \ \ \ \ \ \ \ \ \ \ \ \ \ \ \ \ i\neq j\nonumber
\end{aligned}
\right.\nonumber\\
&\mathbb{E}\{\hat{m}_k\}=\sum_{t=1}^{M_k}\sum_{l=1}^{M_k}\hat{m}_{k,t}\alpha_{k,t,l}{\mathcal I}_{k,l}^{2},\ \ k\!=\!1, ..., K, \label{eq:sub_E_mhat_i_mhat_j_2}
\end{align}
\end{subequations}
where ${\mathcal I}_{k,l}^{2}$ and ${\mathcal I}_{i,j,l_1,l_2}^{3}$ are scalars. We find these integrals (see Appendix \ref{deriving the integrals} for derivations) as below:
\vspace{-0.2cm}
\begin{align} \label{integrals I_1,I_2,I_3}
\boldsymbol{\mathcal I}_{k,l}^{1}&=\frac{\boldsymbol{\mathcal C}_{\boldsymbol{\theta}}\mathbf{a}_k}{\sqrt{2\pi}\sigma_k}(\exp{(-\frac{u_{k,l}^2}{2\sigma_k^2})}-
\exp{(-\frac{u_{k,l+1}^2}{2\sigma_k^2})}),\\
{\mathcal I}_{k,l}^{2}&=Q(\frac{u_{k,l}}{\sigma_k})-Q(\frac{u_{k,l+1}}{\sigma_k}),\nonumber\\
{\mathcal I}_{i,j,l_1,l_2}^{3}&={\mathfrak Q}(\!\frac{u_{i,l_1}}{\sigma_i},\frac{u_{j,l_2}}{\sigma_j};\rho_{ij}\!)\!-\!{\mathfrak Q}(\!\frac{u_{i,l_1}}{\sigma_i},\frac{u_{j,l_2+1}}{\sigma_j};\rho_{ij}\!)\nonumber\\
&-{\mathfrak Q}(\!\frac{u_{i,l_1+1}}{\sigma_i},\frac{u_{j,l_2}}{\sigma_j};\rho_{ij}\!)\!+\!{\mathfrak Q}(\!\frac{u_{i,l_1+1}}{\sigma_i},\frac{u_{j,l_2+1}}{\sigma_j};\rho_{ij}\!),\nonumber
\end{align}
in which:
\vspace{-0.2cm}
\begin{align} \label{definition of sigma_k and rho_ij}
\sigma_k=\sqrt{\sigma_{n_k}^2+\mathbf{a}_k^T\boldsymbol{\mathcal C}_{\boldsymbol{\theta}}\mathbf{a}_k}\ \ ,\ \ \rho_{ij}=\frac{\mathbf{a}_i^T\boldsymbol{\mathcal C}_{\boldsymbol{\theta}}\mathbf{a}_j}{\sigma_i\sigma_j}.
\end{align}
\vspace{-0.1cm}
$\!\!$Substituting \eqref{E_theta_mhat_k_final}-\eqref{definition of sigma_k and rho_ij} in \eqref{LMMSE and its MSE}, the MSE matrix $\boldsymbol{\mathcal D}$ is computed.
%

{\blue For $\boldsymbol{\mathcal D}$ in \eqref{LMMSE and its MSE} there exists two baselines. For the first baseline}, we consider the centralized estimation case in Section \ref{finding FIM final} with the LMMSE estimator at the FC and let $\boldsymbol{\mathcal D}_0$ denote the corresponding MSE matrix. We have:
\vspace{-0.1cm}
\begin{equation} \label{clairvoyant MSE}
\boldsymbol{\mathcal D}_0=\boldsymbol{\mathcal C}_{\boldsymbol{\theta}}-\mathbb{E}\{\boldsymbol{\theta}\boldsymbol{x}^T\}(\mathbb{E}\{\boldsymbol{x}\boldsymbol{x}^T\})^{-1}\mathbb{E}\{\boldsymbol{\theta}\boldsymbol{x}^T\}^{T},
\end{equation}
where $\mathbb{E}\{\boldsymbol{x}\boldsymbol{x}^T\}$ and $\mathbb{E}\{\boldsymbol{\theta}\boldsymbol{x}^T\}$ respectively are, auto-covariance matrix of noisy observations, and cross-covariance matrix between $\boldsymbol{\theta}$ and $\boldsymbol{x}$. For linear observation model in \eqref{obs_model} we get:
\vspace{-0.15cm}
\begin{equation*}
\mathbb{E}\{\boldsymbol{\theta}\boldsymbol{x}^T\}\!=\!\boldsymbol{\mathcal C}_{\boldsymbol{\theta}}\boldsymbol{\mathcal A},\ \ \mathbb{E}\{\boldsymbol{x}\boldsymbol{x}^T\}\!=\!\boldsymbol{\mathcal A}^T\boldsymbol{\mathcal C}_{\boldsymbol{\theta}}\boldsymbol{\mathcal A}\!+\!\text{diag}\left(\sigma_{n_1}^2, ..., \sigma_{n_K}^2\right).
\end{equation*}
{\blue 
For the second baseline, suppose communication channels between sensors and the FC are error-free and hence vector $\boldsymbol{m}$ is available at the FC. Let vector $\boldsymbol{\mathring{m}}\!=\!\boldsymbol{m}\!-\!\mathbb{E}\{\boldsymbol{m}\}$. Then, the corresponding MSE matrix is $\boldsymbol{\mathcal D}^{ideal}\!=\!\boldsymbol{\mathcal C}_{\boldsymbol{\theta}}\!-\!\mathbb{E}\{\boldsymbol{\theta}\boldsymbol{\mathring{m}}^T\}(\mathbb{E}\{\boldsymbol{\mathring{m}}\boldsymbol{\mathring{m}}^T\})^{-1}\mathbb{E}\{\boldsymbol{\theta}\boldsymbol{\mathring{m}}^T\}^{T}$. Since $\boldsymbol{\theta}$ is zero-mean, we obtain $\mathbb{E}\{\boldsymbol{\theta}\boldsymbol{\mathring{m}}^T\}\!=\!\mathbb{E}\{\boldsymbol{\theta}\boldsymbol{m}^T\}$. We let $\mathbb{E}\{\boldsymbol{\theta}m_k\}$ and $[\mathbb{E}\{\boldsymbol{\mathring{m}}\boldsymbol{\mathring{m}}^T\}]_{ij}$, respectively, be the $k$-th column of matrix $\mathbb{E}\{\boldsymbol{\theta}\boldsymbol{m}^T\}$, and the $(i,j)$-th entry of matrix $\mathbb{E}\{\boldsymbol{\mathring{m}}\boldsymbol{\mathring{m}}^T\}$. Taking steps similar to the ones we took to obtain \eqref{E_theta_mhat_k_final}-\eqref{E_mhat_i_mhat_j}, 
we find $\mathbb{E}\{\boldsymbol{\theta}m_k\}\!=\!\sum_{l=1}^{M_k}m_{k,l}\boldsymbol{\mathcal I}_{k,l}^{1}$, $\mathbb{E}\{m_k\}\!=\!\sum_{l=1}^{M_k}m_{k,l}\boldsymbol{\mathcal I}_{k,l}^{2}$, $[\mathbb{E}\{\boldsymbol{\mathring{m}}\boldsymbol{\mathring{m}}^T\}]_{ij}\!=\!\mathbb{E}\{m_{i}m_{j}\}\!-\!\mathbb{E}\{m_{i}\}\mathbb{E}\{m_{j}\},\ i,j=1, ..., K$, in which 
$\mathbb{E}\{m_{i}m_{j}\}\!=\!\sum_{l=1}^{M_k}\!m^2_{k,l}\boldsymbol{\mathcal I}_{k,l}^{2}$ for $i\!=\!j\!=\!k$, and $\mathbb{E}\{m_{i}m_{j}\}\!=\!\sum_{l_1=1}^{M_{i}}\sum_{l_2=1}^{M_{j}}m_{i,l_1}\!m_{j,l_2}{\mathcal I}_{i,j,l_1,l_2}^{3}$ for $i\!\neq \!j$. Clearly, $\boldsymbol{\mathcal D}_0\! \preceq\! \boldsymbol{\mathcal D}^{ideal} \!\preceq\! \boldsymbol{\mathcal D}$.
%
%
}
\vspace{-0.1cm}
\begin{remark}
\textup{If $\boldsymbol{\theta}$ has a known nonzero-mean $\boldsymbol{\mu}_{\theta}$, the expressions for the LMMSE estimator $\hat{\boldsymbol{\theta}}$ and its corresponding MSE matrix $\boldsymbol{\mathcal D}$ change as the following:
\vspace{-0.1cm}
\begin{align*} 
\hat{\boldsymbol{\theta}}&=(\mathbb{E}\{\boldsymbol{\theta}\boldsymbol{\hat{m}}^T\}\!-\!\boldsymbol{\mu}_{\theta}\mathbb{E}\{\boldsymbol{\hat{m}}^T\})(\mathbb{E}\{\boldsymbol{\breve{m}}\boldsymbol{\breve{m}}^T\})^{-1}\boldsymbol{\breve{m}}+\boldsymbol{\mu}_{\theta},\notag\\
\boldsymbol{\mathcal D}&=\boldsymbol{\mathcal C}_{\boldsymbol{\theta}}-(\mathbb{E}\{\boldsymbol{\theta}\boldsymbol{\hat{m}}^T\}\!-\!\boldsymbol{\mu}_{\theta}\mathbb{E}\{\boldsymbol{\hat{m}}^T\})(\mathbb{E}\{\boldsymbol{\breve{m}}\boldsymbol{\breve{m}}^T\})^{-1}(\mathbb{E}\{\boldsymbol{\theta}\boldsymbol{\hat{m}}^T\}\\
&-\boldsymbol{\mu}_{\theta}\mathbb{E}\{\boldsymbol{\hat{m}}^T\})^T,
\end{align*}
where $\boldsymbol{\mathcal C}_{\boldsymbol{\theta}}=\mathbb{E}\{\boldsymbol{\theta}\boldsymbol{\theta}^T\}-\boldsymbol{\mu}_{\theta}{\boldsymbol{\mu}_{\theta}}^T$.}
\end{remark}
{\blue
\section{Discussion on Appropriateness and Achievability of Bayesian CRB}\label{Appropriateness and Achievability}
One may wonder how the FIM-max schemes in Section \ref{Power Allocation} are compared with the power allocation that can be obtained  from constrained minimization of the MSE of the LMMSE estimator derived in Section \ref{MSE}. On the other hand, {\red the literature \cite{WWB} suggests} that the WWB in Section \ref{WWB bound} is a tighter bound (compared to Bayesian CRB). This observation raises the question whether using the WWB as the optimization metric would be a more appropriate choice. This section provides answers to these questions.
\vspace{-0.2cm}
\subsection{Appropriateness of Bayesian FIM as the Optimization Metric} \label{subsection on Appropriateness}
Let $\mathcal D=\text{tr}(\boldsymbol{\mathcal D})$, where $\boldsymbol{\mathcal D}$ is the MSE matrix of the LMMSE estimator given in \eqref{LMMSE and its MSE}. 
We consider the following constrained optimization problem:
%
\vspace{-0.1cm}
\begin{align} \label{minimization problem of MSE}
\mathop{\text{minimize}}_{P_k,\forall k}\ \ \ \ \,&\text{tr}(\boldsymbol{\mathcal D}(\{P_k\}_{k=1}^K))\nonumber\\
\text{s.t.}\ \ \ \ &\sum_{k=1}^{K}P_k\leq P_{tot},\ P_k\in \mathbb{R}^{+},\ \forall k.
\end{align}
In the absence of analytical solution, we resort to exhaustive search method to find the solution of the problem in \eqref{minimization problem of MSE}. Let MSE-min scheme corresponds to this solution. For all three types of receivers, our extensive simulations show that $\frac{\partial\,\text{tr}(\boldsymbol{\mathcal D})}{\partial P_k}>0$, however, the sign of $\frac{\partial^2 \text{tr}(\boldsymbol{\mathcal D})}{\partial P_k^2}$ for various system parameters changes, and hence, tr$(\boldsymbol{\mathcal D})$ is not necessarily a convex function over $P_k$'s. 
Furthermore, the cost function in \eqref{minimization problem of MSE} cannot be decoupled over the optimization variables $P_k$'s and thus $P_k$'s across sensors are related to each other. Because of this, finding MSE-min is computationally complex, and the solution cannot be implemented in a distributed fashion (i.e., sensor $k$ cannot find $P_k$ relying on its own local information only). 
This contrasts FIM-max scheme obtained from solving the problem in \eqref{maximization problem of tr(J)}, where the cost function in \eqref{maximization problem of tr(J)} can be decoupled over $P_k$'s and thus $P_k$'s across sensors are not related to each other. Because of this, finding FIM-max is computationally simple, and the solution can be implemented in a distributed fashion.
Figures \ref{MSE_power_aloc} and \ref{effect_of_network_size} in Section \ref{simulation} illustrate the numerical evaluations of (i) trace of $\boldsymbol{\mathcal D}$ at power allocation obtained from solving the problem in \eqref{minimization problem of MSE}, denoted as ${\cal D}_{m}\!=\!\text{tr}(\boldsymbol{\mathcal D}(\text{MSE-min}))$ and (ii) trace of $\boldsymbol{\mathcal D}$ at power allocation obtained from solving the problem in \eqref{maximization problem of tr(J)}, denoted as ${\cal D}_{t}\!=\!\text{tr}(\boldsymbol{\mathcal D}(\text{FIM-max}))$, given $P_{tot}$. The figures show that:
	%
	\begin{equation}\label{mse-comparison}
	{\cal D}_{m} \!\lesssim\! {\cal D}_{t},
	\end{equation}
where $a\!\lesssim\!b$ means that $a$ is less than $b$, but very close to $b$. Obviously, from the estimation theory we know ${\cal D}_{m}\! < \! {\cal D}_{t}$. What our numerical results reveal is that in our problem they are very close to each other. This indicates the appropriateness of using Bayesian FIM as the optimization metric, since the loss in terms of the MSE performance is not significant. 
\vspace{-.1cm}
\subsection{Tightness and Achievability of Bayesian CRB} \label{subsection on Achievability}
\vspace{-.1cm}
{\red Although} the WWB is a tighter bound (compared to Bayesian CRB){\red \cite{WWB}, we} note that finding the WWB matrix is computationally much more expensive (compared to finding the Bayesian FIM), due to required matrix inversions ${\boldsymbol{G}}^{-1}$ for each test point in \eqref{supremum}. Consequently, finding the power allocation that minimizes the trace or log-determinant of the WWB is computationally {\red much} more expensive than finding the solutions for the problems in \eqref{maximization problem of tr(J)} or \eqref{maximization problem of logdet(J)}. 
Furthermore, \eqref{mse-comparison} indicates that by not using power allocation obtained from minimizing trace of the WWB matrix (which is tighter than Bayesian CRB) we are not in disadvantage, in terms of the MSE performance. 

According to \cite{Van_Trees_estimation_book} Bayesian CRB is attainable if and only if the posterior probability density of $\boldsymbol{\theta}$ given ``observation'' is Gaussian. In that case, the MMSE and MAP estimators coincide and both are efficient (i.e., their MSE matrices are equal to Bayesian CRB matrix) \cite{Van_Trees_estimation_book}. This bound is attained in the limit as $K$ becomes infinite \cite{WWB}. 
In our work, the recovered quantization levels for all sensors at the FC, denoted as vector $\hat{\boldsymbol{m}}$, plays the role of ``observation''. Since the posterior probability density of $\boldsymbol{\theta}$ given $\hat{\boldsymbol{m}}$ is not Gaussian, Bayesian CRB is not attainable. However, as $K$ increases, we expect that the MSE of MMSE estimator approaches to Bayesian CRB. 
Let $\tr(\text{CRB(FIM-max)})$ denote trace of Bayesian CRB matrix evaluated at FIM-max power allocation, and Let $\tr(\text{CRB(MSE-min)})$ denote trace of Bayesian CRB matrix evaluated at MSE-min power allocation. From the estimation theory we know:
	\vspace{-0.15cm}
	\begin{equation}\label{crb-mse-inequality}
	\tr(\text{CRB}({\text{FIM-max}}))\!<\!\tr(\text{CRB(MSE-min)})\!<\!{\cal D}_{m}\! < \! {\cal D}_{t}.
	\end{equation}
Combining (\ref{crb-mse-inequality}) and (\ref{mse-comparison}) we reach:
	\vspace{-0.15cm}
	\begin{equation*}
	\tr(\text{CRB}({\text{FIM-max}}))\!<\!\tr(\text{CRB(MSE-min)})\!<\!{\cal D}_{m}\! \lesssim \! {\cal D}_{t}.
	\end{equation*}
This suggests that, although Bayesian CRB is not attainable, it is still proper to use Bayesian FIM for transmit power optimization, since the loss in terms of the MSE performance is not significant.
}
{\blue
	\vspace{-0.15cm}
\section{Classical CRB and BLUE for Estimating Deterministic Vector $\boldsymbol{\theta}$} \label{est of deterministic theta}
\vspace{-0.1cm}
In this section, we derive the classical FIM (assuming vector $\boldsymbol{\theta}$ to be estimated is deterministic), the BLUE and its corresponding MSE matrix. We also discuss the behavior of the classical FIM and the MSE of BLUE in low-region and high-region of $P_{tot}$. 
Finally, we discuss optimizing transmit power considering the classical FIM and the MSE of BLUE as the optimization metric. 
\vspace{-0.3cm}
\subsection{Characterization of Classical FIM} \label{classical FIM}
\vspace{-0.05cm}
Let ${\boldsymbol{J}}_c$ denote the $q\times q$ classical FIM and represents the $(i,j)$-th entry of ${\boldsymbol{J}}_c$. We have \cite{Van_Trees_estimation_book}:
\vspace{-0.15cm}
\begin{equation} \label{ij-th element of J_c}
[{\boldsymbol{J}}_c]_{ij}\!=\!-\mathbb{E}\{\frac{\partial^2 \ln p(\boldsymbol{\hat{m}};\boldsymbol{\theta})}{\partial \theta_i\partial \theta_j}\},\ i,j\!=\!1,...,q,
\end{equation}
where $p(\boldsymbol{\hat{m}};\boldsymbol{\theta})$ is the joint probability distribution of $\hat{m}_1, ..., \hat{m}_K$ parameterized by $\boldsymbol{\theta}$. Notice that $[{\boldsymbol{J}}_c]_{ij}$ in \eqref{ij-th element of J_c} is similar to $[\boldsymbol{\Lambda}(\boldsymbol{\theta})]_{ij}$ in \eqref{ij-th element of second term of FIM}, with the difference that for Bayesian FIM we deal with the conditional pdf $p(\boldsymbol{\hat{m}}| \boldsymbol{\theta})$.
Therefore, ${\boldsymbol{J}}_c$ has the same expression as $\boldsymbol{J}$ in \eqref{final formula for FIM}, which depends on $\boldsymbol{\theta}$. That is:
\vspace{+.05cm}
\begin{align} \label{final formula for J_c}
\ \ \ {\boldsymbol{J}}_c=\frac{1}{2\pi}{\boldsymbol{\mathcal A}}\,\text{diag}(\frac{G_1(\boldsymbol{\theta})}{\sigma_{n_1}^2}, ..., \frac{G_K(\boldsymbol{\theta})}{\sigma_{n_K}^2}){\boldsymbol{\mathcal A}}^T, 
\end{align}
in which $G_k(\boldsymbol{\theta})$ is defined in \eqref{G_k(theta)}, and the probabilities $\alpha_{k,t,l}$ and $\beta_{k,l}(\boldsymbol{\theta})$ have the same expressions as for Bayesian FIM.  
\vspace{-0.1cm}
\subsection{Characterization of BLUE and its MSE Matrix} \label{subsection BLUE and its MSE}
%
Recall $\hat{\boldsymbol{m}}$ is the data at the FC based on which we wish to form the BLUE. To satisfy the unbiasedness requirement for BLUE, we need to have $\mathbb{E}\{\hat{\boldsymbol{m}}\}=\boldsymbol{H}\boldsymbol{\theta}$, for a known matrix $\boldsymbol{H}$ \cite{Kay_SSP_book}. 
The unbiasedness requirement is not satisfied in general for our system model. 
However, under three conditions (coherent receiver at the FC, uniform quantizer}\footnote{\blue
		For sensor $k$, we define the quantization noise $\epsilon_k\!=\!x_k-m_k$. Since $n_k$'s in \eqref{obs_model} are uncorrelated Gaussian, $x_k$'s are uncorrelated Gaussian. \cite{Widrow} shows that when uncorrelated Gaussian are quantized with uniform quantizers of quantization step sizes $\Delta_k$'s, $\epsilon_k$'s are independent zero mean uniform random variables with variance $\sigma^2_{\epsilon_k}\!\approx\!\frac{\Delta_k^2}{12}$. Also, $\epsilon_k$'s and $x_k$'s are uncorrelated.}{\blue , and natural binary encoder at the sensors to map quantization levels to information bits), we can establish a linear relationship between $\hat{\boldsymbol{m}}$ and $\boldsymbol{\theta}$, that is $\hat{\boldsymbol{m}}=\boldsymbol{H}\boldsymbol{\theta}+\boldsymbol{\nu}$, where $\boldsymbol{\nu}$ is a zero-mean vector with covariance $\boldsymbol{\mathcal C}_{\boldsymbol{\nu}}$, and show that for this linear model the unbiasedness requirement is met, i.e., $\mathbb{E}\{\hat{\boldsymbol{m}}\}=\boldsymbol{H}\boldsymbol{\theta}$. Then using this linear model, we derive BLUE and its corresponding MSE matrix as the following \cite{Kay_SSP_book}:
%
%
%
%
\vspace{-0.1cm}
\begin{eqnarray} \label{BLUE and its MSE}
{\hat{\boldsymbol{\theta}}}_{BLUE}&=&{({\boldsymbol{H}}^T{\boldsymbol{\mathcal C}_{\boldsymbol{\nu}}}^{-1}\boldsymbol{H})}^{-1}{\boldsymbol{H}}^T{\boldsymbol{\mathcal C}_{\boldsymbol{\nu}}}^{-1}\hat{\boldsymbol{m}},\nonumber\\
{\boldsymbol{\mathcal D}}_{BLUE}&=&{({\boldsymbol{H}}^T{\boldsymbol{\mathcal C}_{\boldsymbol{\nu}}}^{-1}\boldsymbol{H})}^{-1}.
\end{eqnarray}
First we verify the unbiasedness requirement under the three stated condition. Under these three conditions,
we can use the approximations given in \cite{Leung_TSP_2015} and write:
\vspace{-0.1cm}
\begin{align} \label{mean of m_hat}
&\mathbb{E}\{\hat{m}_k|m_k\}=(1-2{\cal E}_k)m_k \Rightarrow \mathbb{E}\{\hat{m}_k\}=\mathbb{E}\{\mathbb{E}\{\hat{m}_k|m_k\}\}=\nonumber\\
&(1\!-\!2{\cal E}_k)\mathbb{E}\{{m_k}\}\!=\!(1\!-\!2{\cal E}_k)\mathbb{E}\{{x_k\!+\!\epsilon_k}\}\!=\!(1\!-\!2{\cal E}_k)\mathbf{a}_k^T \boldsymbol{\theta} \Rightarrow \nonumber \\
&\mathbb{E}\{\hat{\boldsymbol{m}}\}=\underbrace{\text{diag}(1\!-\!2{\cal E}_1, ..., 1\!-\!2{\cal E}_K){\boldsymbol{\mathcal A}}^T}_{=\boldsymbol{H}}\boldsymbol{\theta}. 
\end{align}
Equation \eqref{mean of m_hat} shows that the unbiasedness constraint is satisfied. 
%
Next, we establish the linear relationship  $\hat{\boldsymbol{m}}=\boldsymbol{H}\boldsymbol{\theta}+\boldsymbol{\nu}$, where $\boldsymbol{\nu}$ is a zero-mean vector with covariance $\boldsymbol{\mathcal C}_{\boldsymbol{\nu}}$, and we find $\boldsymbol{\mathcal C}_{\boldsymbol{\nu}}$. Knowing $\boldsymbol{\mathcal C}_{\boldsymbol{\nu}}$ and $\boldsymbol{H}$ we can then use (\ref{BLUE and its MSE}) to express BLUE and its corresponding MSE. To establish the linear relationship, suppose: 
\vspace{-0.1cm}
\begin{equation}\label{m_hat=m_plus_nu}
\hat{m}_k=\mathbb{E}\{\hat{m}_k\}+\nu_k, ~~~\mbox{for}~ k=1,...,K,
\end{equation}
where $\nu_k$ is zero-mean with variance $var(\nu_k)=var(\hat{m}_k)$. The equivalent vector-matrix representation of (\ref{m_hat=m_plus_nu}) becomes $\hat{\boldsymbol{m}}=\boldsymbol{H}\boldsymbol{\theta}+\boldsymbol{\nu}$, in which $\boldsymbol{\nu}=[\nu_1, ..., \nu_K]^T$, $\boldsymbol{\mathcal C}_{\boldsymbol{\nu}}=\boldsymbol{\mathcal C}_{\hat{\boldsymbol{m}}}$, and $\boldsymbol{\mathcal C}_{\hat{\boldsymbol{m}}}$ denotes the covariance matrix of vector $\hat{\boldsymbol{m}}$. Hence, to find $\boldsymbol{\mathcal C}_{\boldsymbol{\nu}}$ we need to find $\boldsymbol{\mathcal C}_{\hat{\boldsymbol{m}}}$.
Let $[\boldsymbol{\mathcal C}_{\hat{\boldsymbol{m}}}]_{kl}$ be the $(k,l)$-th entry of matrix 
$\boldsymbol{\mathcal C}_{\hat{\boldsymbol{m}}}$.
Starting with the diagonal entries of $\boldsymbol{\mathcal C}_{\hat{\boldsymbol{m}}}$, we find $[\boldsymbol{\mathcal C}_{\hat{\boldsymbol{m}}}]_{kk}=var(\hat{m}_k)$. 
Under the three stated conditions,
we can use the approximations given in \cite{Leung_TSP_2015} and write:
%
%
\vspace{-0.1cm}
\begin{eqnarray} \label{E{m_hat|m}}
\!\!\!\!\!var\{\hat{m}_k|m_k\} \!\!& \leq &\!\! \chi_k{\cal E}_k~~~\mbox{where}~ \chi_k=\frac{4\tau_k^2(2^{L_k}\!+\!1)}{3(2^{L_k}\!-\!1)} \Rightarrow \nonumber \\
var(\hat{m}_k)  \!\!&=&\!\! \mathbb{E}\{var\{\hat{m}_k|m_k\}\}\!+\!var(\mathbb{E}\{\hat{m}_k|m_k\}) \nonumber \\
\!\!&\leq&\!\!\chi_k{\cal E}_k\!+\!(1\!-\!2{\cal E}_k)^2(\sigma_{n_k}^2\!+\!\frac{\Delta_k^2}{12})\!=\!\Upsilon_k, \label{var of m_hat}
\end{eqnarray}
%
\begin{figure*}[b]
	{\blue
		\begin{align} \label{QBLUE}
		\hat{\boldsymbol{\theta}}_{QBLUE}\!=\!{(\sum_{k=1}^{K}\!\frac{\mathbf{a}_k\mathbf{a}_k^T}{\frac{\chi_k{\cal E}_k}{(1\!-\!2{\cal E}_k)^2}\!+\!\sigma_{n_k}^2\!+\!\frac{\Delta_k^2}{12}})}^{-1}\!(\sum_{k=1}^{K}\!\frac{\hat{m}_k\mathbf{a}_k}{\frac{\chi_k{\cal E}_k}{1\!-\!2{\cal E}_k}\!+\!(1\!-\!2{\cal E}_k)(\sigma_{n_k}^2\!+\!\frac{\Delta_k^2}{12})}),{\boldsymbol{\mathcal D}}_{QBLUE}\!=\!{(\sum_{k=1}^{K}\!\frac{\mathbf{a}_k\mathbf{a}_k^T}{\frac{\chi_k{\cal E}_k}{(1\!-\!2{\cal E}_k)^2}\!+\!\sigma_{n_k}^2\!+\!\frac{\Delta_k^2}{12}})}^{-1}
		\end{align}
	}
\end{figure*}
$\!\!\!$where ${\Delta_k}\!=\!\frac{2\tau_k}{(2^{L_k}-1)}$. Next, we compute the non-diagonal elements  ${[\boldsymbol{\mathcal C}_{\hat{\boldsymbol{m}}}]}_{kl}\!=\!\mathbb{E}\{\hat{m}_{k}\hat{m}_{l}\}\!-\!\mathbb{E}\{\hat{m}_{k}\}\mathbb{E}\{\hat{m}_{l}\}$, where the mean $ \mathbb{E}\{\hat{m}_{k}\}$ is given in (\ref{mean of m_hat}). Hence, we need to find 
$\mathbb{E}\{\hat{m}_{k}\hat{m}_{l}\}$ as the following:
\vspace{-0.15cm}
\begin{align*} 
&\mathbb{E}\{\!\hat{m}_{k}\hat{m}_{l}\!\}\!=\!\mathbb{E}\{\!\mathbb{E}\{\hat{m}_{k}\hat{m}_{l}|m_k,m_l\}\!\}\!\overset{(a)}{=}\!\mathbb{E}\{\!\mathbb{E}\{\hat{m}_{k}|m_k\}\mathbb{E}\{\hat{m}_{l}|m_l\}\!\}\nonumber\\
&\!\!\overset{(b)}{=}(1\!-\!2{\cal E}_k)(1\!-\!2{\cal E}_l)\mathbb{E}\{m_k m_l\}\overset{(c)}{=}(1\!-\!2{\cal E}_k)(1\!-\!2{\cal E}_l)\mathbb{E}\{x_k x_l\}\nonumber\\
&\!\!=((1\!-\!2{\cal E}_k)\mathbf{a}_k^T \boldsymbol{\theta})((1\!-\!2{\cal E}_l)\mathbf{a}_l^T \boldsymbol{\theta})\overset{(d)}{=}\mathbb{E}\{\hat{m}_{k}\}\mathbb{E}\{\hat{m}_{l}\} \nonumber \\
&\!\! \Rightarrow {[\boldsymbol{\mathcal C}_{\hat{\boldsymbol{m}}}]}_{kl}=\mathbb{E}\{\hat{m}_{k}\hat{m}_{l}\}-\mathbb{E}\{\hat{m}_{k}\}\mathbb{E}\{\hat{m}_{l}\}=0~ \mbox{for} ~ k \neq  l \nonumber\\ 
&\!\! \Rightarrow \boldsymbol{\mathcal C}_{\hat{\boldsymbol{m}}} ~\mbox{is~diagonal}
\end{align*}
%
%
%
%
%
in which ($a$) follows from the fact that, given $m_k, m_l$, then $\hat{m}_{k}, \hat{m}_{l}$ are independent, ($b$) comes from \eqref{mean of m_hat}, ($c$) is obtained from the fact that the quantization noises $\epsilon_k$’s are 
uncorrelated from each other, and $\epsilon_k$’s and $x_k$'s are  uncorrelated, and ($d$) follows from \eqref{mean of m_hat}. 
%
Recall according to  (\ref{E{m_hat|m}}) $var(\hat{m}_k)\leq\Upsilon_k$. 
Let $\boldsymbol{\mathcal C}_{\boldsymbol{Q}}=\text{diag}(\Upsilon_1, ..., \Upsilon_K)$ be a diagonal matrix. Clearly by the construction of $\boldsymbol{\mathcal C}_{\boldsymbol{Q}}$ we have
$\boldsymbol{\mathcal C}_{\hat{\boldsymbol{m}}}\preceq\boldsymbol{\mathcal C}_{\boldsymbol{Q}}$ and thus $\boldsymbol{\mathcal C}_{\boldsymbol{\nu}}\preceq\boldsymbol{\mathcal C}_{\boldsymbol{Q}}$. 
Replacing $\boldsymbol{\mathcal C}_{\boldsymbol{\nu}}$ with its upper bound $\boldsymbol{\mathcal C}_{\boldsymbol{Q}}$ and substituting $\boldsymbol{H}$ in \eqref{BLUE and its MSE}, we find quasi BLUE and its corresponding MSE matrix as shown in \eqref{QBLUE}.}
%
%
{\blue
The notion of quasi BLUE in the context of distributed estimation of an unknown deterministic scalar has been used before in \cite{Goldsmith_2006,AlRegib_2009}, where an upper bound on the variance of the data at the FC (based on which BLUE is formed) is utilized, instead of the variance of the data itself, to derive the unbiased estimator and its corresponding MSE.
%

%
%
%
%
%
\vspace{-0.1cm}
\subsection{Behavior of the classical FIM and the MSE of BLUE in low-region and high-region of $P_{tot}$} \label{behavior of J_c and D_Qblue}
Consider coherent receiver where we model the channel between sensor $k$ and the FC as a BSC with the probability of flipping a bit ${\cal E}_k=Q(2 \gamma_k)$ and $\gamma_k$, defined in \eqref{SNR for k-th channel}, depends on $P_k$. In low-region of $P_{tot}$ (when $P_k\rightarrow 0$) we have ${\cal E}_k\rightarrow \frac{1}{2}$ (worst communication channel effect). Then \eqref{alpha_{k,t,l} for coherent} implies that $\alpha_{k,t,l}\approx\frac{1}{2^{L_k}}$ and one can show that $G_k(\boldsymbol{\theta}) \rightarrow 0$.
Therefore ${\boldsymbol{J}}_c\rightarrow \boldsymbol{0}$. 
On the contrary, in high-region of $P_{tot}$ (when $P_k\rightarrow\infty$) we have ${\cal E}_k\rightarrow 0$. This implies that 
%
\begin{equation*} \label{asymptotic alpha_ktl}
\alpha_{k,t,l}\!\approx\!\left\{
\begin{array}{lr}
\!\!0,&t\!\neq\!l\ (\text{or equivalently}\ N_{e_{k,t,l}}\!\neq\!0~ \mbox{in \eqref{alpha_{k,t,l} for coherent}})\\
\!\!1.&t\!=\!l\ (\text{or equivalently}\ N_{e_{k,t,l}}\!=\!0~ \mbox{in \eqref{alpha_{k,t,l} for coherent}})
\end{array}
\right.
\end{equation*}
Then one can show that  $G_k(\boldsymbol{\theta})\!\rightarrow\!G_k^{ideal}(\boldsymbol{\theta})$ and ${\boldsymbol{J}}_c\!\rightarrow\!{\boldsymbol{J}}_c^{ideal}$, where $G_k^{ideal}(\boldsymbol{\theta})$ is given in Section \ref{finding FIM final} and ${\boldsymbol{J}}_c^{ideal}$ is obtained from 
(\ref{final formula for J_c}) after substituting $G_k(\boldsymbol{\theta})$ with $G_k^{ideal}(\boldsymbol{\theta})$.
Similar discussions can be made and similar conclusions can be reached for both types of noncoherent receivers. 
For coherent receiver in low-region of $P_{tot}$ (when $P_k\rightarrow 0$) we have ${\cal E}_k\rightarrow \frac{1}{2}$. Examining (\ref{QBLUE}) we realize that this implies ${\boldsymbol{\mathcal D}}_{QBLUE}\rightarrow \boldsymbol{\infty}$. 
On the contrary, in high-region of $P_{tot}$ (when $P_k\rightarrow\infty$) we have  ${\cal E}_k\rightarrow 0$ and ${\boldsymbol{\mathcal D}}_{QBLUE}\rightarrow {\boldsymbol{\mathcal D}}_{QBLUE}^{ideal}= {(\sum_{k=1}^{K}\frac{\mathbf{a}_k\mathbf{a}_k^T}{\sigma_{n_k}^2+\frac{\Delta_k^2}{12}})}^{-1}$, where ${\boldsymbol{\mathcal D}}_{QBLUE}^{ideal}$ denotes ${\boldsymbol{\mathcal D}}_{QBLUE}$ when communication channels between sensors and the FC are error free. 
%
\vspace{-0.2cm}
\subsection{Transmit Power Optimization Using MSE of Quasi BLUE and Classical FIM}
\vspace{-0.1cm}
One can consider the following constrained transmit power optimization problem, where trace (or log-determinant) of ${\boldsymbol{\mathcal D}}_{QBLUE}$ is minimized, subject to the network transmit power constraint as follows:
\vspace{-0.2cm}
\begin{align} \label{minimization problem of MSE_BLUE_ub}
\mathop{\text{minimize}}_{P_k,\forall k}\ \ \ \ \,&\text{tr}({\boldsymbol{\mathcal D}}_{QBLUE}(\{P_k\}_{k=1}^K))\\
\text{s.t.}\ \ \ \ &\sum_{k=1}^{K}P_k\leq P_{tot},\ P_k\in \mathbb{R}^{+},\ \forall k.\nonumber
\end{align}
It is straightforward to show $\frac{\partial\,\text{tr}({\boldsymbol{\mathcal D}}_{QBLUE})}{\partial P_k}\!<\!0$. This implies tr$({\boldsymbol{\mathcal D}}_{QBLUE})$ is a decreasing function of $P_k$'s and the constraint holds with equality. Furthermore, we have $\frac{\partial^2 \text{tr}({\boldsymbol{\mathcal D}}_{QBLUE})}{\partial P_k^2}\!>\!0$, implying that the Hessian is a positive definite matrix and $\text{tr}({\boldsymbol{\mathcal D}}_{QBLUE})$ is jointly convex over $P_k$'s. Moreover, the constraints are linear, and thus, the problem in \eqref{minimization problem of MSE_BLUE_ub} is convex. We could not find a closed-form solution for $P_k$'s. One needs to solve \eqref{minimization problem of MSE_BLUE_ub} numerically to find the optimal $P_k$'s. Since the problem is convex, it is guaranteed that the numerical solution (obtained via the numerical search algorithm) is globally optimal. Since the cost function in \eqref{minimization problem of MSE_BLUE_ub} can be decoupled over $P_k$'s the solution can be implemented in a distributed fashion.

On the other hand, a constrained optimization problem based on maximizing tarce (or log-determinant) of classical FIM ${\boldsymbol{J}}_c$ in \eqref{final formula for J_c} is not meaningful, since ${\boldsymbol{J}}_c$ depends on $\boldsymbol{\theta}$ and thus the power allocation is not realizable.
}
\vspace{-.1cm}
\section{Numerical Results} \label{simulation}
\vspace{-.1cm}
In this section through simulations we corroborate our analytical results. Our analytical results are valid as long as sensors use symmetric mid-rise quantizers. We consider uniform quantizer \cite{Vandendorpe_2012,Vosoughi_Sani_2016,Chang_TSP_2011}, and Lloyd-Max quantizer \cite{Lloyd_Max_quantizer}. 
For the uniform quantizer, quantization levels are $m_{k,l}\!=\!\frac{(2l-1-M_k)\Delta_k}{2}$ for $l\!=\!1,...,M_k$ and quantization boundaries are $u_{k,l}\!=\!\frac{(2l-2-M_k)\Delta_k}{2}$ for $l\!=\!2,...,M_k$, where $\Delta_k$ denotes the quantization step size. Similar to \cite{Vandendorpe_2012}, we assume $x_k$ lies in the interval $[-\tau_k, \tau_k]$ with a high probability for some reasonably large\footnote{Consider quantizing a zero-mean Gaussian $x_k$. For $\tau_k\!=\!3\sigma_{x_k}$ we have
$p(|x_k|\!\geq\!\tau_k)\!=\!2\Phi(-3)\!=\!2.6\times{10}^{-3}$ and for $\tau_k\!=\!5\sigma_{x_k}$ we have $p(|x_k|\!\geq\!\tau_k)\!=\!2\Phi(-5)\!=\!2.86\times{10}^{-5}$, where $\Phi(.)$ is the cumulative distribution function of the standard Gaussian random variable. On the other hand, $\tau_k$ can be decided by the sensor's sensing dynamic range, considering its hardware limitation and sensing capability \cite{Goldsmith_2006}.
} $\tau_k$, i.e., $p(|x_k|\!\geq\!\tau_k)\! \approx \! 0$. To this end, we assume $\tau_k=3\sigma_k$ where $\sigma_k$ is defined in \eqref{definition of sigma_k and rho_ij}. Hence, we choose ${\Delta_k} \!= \! \frac{2\tau_k}{(2^{L_k}-1)}$ \cite{Vandendorpe_2012,Vosoughi_Sani_2016}. 
For the Lloyd-Max quantizer, quantization levels are $m_{k,l}\!=\!\frac{\int_{u_{k,l}}^{u_{k,l+1}}x_kf(x_k)dx_k}{\int_{u_{k,l}}^{u_{k,l+1}}f(x_k)dx_k}$ for $l\!=\!1,...,M_k$ and quantization boundaries are $u_{k,l}\!=\!\frac{m_{k,l-1}+m_{k,l}}{2}$ for $l\!=\!2,...,M_k$ that can be found via iterative design. 
%
\begin{figure}[t]
	\centering
	\includegraphics[width=3.5in,height=1.4in]{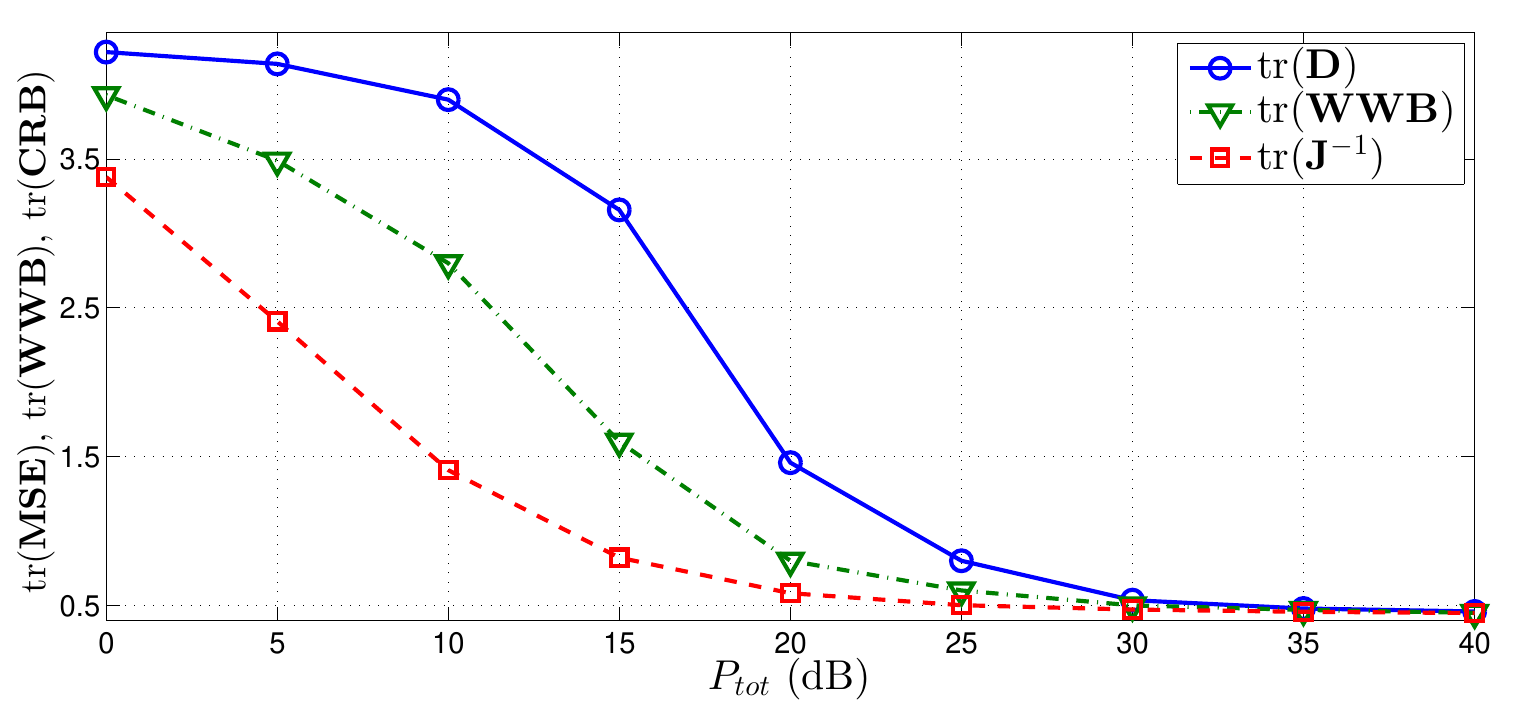}
	\caption{\blue tr(\textbf{MSE}), tr(\textbf{WWB}), and tr(\textbf{CRB}) versus $P_{tot}$ (dB).}
	\label{MSE_WWB_CRB_tr}
\end{figure}
%
{\blue
	\vspace{-0.2cm}
\subsection{Comparison of WWB, Bayesian CRB, and MSE of LMMSE Estimator} \label{wwb, crb, mse}
\vspace{-0.1cm}
We numerically compare {\red traces} of {\red the MSE matrix} of LMMSE estimator{\red ,} the WWB {\red matrix} and {\red the} Bayesian CRB {\red matrix} in Fig.~\ref{MSE_WWB_CRB_tr} 
for various $P_{tot}$, assuming $P_{tot}$ is uniformly distributed among sensors, and uniform quantization and coherent receiver are employed. 
The figure suggests that the WWB is a tighter bound, compared to the Bayesian CRB. 
Similar observations can be made for two types of noncoherent receivers, and also when we compare the determinant of these three matrices. Due to lack of space, we have omitted those plots.}
{\blue
	\vspace{-0.2cm}
\subsection{Behavior of tr$(\boldsymbol{J})$ and $|\boldsymbol{J}|$ in terms of $P_{tot}$ and Quantizer} \label{tr_J and |J| for diff quantizers}} 
Without loss of generality and for the simplicity of presentation, we let $K\!\!=\!\!2$ and consider a zero-mean Gaussian vector
$\boldsymbol{\theta}=\left[\theta_1,\theta_2\right]^T$ with $\boldsymbol{\cal C}_{\boldsymbol{\theta}}=[4,0.5;0.5,0.25]$. We assume $\mathbf{a}_k=[0.6,0.8]^T$, $\sigma_{n_k}\!=\!1$, $\sigma_{w_k}\!=\!1, L_k=3$ bits, $\forall k$. 
%
\begin{figure}[t]
\centering
\includegraphics[width=3.5in,height=1.4in]{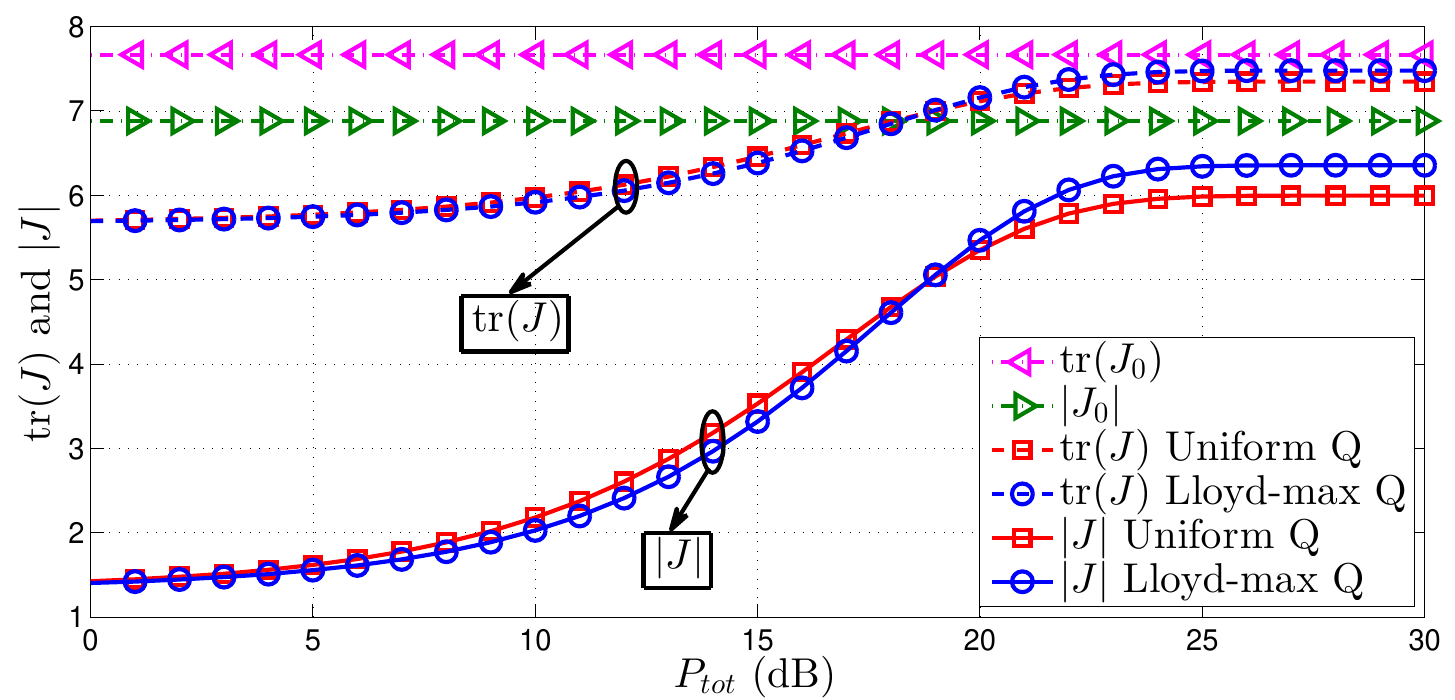}
\caption{tr$(\boldsymbol{J})$ {\blue and} $|\boldsymbol{J}|$ {\blue for uniform and Lloyd-Max quantizers}, baseline tr$(\boldsymbol{J}_0)$ and baseline $|\boldsymbol{J}_0|$ versus $P_{tot}$ (dB).}
\label{all_metrics}
\end{figure}

\begin{figure}[!t]
	
	\hspace{-.2cm}
	\begin{subfigure}[b]{0.25\textwidth}
		
		\centering
		\subcaptionbox{\scriptsize{\blue $\sigma_{n_1}\!=\!\sigma_{n_2}\!=\!1$, $\overline{\delta}_1\!=\!2\!$ dB, $\overline{\delta}_2\!=\!14\!$ dB} \label{tr(J)_power_aloc-1}}{\vspace{-.2 cm}\includegraphics[width=1.8in,height=1.1in]{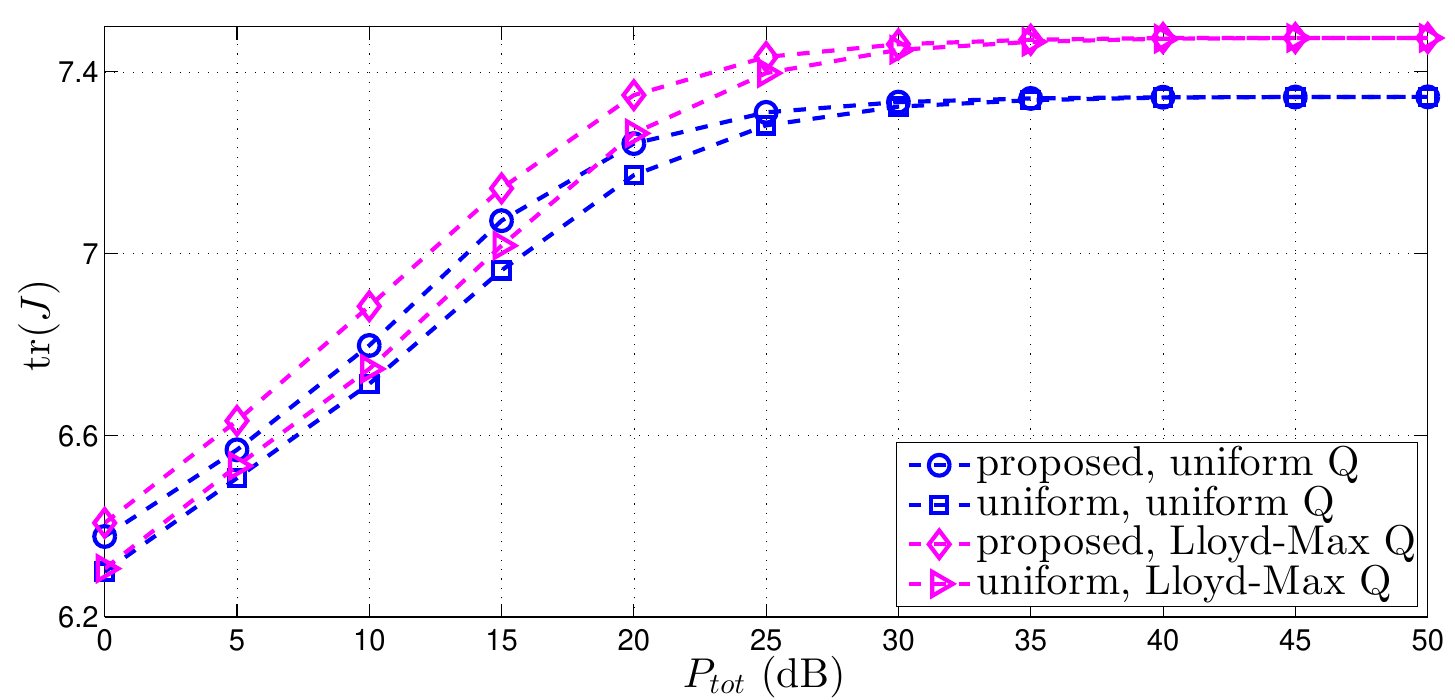}}
		
	\end{subfigure}%
	\begin{subfigure}[b]{0.25\textwidth}
		
		\centering
		\subcaptionbox{\scriptsize{$\sigma_{n_1}\!=\!4, \sigma_{n_2}\!=\!0.5$, $\overline{\delta}_1\!=\!\overline{\delta}_2\!=\!4\!$ dB} \label{tr(J)_power_aloc-2}}{\vspace{-.2 cm}\includegraphics[width=1.8in,height=1.1in]{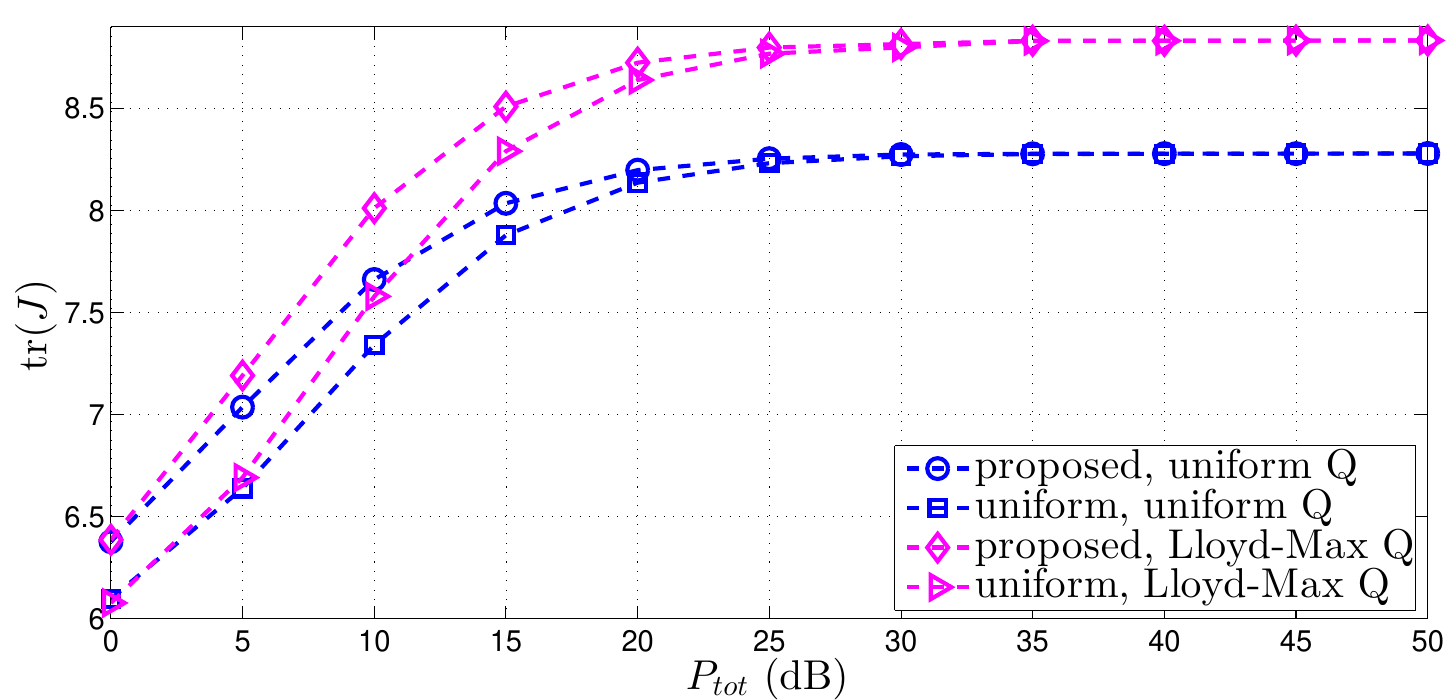}}
		
	\end{subfigure} \\
	
	\caption{{\blue tr$(\boldsymbol{J})$} versus $P_{tot}$ (dB) {\red using proposed and uniform power allocation} for coherent receiver {\blue and uniform and Lloyd-max quantizers}.}  
	\label{tr(J)_power_aloc}
	
\end{figure}

Assuming $|h_k|\!=\!0.5$, Fig.~\ref{all_metrics} depicts tr$(\boldsymbol{J})$ and $|\boldsymbol{J}|$ versus $P_{tot}$ for coherent receiver, considering both uniform and Lloyd-Max quantizers. Fig.~\ref{all_metrics} shows as $P_{tot}$ increases, both metrics increase and asymptotically approach their corresponding baseline (i.e., centralized estimation when full precision observations are used to derive {\blue Bayesian} FIM and form $\hat{\boldsymbol{\theta}}$). There is also a gap between each metric and its corresponding baseline, which is due to quantization. Note that this gap for Lloyd-Max quantizer is smaller than that of uniform quantizer. Comparing Lloyd-Max and uniform quantizers, we observe that when $P_{tot}$ is less than a certain threshold (which depends on the network setup parameters), the latter slightly outperforms the former, and when $P_{tot}$ is greater than the threshold, the former outperforms the latter. As $L_k$ increases, this threshold becomes larger and the performance of both quantizers get closer to each other. The behaviors of tr$(\boldsymbol{J})$ and $|\boldsymbol{J}|$ for noncoherent receivers are the same as those of coherent receiver, hence are omitted due to lack of space. Regarding the behaviors of the two metrics with respect to the observation model parameters, we state that tr$(\boldsymbol{J})$ and $|\boldsymbol{J}|$ increase as the variance of observation noise $\sigma_{n_k}^2$ decreases. 
{\blue
	\vspace{-0.25cm}
\subsection{FIM-max vs. Uniform Power Allocation} \label{Behavior of FIM}}
{\blue We investigate how the behavior of tr$(\boldsymbol{J})$ changes as communication channel and observation model parameters vary. 
Let $\overline{\delta}_k=\frac{\sigma_{h_k}^2}{\sigma_{w_k}^2}$. For coherent receiver, Fig.~\ref{tr(J)_power_aloc-1} plots tr$(\boldsymbol{J})$ evaluated at the corresponding optimal power allocation (i.e., $P_k$'s are the solutions of the problem in \eqref{maximization problem of tr(J)}) versus $P_{tot}$, for both uniform and Lloyd-Max quantizers, when $\sigma_{n_1}\!=\!\sigma_{n_2}\!=\!1$, $\overline{\delta}_1\!=\!2$ dB, $\overline{\delta}_2\!=\!14$ dB.} 
{\blue Fig.~\ref{tr(J)_power_aloc-2} plots the same, with the difference that $\sigma_{n_1}\!=\!4, \sigma_{n_2}\!=\!0.5$, $\overline{\delta}_1\!=\!\overline{\delta}_2\!=\!4$ dB.}   
{\blue To demonstrate the effectiveness of the proposed FIM-max schemes, we also include tr$(\boldsymbol{J})$ evaluated at uniform power allocation $P_k\!=\!P_{tot}/K$ in these figures. Overall,} Fig.~\ref{tr(J)_power_aloc-1}, Fig.~\ref{tr(J)_power_aloc-2} show {\blue that for coherent receiver} the proposed FIM-max schemes outperform uniform power allocation, {\blue for both quantizers and} for all ranges of $P_{tot}$.
Moreover, it is evident that Lloyd-Max quantizer outperforms uniform quantizer in moderate-region to high-region of $P_{tot}$. 
{\blue Similar observations can be made for two types of noncoherent receivers, and also when the optimization metric is $|\boldsymbol{J}|$ (i.e., $P_k$'s are the solutions of the problem in \eqref{maximization problem of logdet(J)}). Due to lack of space, we have omitted those plots. Comparing three types of receivers, our simulations demonstrate that for a given $P_{tot}$, coherent receiver and noncoherent receiver with known channel statistics have the best and the worst performance in terms of tr$(\boldsymbol{J})$ and $|\boldsymbol{J}|$.}
\vspace{-0.25cm}
\subsection{Behavior of FIM-max Power Allocation Across Sensors} \label{Behavior of Power Allocation}
\begin{figure}[t]
	\centering
	{\vspace{-.2 cm}\includegraphics[width=3.5in]{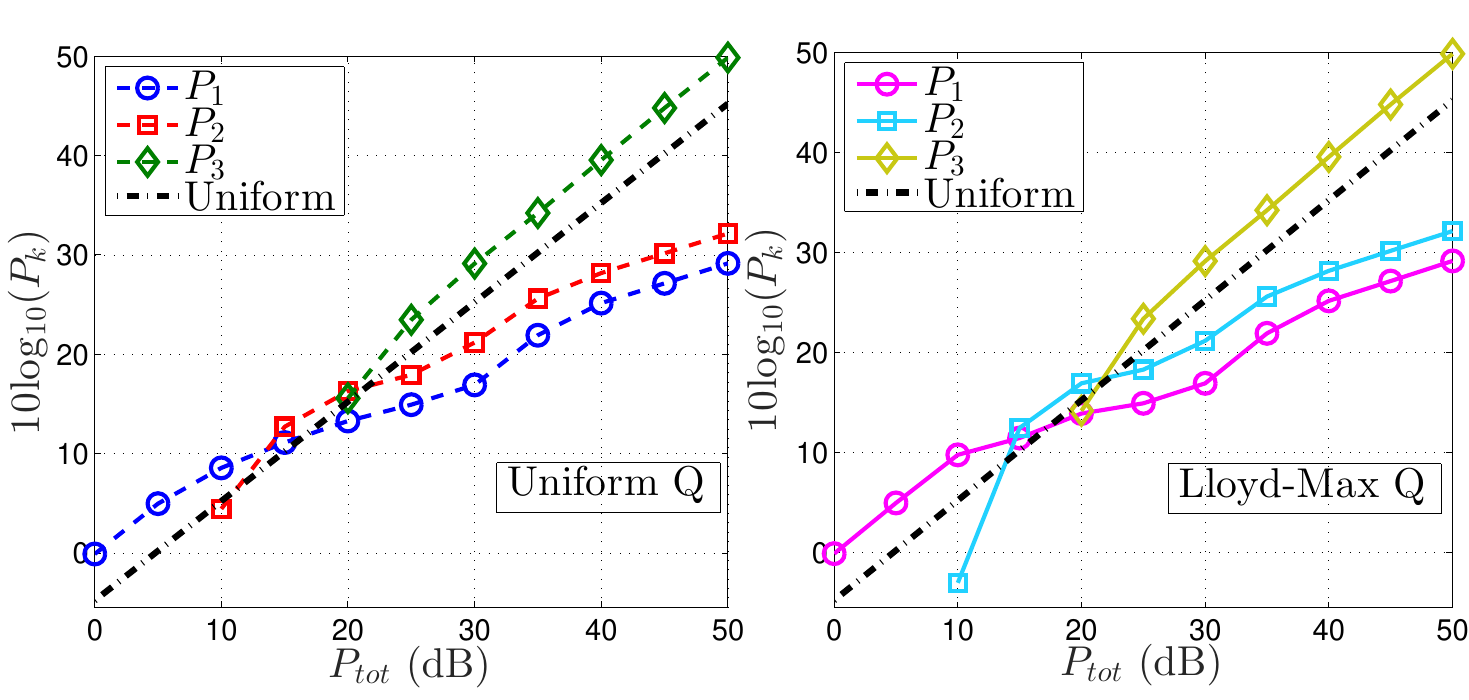}}
	\caption{$\{10\text{log}_{10}(P_k)\}_{k=1}^3$ versus $P_{tot}$ (dB) to maximize tr$(\boldsymbol{J})$ for coherent receiver {\red and uniform and Lloyd-max quantizers. As benchmark uniform power allocation is included}.}  
	\label{P_k_tr_unif_vs_Lloyd}
\end{figure}

{\blue We study the behavior of the FIM-max power allocation across sensors as $P_{tot}$ increases.} 
Recall $\delta_k=\frac{|h_k|^2}{2\sigma_{w_k}^2}$. We let $K\!=\!3$, $\delta_1\!=\!14, \delta_2\!=\!8, \delta_3\!=\!2$, $\mathbf{a}_k=[0.6,0.8]^T$, $\sigma_{n_k}\!=\!1$, $L_k=3$ bits, $\forall k$. Fig.~\ref{P_k_tr_unif_vs_Lloyd} illustrates $\{10\text{log}_{10}(P_k)\}_{k=1}^3$ versus $P_{tot}$ for {\blue coherent receiver}, where $P_k$ 's are the solutions of the problem in \eqref{maximization problem of tr(J)}, for both uniform and Lloyd-Max quantizers. {\blue Regarding Fig.~\ref{P_k_tr_unif_vs_Lloyd} we make the following four observations}: 1) $P_k$ increases as $P_{tot}$ increases, 2) the power allocations obtained for Lloyd-Max quantizer are very close to those obtained for uniform quantizer, 3) when $P_{tot}$ is small, only sensor 1 is active, and as $P_{tot}$ increases, sensors 2 and 3 become active in a sequential order, 4) in low-region of $P_{tot}$, a sensor with a larger $\delta_k$ is allotted a larger $P_k$ (water filling), and in high-region of $P_{tot}$, a sensor with a smaller $\delta_k$ is allotted a larger $P_k$ (inverse of water filling). 
{\blue
Although we don't have a closed-form solution for $P_k$'s, our conjecture is that its change of behavior in terms of $P_{tot}$, can be explained by examining the $P_k$'s solution provided in \cite{Vosoughi_Sani_2016}, where the authors have considered a related problem.
In particular, \cite{Vosoughi_Sani_2016} considered minimizing an upper bound on the MSE of LMMSE estimator, subject to a network transmit power constraint, given quantization bits. 
%
%
For coherent receiver, based on the closed-form solutions of $P_k$'s the authors in \cite{Vosoughi_Sani_2016} found the following:
%
\vspace{-0.1cm}
\begin{eqnarray} \label{gamme-water-filling}
\!\left\{
\begin{array}{lr}
\!\mbox{when}~\delta_k<\frac{e{\lambda}^{*}}{\alpha_k},~~~~&\mbox{as}~ \delta_k ~\mbox{increases,} ~P_k ~{\mbox{increases}},\\
\!\mbox{when}~\delta_k >\frac{e{\lambda}^{*}}{\alpha_k},~~~~&\mbox{as}~ \delta_k ~\mbox{increases,} ~P_k ~{\mbox{decrease}}.
\end{array}
\right.
\end{eqnarray}
%
%
\vspace{+.01cm}
$\!\!$Equation (\ref{gamme-water-filling}) shows that the behavior of $P_k$'s can change, depending on whether $\delta_k$ is larger or smaller than the threshold  $\delta^{th}_k=\frac{e{\lambda}^{*}}{\alpha_k}$.
%
The parameter $\alpha_k$ in (\ref{gamme-water-filling}) depends on the observation vectors and quantization. The optimal value of Lagrange multiplier ${\lambda}^{*}$ in (\ref{gamme-water-filling}) is related to $P_{tot}$ according to ${\lambda}^{*}=e^{a(-P_{tot}+b)}$ where $a>0, b$ are common terms among sensors. 
Revisiting the results in \cite{Vosoughi_Sani_2016}, now we return to Fig.~\ref{P_k_tr_unif_vs_Lloyd}. Given the observation vectors and quantization (given $\alpha_k$) and given $\delta_k$, suppose $P_{tot}$ increases. Increasing $P_{tot}$ implies that ${\lambda}^{*}$ and thus the thresholds $\delta^{th}_k$'s decrease. Therefore, $\delta_k$'s are being compared against smaller thresholds $\delta^{th}_k$'s. In high-region of $P_{tot}$ the thresholds $\delta^{th}_k$'s are so small that each $\delta_k$ exceeds $\delta^{th}_k$ (all channels can be viewed as ``strong''). In this case, the allocation of power among sensors is such that, if $\delta_1 \!<\! \delta_2 \!<\! \delta_3$ then $P_3 \!<\! P_2\!<\! P_1,$ (the sensor with a less stronger channel is allocated more transmit power). 
In contrary, given $\alpha_k$ and given $\delta_k$ suppose $P_{tot}$ decreases. Decreasing $P_{tot}$ implies that ${\lambda}^{*}$ and thus the thresholds $\delta^{th}_k$'s increase. Hence, $\delta_k$'s are being compared against larger thresholds $\delta^{th}_k$'s. 
In low-region of $P_{tot}$ the thresholds $\delta^{th}_k$'s are so large that each $\delta_k$ is below $\delta^{th}_k$ (all channels can be viewed as ``weak''). In this case, the allocation of power among sensors is such that, if $\delta_1 \!<\! \delta_2 \!<\! \delta_3$ then $P_1 \!<\!P_2\!<\! P_3,$ (the sensor with a less weaker channel is allocated more transmit power).
}\\
Note that the behavior of $P_k$'s as the solutions of the problem in \eqref{maximization problem of logdet(J)} with respect to $P_{tot}$ is analogous to that depicted in Fig.~\ref{P_k_tr_unif_vs_Lloyd}. 
{\blue Moreover, the behavior of $P_k$'s for two types of noncoherent receivers are similar to that of coherent receiver. Due to lack of space, we have omitted those plots.}
%
\begin{figure}[!t]
	
		\hspace{-.2cm}
	\begin{subfigure}[b]{0.25\textwidth}
		
		\centering
		\subcaptionbox{$\sigma_{n_1}\!=\!\sigma_{n_2}\!=\!1, \overline{\delta}_1\!=\!2\text{dB}, \overline{\delta}_2\!=\!14\text{dB}$  \label{MSE_power_aloc-1a}}{\vspace{-.2 cm}\includegraphics[width=1.8in,height=1.04in]{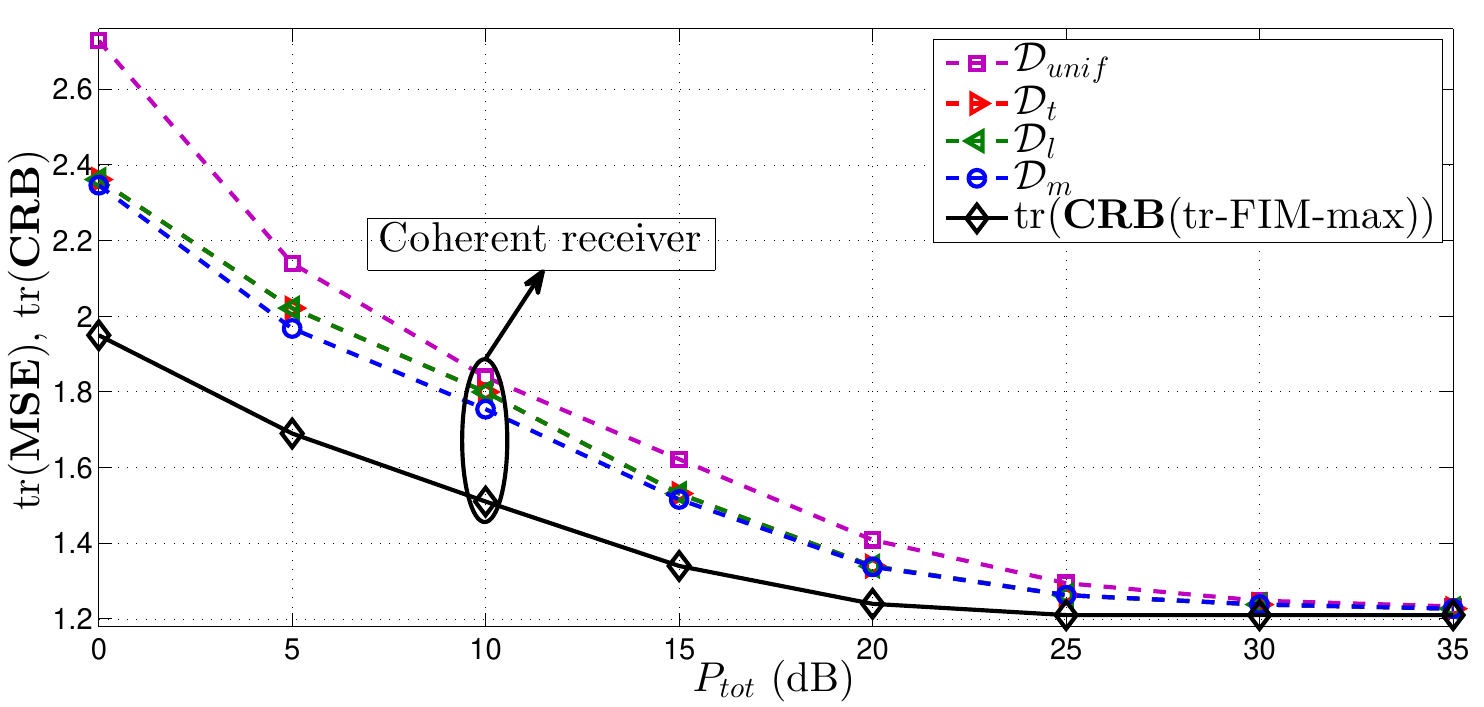}}
		
	\end{subfigure}%
	\begin{subfigure}[b]{0.25\textwidth}
		
		\centering
		\subcaptionbox{$\sigma_{n_1}\!=\!\sigma_{n_2}\!=\!1, \overline{\delta}_1\!=\!2\text{dB}, \overline{\delta}_2\!=\!14\text{dB}$ \label{MSE_power_aloc-1b}}{\vspace{-.2 cm}\includegraphics[width=1.8in,height=1.04in]{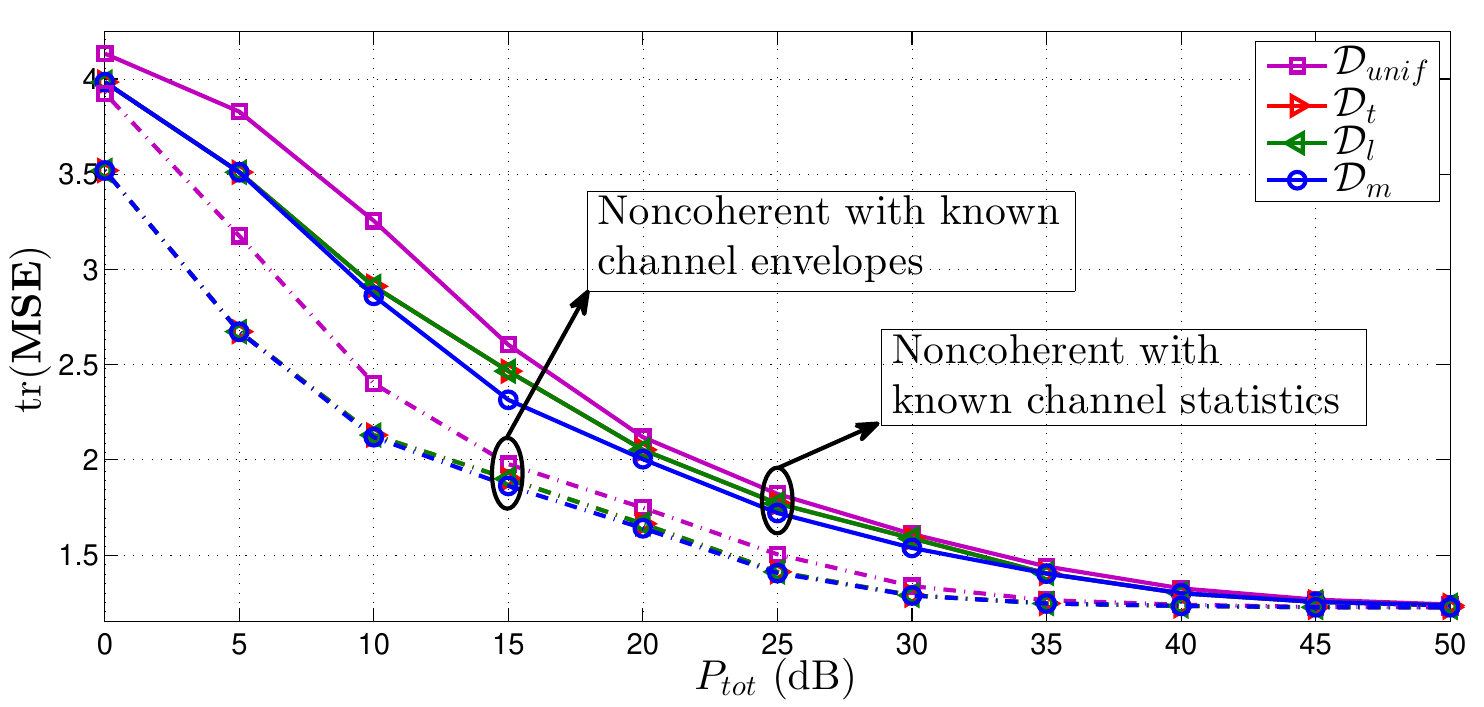}}
		
	\end{subfigure} 
	%
	\vspace{.1cm}
	\begin{subfigure}[b]{0.25\textwidth}
		\hspace{-.3cm}
		\centering
		\subcaptionbox{$\sigma_{n_1}\!=\!4, \sigma_{n_2}\!=\!0.5, \overline{\delta}_1\!=\!\overline{\delta}_2\!=\!4\text{dB}$ \label{MSE_power_aloc-2a}}{\vspace{-.2 cm}\includegraphics[width=1.8in,height=1.04in]{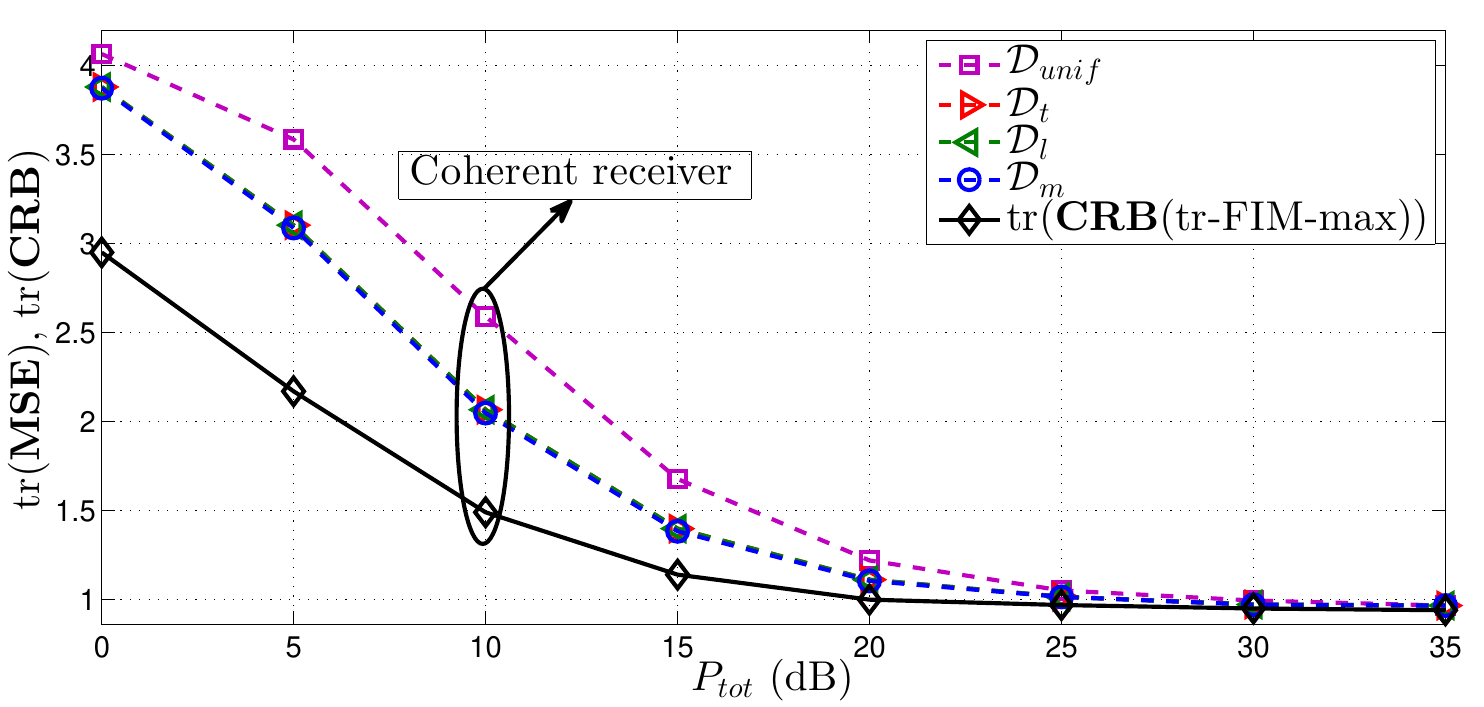}}
		
	\end{subfigure}%
	\begin{subfigure}[b]{0.25\textwidth}
		\hspace{-.3cm}
		\centering
		\subcaptionbox{$\sigma_{n_1}\!=\!4, \sigma_{n_2}\!=\!0.5, \overline{\delta}_1\!=\!\overline{\delta}_2\!=\!4\text{dB}$ \label{MSE_power_aloc-2b}}{\vspace{-.2 cm}\includegraphics[width=1.8in,height=1.04in]{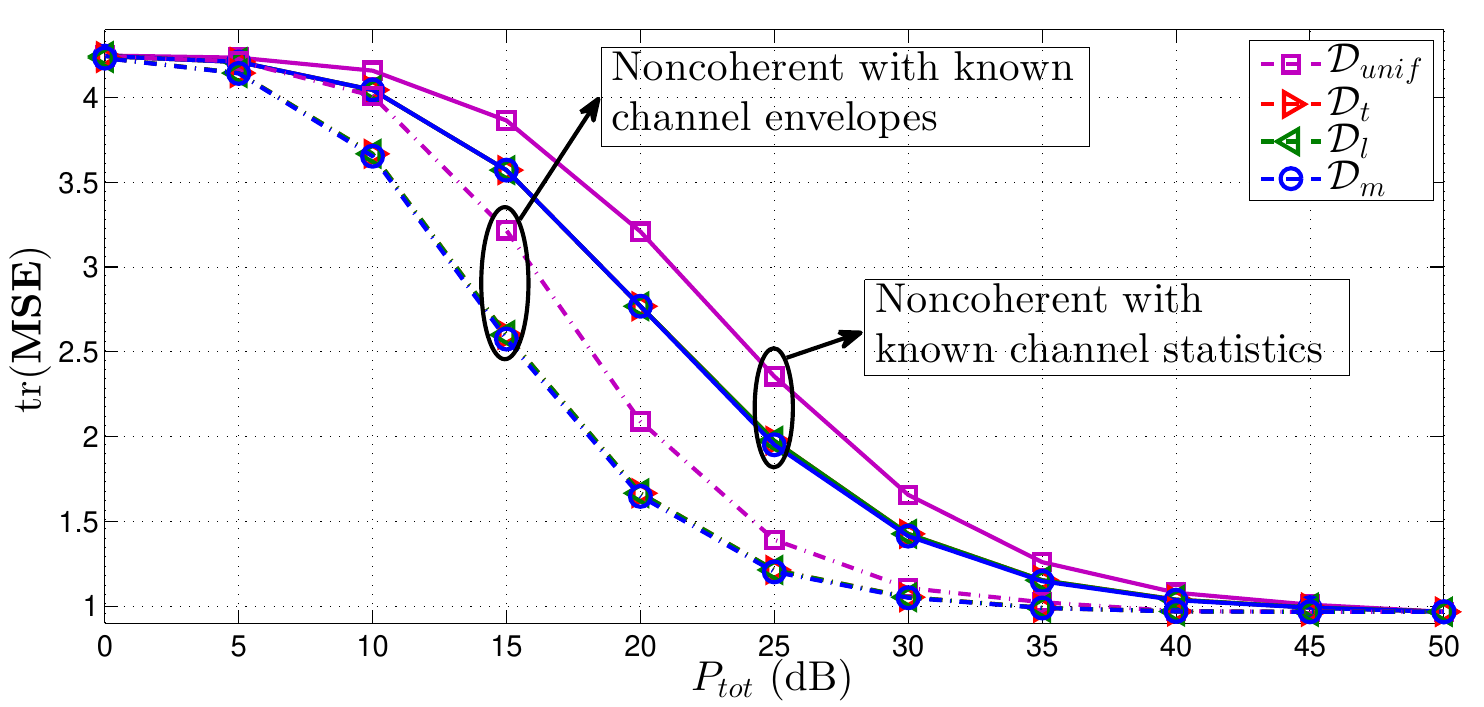}}
		
	\end{subfigure}
	
	\caption{{\blue ${\cal D}_{m}$, ${\cal D}_{l}$, ${\cal D}_{t}$, ${\cal D}_{unif}$} versus $P_{tot}$ (dB) for uniform quantizer. 
		{\red (a) and (c) coherent receiver, as benchmark tr(CRB(tr-FIM-max)) is included, (b) and (d) two types of noncoherent receivers.} 
        {\blue For all three types of receivers,} {\red tr(CRB(tr-FIM-max)) $<$} {\blue ${\cal D}_{m}\leq{\cal D}_{l}\approx{\cal D}_{t}\leq{\cal D}_{unif}$.}}  
	\label{MSE_power_aloc}
	
\end{figure}
\vspace{-0.3cm}
\subsection{FIM-max vs. MSE-min Power Allocation} \label{Performance in the MSE Sense}
\vspace{-0.1cm}
{\blue We explore how the FIM-max schemes are compared with the power allocation that can be obtained from constrained minimization of the MSE of the LMMSE estimator derived in Section \ref{MSE}. Let ${\cal D}_{l}\!=\!\text{tr}(\boldsymbol{\mathcal D}(\text{FIM-max}))$ and ${\cal D}_{unif}=\text{tr}(\boldsymbol{\mathcal D}(\{P_k\!=\!P_{tot}/K\}_{k=1}^K))$, denote trace of $\boldsymbol{\mathcal D}$ at $P_k$'s obtained from solving the problem in \eqref{maximization problem of logdet(J)} and uniform power allocation, respectively.} {\red Fig.~\ref{MSE_power_aloc-1a} and Fig.~\ref{MSE_power_aloc-1b} illustrate}
{\blue the numerical evaluations of ${\cal D}_{m}$, ${\cal D}_{t}$ defined in Section \ref{subsection on Appropriateness}, as well as ${\cal D}_{l}$, ${\cal D}_{unif}$, versus $P_{tot}$ for} 
{\red coherent receiver and two types of noncoherent receivers, respectively,} {\blue and for the same setup parameters as Fig.~\ref{tr(J)_power_aloc-1}. To fairly compare the performance of different receivers, we obtain the numerical results for coherent receiver and noncoherent receiver with known channel envelopes by taking expectation over fading channel envelope vector $\boldsymbol{|h|}$, such that $\mathbb{E}\left[|h_k|^2\right]=2\sigma_{h_k}^2, \forall k$.}
{\red Fig.~\ref{MSE_power_aloc-2a} and Fig.~\ref{MSE_power_aloc-2b} plot the same as Fig.~\ref{MSE_power_aloc-1a} and Fig.~\ref{MSE_power_aloc-1b},}
{\blue with different setup parameters though (the same parameters as Fig.~\ref{tr(J)_power_aloc-2}).}
These figures show ${\cal D}_{m}\leq{\cal D}_{l}{\blue \approx}{\cal D}_{t}\leq{\cal D}_{unif}$ for all three receivers and all ranges of $P_{tot}$, i.e., performance of both FIM-max schemes are very close to that of MSE-min scheme (when we average over $\boldsymbol{|h|}$). 
{\red We also plot tr(CRB(tr-FIM-max)) versus $P_{tot}$ for coherent receiver. Fig.~\ref{MSE_power_aloc-1a} and Fig.~\ref{MSE_power_aloc-2a} illustrate the inequality $\tr(\text{CRB}({\text{FIM-max}}))\!<\!{\cal D}_{m}$ in \eqref{crb-mse-inequality}. The same observation is made for two types of noncoherent receivers. Due to lack of space, these plots are omitted. 
}
{\blue It is worth mentioning that from the estimation theory we know ${\cal D}_{m}\!<\!{\cal D}_{t}$ and ${\cal D}_{m}\!<\!{\cal D}_{l}$. What our simulations suggest is that in our problem they are indeed very close to each other. This observation is very important since it indicates that, although Bayesian CRB is not attainable in our problem and the WWB is tighter than Bayesian CRB, it is still proper to use FIM-max power allocation (instead of power allocation that minimizes the WWB or the MSE of the LMMSE estimator), since the {\red differences} ${\cal D}_{m}\!-\!{\cal D}_{t}$ {\red and} ${\cal D}_{m}\!-\!{\cal D}_{l}$ {\red are} small and not significant.} 
While in low-region and high-region of $P_{tot}$, ${\cal D}_{t}$ and ${\cal D}_{l}$ are much closer to ${\cal D}_{m}$, in moderate-region of $P_{tot}$, there is a small gap between them. 
{\blue Comparing three types of receivers for a given $P_{tot}$, coherent receiver and noncoherent receiver with known channel statistics have the best and the worst performance.} 
{\blue Similar observations can be made for Lloyd-Max quantizers. Due to lack of space we have omitted those plots.}
\vspace{-0.3cm}
\subsection{Estimation Performance of a Randomly Deployed Network} \label{effect of K}
\vspace{-0.1cm}
{\blue We investigate the impact of network size $K$ on the MSE performance and compare tr(MSE) that is evaluated at different transmit power allocation.}
%
We assume $K\!=\!20$ sensors are randomly deployed in a $2m\times2m$ field, where the origin is the center of the field, and compare the numerical results with $K\!=\!2$ sensors. 
We consider a zero-mean Gaussian vector $\boldsymbol{\theta}=\left[\theta_1,\theta_2\right]^T$ with $\boldsymbol{\cal C}_{\boldsymbol{\theta}}=[4,0.5;0.5,0.25]$. The distance between each external signal source $\theta_i$ located at $(x_{t_i},y_{t_i})$ and sensor $k$ located at $(x_{s_k},y_{s_k})$ is: 
\vspace{-0.1cm}
\begin{equation*}
d_{ki}=\sqrt{(x_{s_k}-x_{t_i})^2+(y_{s_k}-y_{t_i})^2}, \ \ \ k=1,..., 20, \ \ i=1,2
\end{equation*}
Let $d_{0i}$ be the distance of source $\theta_i$ from the origin. Without loss of generality, we assume $d_{01}=d_{02}=1m$. To characterize the observation gain vectors $\mathbf{a}_k, \forall k$ in \eqref{obs_model} we adopt an isotropic intensity attenuation model,
%
%
where $\mathbf{a}_k\!=\![(\frac{d_{01}}{d_{k1}})^n, (\frac{d_{02}}{d_{k2}})^n]^T$ and $n$ is the signal decay exponent which is approximately 2 for distances $\leq\!1km$ \cite{Li_Eurasip_2003}. We assume $\sigma_{w_k}=1$. For coherent receiver and noncoherent receiver with known channel envelopes we let $|h_k|\!=\!1, \forall k$, and for noncoherent receiver with known channel statistics we let $\sigma_{h_k}\!=\!1, \forall k$.
\begin{figure}[t]
	\centering
	{\vspace{-.2 cm}\includegraphics[width=3.5in]{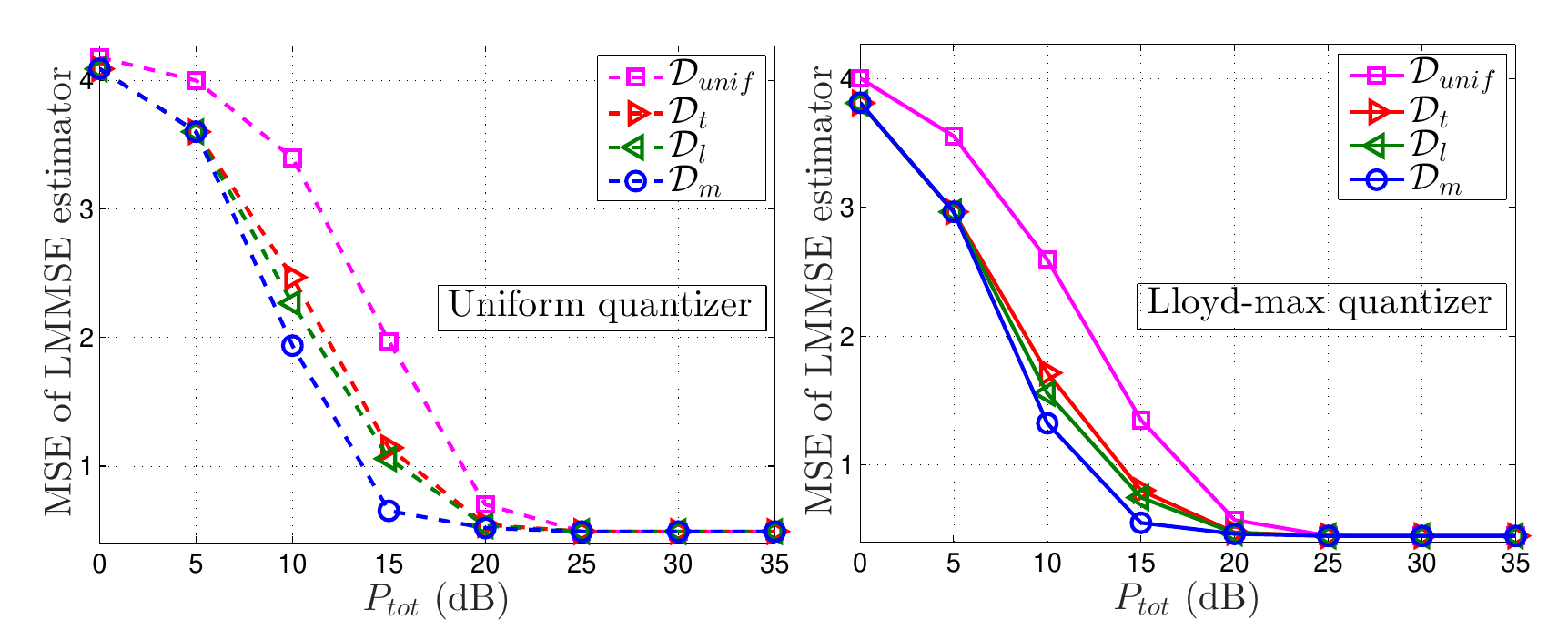}}
	\caption{{\blue ${\cal D}_{m}$, ${\cal D}_{l}$, ${\cal D}_{t}$, ${\cal D}_{unif}$} versus $P_{tot}$ (dB) for coherent receiver, $K\!=\!20$.} 
	\label{effect_of_network_size} 
\end{figure}

Fig.~\ref{effect_of_network_size} plots {\blue ${\cal D}_{m}, {\cal D}_{l}, {\cal D}_{t}, {\cal D}_{unif}$} versus $P_{tot}$ for coherent receiver, and both uniform and Lloyd-Max quantizers. Fig.~\ref{effect_of_network_size} demonstrates the superiority of FIM-max schemes, compared to uniform power allocation for all ranges of $P_{tot}$. 
Furthermore, the observation ${\cal D}_{l}\leq{\cal D}_{t}$, suggests that {\blue log-det}-FIM-max power allocation is closer to MSE-min power allocation, compared to tr-FIM-max power allocation (for a given realization of $\boldsymbol{|h|}$). This is intuitively appealing, since the Bayesian FIM $\boldsymbol{J}$ is not a diagonal matrix and {\blue log-det}-FIM-max power allocation extracts and utilizes more information from $\boldsymbol{J}$, compared to tr-FIM-max power allocation. 
{\blue Similar observations can be made for two types of noncoherent receivers. Due to lack of space we have omitted those plots.}
\vspace{-.1cm}
\section{Conclusions} \label{conclusions}
We derived the Bayesian FIM $\boldsymbol{J}$ {\blue and the WWB} for distributed estimation of a Gaussian vector, when sensors transmit their digitally modulated quantized observations to the FC over power-constrained orthogonal noisy fading channels. We formulated and addressed constrained maximization of tr$(\boldsymbol{J})$ and log$_2(|\boldsymbol{J}|)$ under the constraint on $P_{tot}$. We also derived the LMMSE estimator and its corresponding MSE. Through simulations we observed that both tr$(\boldsymbol{J})$ and $|\boldsymbol{J}|$ increase as $P_{tot}$ increases. Regarding the solutions of the formulated constrained maximization problems, we noticed that in low-region and high-region of $P_{tot}$, $P_{tot}$ is alloted among sensors in a water filling and inverse of water filling fashion, respectively. 
We also considered the power allocation solution obtained from minimizing the MSE of the LMMSE estimator (MSE-min scheme). Numerical results demonstrated the {\blue effectiveness} of FIM-max schemes for different network setup parameters, as the MSE associated with FIM-max schemes are very close to that of MSE-min scheme and outperform that of uniform power allocation in all simulation scenarios. 
{\blue These suggest that, although the WWB is tighter than the Bayesian CRB in our problem (and Bayesian CRB is not attainable), it is still appropriate to use FIM-max schemes, since the performance loss in terms of the MSE of the LMMSE estimator is not significant.} 
Comparing the performance of three types of receivers, our numerical results revealed that coherent receiver and noncoherent receiver with known channel statistics have the best and the worst performance, respectively. 
Comparing uniform and Lloyd-Max quantizers, we observed that the latter outperforms the former in moderate-region to high-region of $P_{tot}$ for all receivers.
\begin{figure}[t]
	\centering
	\includegraphics[width=3.4in,height=1.5in]{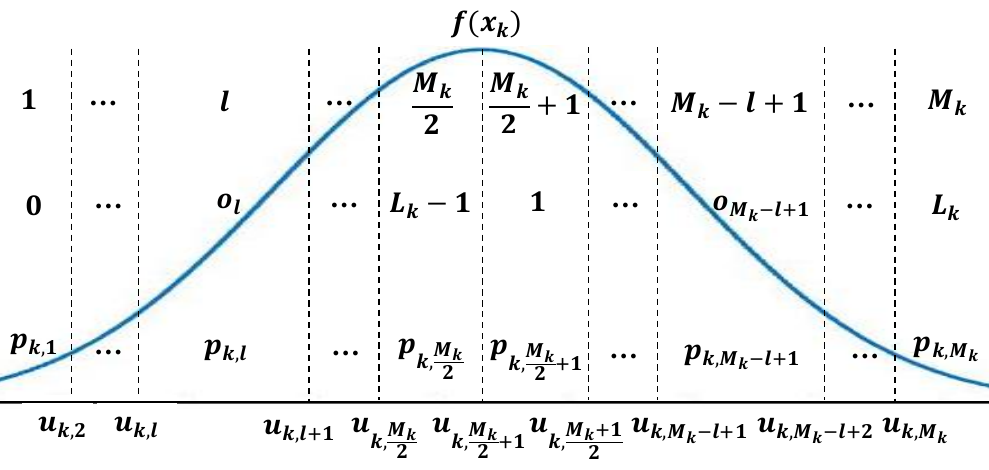}
	\caption{Quantization and encoding of $x_k$ with symmetric pdf.}
	\label{lemma p(H_0)=p(H_1)}
\end{figure}
%
\vspace{-0.2cm}
\appendix
\section{Appendix} \label{appendix}
{\blue
\subsection{Proof of Lemma \ref{p(m_hat|theta)}} \label{Proof of Lemma p(m_hat|theta)}}
\vspace{-0.1cm}
By using the Bayes' rule, we have:
\vspace{-0.1cm}
\begin{equation} \label{expanding p(m_hat_vec|theta)} 
p(\boldsymbol{\hat{m}}\arrowvert\boldsymbol{\theta})=\sum\limits_{m_1}\dots\sum\limits_{m_K}p(\boldsymbol{\hat{m}}\arrowvert\boldsymbol{m},\boldsymbol{\theta})p(\boldsymbol{m}\arrowvert\boldsymbol{\theta}).
\end{equation}
Since the communication channels are orthogonal and communication channel noises are independent, we can write:
\vspace{-0.15cm}
\begin{equation} \label{p(m_hat_k|m_k)}
p(\boldsymbol{\hat{m}}\arrowvert\boldsymbol{m})=\prod_{k=1}^{K}p(\hat{m}_k\arrowvert m_k). 
\end{equation}
Moreover, given $\boldsymbol{m}$, $\boldsymbol{\hat{m}}$ depends on communication channel noises and $\boldsymbol{\theta}$ depends on observation noises. However, observation and channel noises are two independent random processes. Hence, given $\boldsymbol{m}$, $\boldsymbol{\theta}$ and $\boldsymbol{\hat{m}}$ are conditionally independent. That is, $\boldsymbol{\theta}$, $\boldsymbol{m}$ and $\boldsymbol{\hat{m}}$ form a Markov chain\footnote{We say that random variables $x,y,z$ form a Markov chain, denoted by $x \rightarrow y \rightarrow z$, if Markov property holds $p(z|x,y)=p(z|y)$ \cite{Cover_Book}.} and we conclude:
\vspace{-0.1cm}
\begin{equation} \label{m_hat|m,theta using markov property}
p(\boldsymbol{\hat{m}}\arrowvert\boldsymbol{m},\boldsymbol{\theta})=p(\boldsymbol{\hat{m}}\arrowvert\boldsymbol{m}).
\end{equation}
Combining \eqref{p(m_hat_k|m_k)} and \eqref{m_hat|m,theta using markov property}, $p(\boldsymbol{\hat{m}}\arrowvert\boldsymbol{m},\boldsymbol{\theta})$ in \eqref{expanding p(m_hat_vec|theta)} {\blue becomes}:
\vspace{-0.2cm}
\begin{equation} \label{m_hat|m,theta equals prod m_hat_k|m_k}
p(\boldsymbol{\hat{m}}\arrowvert\boldsymbol{m},\boldsymbol{\theta})=\prod_{k=1}^{K}p(\hat{m}_k\arrowvert m_k).
\end{equation}
%
Let $\boldsymbol{x}\!=\![x_1,...,x_K]^T$ be the observation vector. Since Gaussian observation noises $n_k$'s are uncorrelated across the sensors and also uncorrelated with Gaussian {\boldmath$\theta$}, we have $f(\boldsymbol{x}\arrowvert\boldsymbol{\theta})=\prod_{k=1}^{K}f(x_k\arrowvert\boldsymbol{\theta})$.
This implies:
\vspace{-0.25cm}
\begin{equation} \label{pdf of m_vec given theta}
p(\boldsymbol{m}\arrowvert\boldsymbol{\theta})\!=\!\prod_{k=1}^{K}p(m_k\arrowvert\boldsymbol{\theta}),
\end{equation}
%
\begin{figure*}[b]
	\begin{equation} \label{combining all to obtain p(m_hat_vec|theta)}
	p(\boldsymbol{\hat{m}}\arrowvert\boldsymbol{\theta})=\sum\limits_{m_1}\dots\sum\limits_{m_K}\prod_{k=1}^{K}\left[p(\hat{m}_k|m_k)p(m_k|\boldsymbol{\theta})\right]\overset{(a)}{=}\prod_{k=1}^{K}[\sum_{m_k}p(\hat{m}_k|m_k)p(m_k|\boldsymbol{\theta})]\overset{(b)}{=}\prod_{k=1}^{K}p(\hat{m}_k\arrowvert\boldsymbol{\theta}),
	\end{equation}
\end{figure*}
$\!\!$Substituting \eqref{m_hat|m,theta equals prod m_hat_k|m_k} and \eqref{pdf of m_vec given theta} in \eqref{expanding p(m_hat_vec|theta)}, we reach \eqref{combining all to obtain p(m_hat_vec|theta)} bellow in which ($a$) is obtained from some straightforward mathematical manipulations and $(b)$ is obtained using the Bayes' rule and the fact that $\boldsymbol{\theta}, m_k, \hat{m}_k$ form a Markov chain.
\subsection{Proof of Lemma \ref{p(H_0)=p(H_1)}} \label{Proof of Lemma p(H_0)=p(H_1)}
\begin{figure*}[b]
\begin{align} \label{I2 integral after some manipulation}
{\mathcal I}_{k,l}^{2}\!=\!s_{1k}\!\!\!\!\!\int\limits_{u_{k,l}}^{u_{k,l+1}}\!\!\!\int\limits_{\boldsymbol{v}\in V_v}\!\!\!\!\exp\{\!-\frac{1}{2}\![(\boldsymbol{v}\!-\!\boldsymbol{\omega}_k)^T\!\!{\boldsymbol{Q}}_k^{-1}\!(\boldsymbol{v}\!-\!\boldsymbol{\omega}_k)]\!-\!\frac{1}{2}[\!\frac{x_k^2}{\sigma_{n_k}^2}\!-\boldsymbol{\omega}_k^T\!{\boldsymbol{Q}}_k^{-1}\boldsymbol{\omega}_k]\}d\boldsymbol{v}dx_k\!=\!s_{2k}\!\!\!\!\int\limits_{u_{k,l}}^{u_{k,l+1}}\!\!\!\!\exp\{\!-\frac{1}{2}[\!\frac{x_k^2}{\sigma_{n_k}^2}\!-\!\boldsymbol{\omega}_k^T{\boldsymbol{Q}}_k^{-1}\boldsymbol{\omega}_k]\}dx_k,
\end{align}
\end{figure*}
Given the assumptions made in lemma \ref{p(H_0)=p(H_1)} and the number of quantization bits $L_k$, Fig.~\ref{lemma p(H_0)=p(H_1)} illustrates how the noisy observation $x_k$ is quantized and encoded. Define $p\left(u_{k,l}<x_k\leq u_{k,l+1}\right)\!=\!p_{k,l}$, where $u_{k,l}$'s are the quantization boundaries specified in Section \ref{System Model}. Since $x_k$ has a symmetric pdf and the quantizer is symmetric, we have:
\vspace{-0.1cm}
\begin{equation*}
p_{k,l}=p_{k,M_k-l+1},\ \ \ \ \ \ l=1,...,M_k.
\end{equation*} 
Define $o_{k,l}$ as the number of ones in encoded quantization index $l$. When the quantization indices are encoded using natural binary coding we can show {\blue that} $o_{k,l}=L_k-o_{k,M_k-l+1}$. Therefore, the prior probability $p({\cal H}_{1,i}),\ i=1,...,L_k$ can be computed as:
\vspace{-0.15cm}
\begin{equation*}
p({\cal H}_{1,i})\!=\!\frac{\sum_{l=1}^{M_k}\!o_{k,l}p_{k,l}}{L_k}\!=\!\frac{\sum_{l=1}^{\frac{M_k}{2}}\!\left[o_{k,l}p_{k,l}\!+\!(L_k\!-\!o_{k,l})p_{k,l}\right]}{L_k}\!=\!\frac{1}{2}.
\end{equation*} 
Similarly, we can show {\blue that} $p({\cal H}_{0,i})=1/2$.
\vspace{-0.1cm}
\subsection{Calculation of $\boldsymbol{\mathcal I}_{k,l}^{1}$ in \eqref{E_theta_mhat_k_final}, and ${\mathcal I}_{k,l}^{2}$ and ${\mathcal I}_{i,j,l_1,l_2}^{3}$ in \eqref{E_mhat_i_mhat_j}} \label{deriving the integrals}
\vspace{-0.1cm}
We first calculate ${\mathcal I}_{k,l}^{2}$. We consider the eigenvalue decomposition of $\boldsymbol{\mathcal C}_{\boldsymbol{\theta}}\!=\!\boldsymbol{U}\boldsymbol{\Sigma}\boldsymbol{U}^T$ where $|\boldsymbol{\mathcal C}_{\boldsymbol{\theta}}|\!=\!|\boldsymbol{\Sigma}|$, $|U|\!=\!\pm1$. We define $\boldsymbol{v}\!=\!\boldsymbol{U}^T\boldsymbol{\theta}$ and therefore $d\boldsymbol{v}\!=\!{|\boldsymbol{U}|}^qd\boldsymbol{\theta}$ \cite{Vosoughi_2007}, and also $\boldsymbol{\psi}_k\!=\!\boldsymbol{U}^T\mathbf{a}_k$ in which $\mathbf{a}_k$ is sensor $k$ observation gain vector. Using these definitions and changes of variables along with the definition of $\beta_{k,l}(\boldsymbol{\theta})$ in (\ref{beta_theta}), ${\mathcal I}_{k,l}^{2}$ becomes:
\vspace{-0.2cm}
\begin{equation*} 
{\mathcal I}_{k,l}^{2}\!=\!s_{1k}\!\!\!\!\int\limits_{u_{k,l}}^{u_{k,l+1}}\!\!\!\int\limits_{\boldsymbol{v}\in V_v}\!\!\!\!\exp\{-\frac{1}{2}[\frac{\left(x_k\!-\!\boldsymbol{\psi}_k^T\boldsymbol{v}\right)^2}{\sigma_{n_k}^2}\!+\!\boldsymbol{v}^T{\boldsymbol{\Sigma}}^{-1}\boldsymbol{v}]\}d\boldsymbol{v}dx_k,
\end{equation*}
where $s_{1k}\!=\!\frac{1}{\sqrt{(2\pi)^{q+1}|\boldsymbol{\Sigma}|}\sigma_n{|U|}^q}$ and $V_v$ denotes the $q$-dimensional volume over which we take integral in the new coordinate. After expanding the argument of exponential function of the integrand and using completing square, and defining:
%
%
\vspace{-0.1cm}
\begin{equation} \label{definition of inv of matrix Q and omega}
{\boldsymbol{Q}}_k^{-1}\!=\!{\boldsymbol{\Sigma}}^{-1}+\boldsymbol{\psi}_k{\boldsymbol{\psi}}_k^T/\sigma_{n_k}^2,\ \ \boldsymbol{\omega}_k\!=\!\frac{x_k}{\sigma_{n_k}^2}{\boldsymbol{Q}_k}\boldsymbol{\psi}_k,
\end{equation} 
${\mathcal I}_{k,l}^{2}$ can be obtained as in \eqref{I2 integral after some manipulation}, in which $s_{2k}\!=\!\frac{\sqrt{\left|{\boldsymbol{Q}}_k\right|}}{\sqrt{2\pi|\boldsymbol{\Sigma}|}\sigma_{n_k}}$, and for the second equality, we have used the fact that integral of pdf of Gaussian random vector $\boldsymbol{v}$ over $V_v$ is equal to 1. The term ${|U|}^q\!=\!\pm1$ in the denominator of $s_{1k}$ is absorbed in the integration over $\boldsymbol{v}$, because the effects of change of variable from $\boldsymbol{\theta}$ to $\boldsymbol{v}$ on $V_{\theta}$ to $V_v$ and $d\boldsymbol{\theta}$ to  $d\boldsymbol{v}$ cancel each other. Since $\left|{\boldsymbol{Q}}_k\right|\!=\!1\big/\left|{\boldsymbol{Q}}_k^{-1}\right|$, using the Matrix Determinant Lemma which performs a rank-1 update to a determinant \cite{Matrix_Analysis}, we obtain:
\vspace{-0.1cm}
\begin{equation*} 
\left|{\boldsymbol{Q}}_k\right|=\frac{\sigma_{n_k}^2|\boldsymbol{\Sigma}|}{\sigma_{n_k}^2+{\boldsymbol{\psi}}_k^T\boldsymbol{\Sigma}\boldsymbol{\psi}_k},
\end{equation*}
and therefore $s_{2k}\!=\!\frac{1}{\sqrt{2\pi\left(\sigma_{n_k}^2+{\boldsymbol{\psi}}_k^T\boldsymbol{\Sigma}\boldsymbol{\psi}_k\right)}}$. One can also use the Binomial Inversion Lemma \cite{Matrix_Analysis} to compute ${\boldsymbol{Q}}_k$ in \eqref{definition of inv of matrix Q and omega} as:
\vspace{-.1cm}
\begin{equation} \label{simplified Q_k}
{\boldsymbol{Q}}_k=\boldsymbol{\Sigma}-\frac{\boldsymbol{\Sigma}\boldsymbol{\psi}_k{\boldsymbol{\psi}}_k^T\boldsymbol{\Sigma}}{\sigma_{n_k}^2+{\boldsymbol{\psi}}_k^T\boldsymbol{\Sigma}\boldsymbol{\psi}_k}.
\end{equation}
Substituting \eqref{simplified Q_k} in \eqref{definition of inv of matrix Q and omega} and \eqref{I2 integral after some manipulation}, we obtain:
\vspace{-0.2cm}
\begin{equation*} 
{\mathcal I}_{k,l}^{2}=s_{2k}\int\limits_{u_{k,l}}^{u_{k,l+1}}\exp\{-\frac{x_k^2}{2(\sigma_{n_k}^2+{\boldsymbol{\psi}}_k^T\boldsymbol{\Sigma}\boldsymbol{\psi}_k)}\}dx_k.
\end{equation*}
From the definition of $\boldsymbol{\psi}_k$ we have
${\boldsymbol{\psi}}_k^T\boldsymbol{\Sigma}\boldsymbol{\psi}_k\!=\!\mathbf{a}_k^T\boldsymbol{\mathcal C}_{\boldsymbol{\theta}}\mathbf{a}_k$. Having 
$\sigma_k$
from \eqref{definition of sigma_k and rho_ij}, we conclude:
%
\begin{equation*} 
{\mathcal I}_{k,l}^{2}=Q(\frac{u_{k,l}}{\sigma_k})-Q(\frac{u_{k,l+1}}{\sigma_k}).
\end{equation*}
Taking a similar approach, we can calculate $\boldsymbol{\mathcal I}_{k,l}^{1}$ in \eqref{E_theta_mhat_k_final} and ${\mathcal I}_{i,j,l_1,l_2}^{3}$ in \eqref{E_mhat_i_mhat_j}.
\bibliographystyle{IEEETran}
\bibliography{myref_26}

\vspace{-1.2cm}
\begin{IEEEbiography}[{\includegraphics[width=1in,height=1in,clip,keepaspectratio]{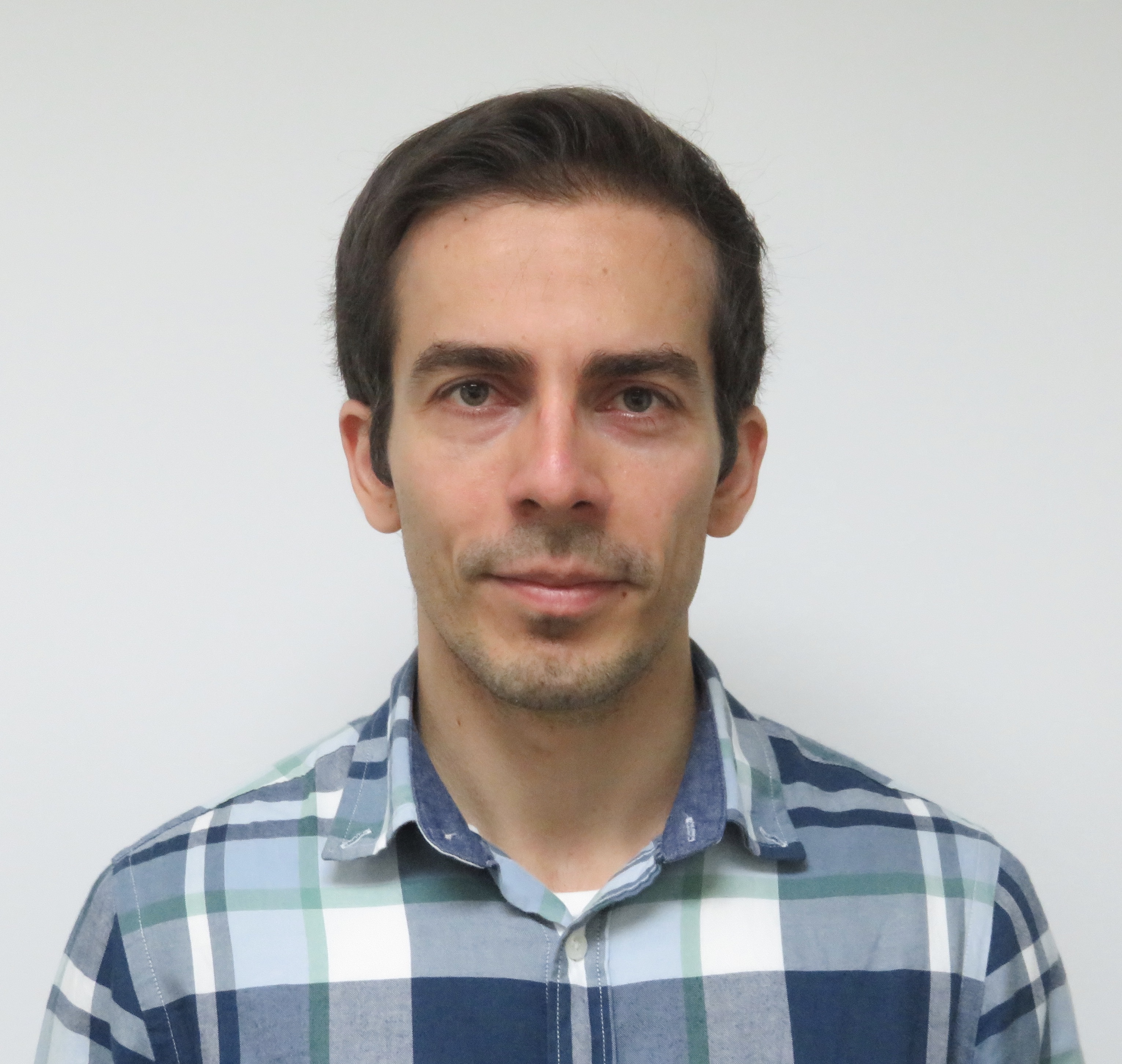}}]{Mojtaba Shirazi}
	received his B.Sc. degree
	from the Shiraz University of Technology, Shiraz, Iran, in 2009 
	and his M.Sc. degree from the Amirkabir University of Technology, Tehran, Iran, in 2012, both in electrical engineering.
	He is currently working toward his Ph.D. in electrical engineering at the University of Central Florida.
	Mr. Shirazi's current research interests include statistical signal processing, distributed estimation and detection in wireless networks. 
\end{IEEEbiography}
\vspace{-1.6cm}
\begin{IEEEbiography}[{\includegraphics[width=1.8in,height=1.3in,clip,keepaspectratio]{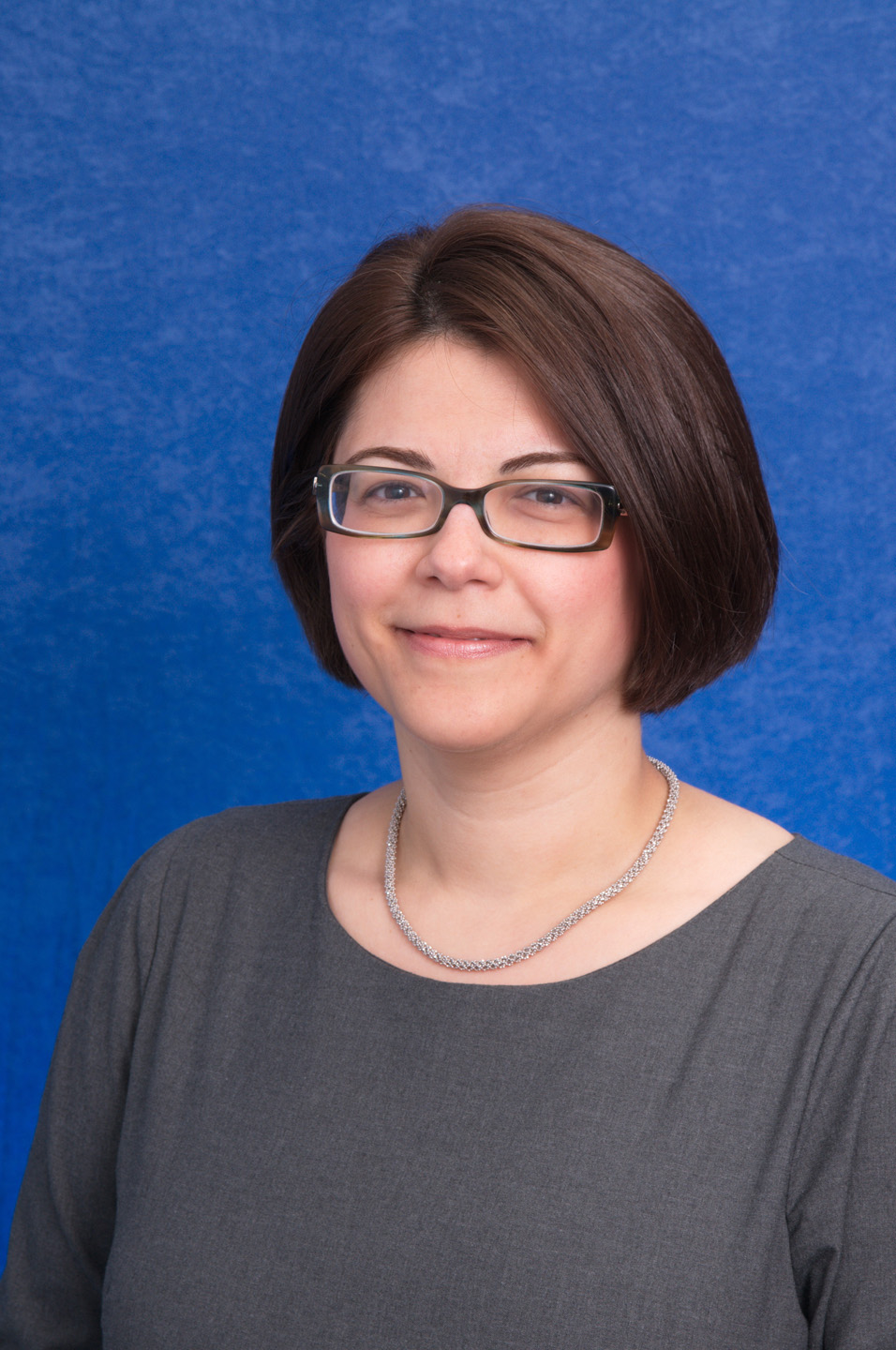}}]{Azadeh Vosoughi}
	(M' 06, SM' 14) is an Associate Professor in the Department of Electrical and Computer Engineering at the University of Central Florida. She received her B.S. degree from Sharif
	University of Technology, Tehran, Iran, in 1997; her MS degree from Worcester Polytechnic Institute, Worcester, MA, in 2001; and her Ph.D. degree from Cornell University, Ithaca, NY, in 2006, all in Electrical Engineering. Her research interests lie in the general areas of wireless communications, statistical signal processing, distributed detection and estimation theory, and brain signal processing. Dr. Vosoughi received the NSF CAREER award in 2011.  
\end{IEEEbiography}
\end{document}